# The Pale Orange Dot: The Spectrum and Habitability of Hazy Archean Earth

Giada Arney[1,2,3,4], Shawn D. Domagal-Goldman[2,5], Victoria S. Meadows[1,2,3], Eric T. Wolf[6], Edward Schwieterman[2,7], Benjamin Charnay[8], Mark Claire[2,9,10], Eric Hébrard[11], Melissa G. Trainer[5]

**Affiliations:**
[1]University of Washington Astronomy Department, Box 351580, U.W. Seattle, WA 98195

[2]NASA Astrobiology Institute Virtual Planetary Laboratory, Box 351580, U.W. Seattle, WA 98195

[3]University of Washington Astrobiology Program, Box 351580, U.W. Seattle, WA 98195

[4]Now at: NASA Goddard Space Flight Center, 8800 Greenbelt Road, Greenbelt, MD 20771

[5]NASA Goddard Space Flight Center, 8800 Greenbelt Road, Greenbelt, MD 20771

[6]University of Colorado at Boulder, Department of Atmospheric and Oceanic Sciences, Laboratory for Atmospheric and Space Physics, 1234 Innovation Drive, Boulder, CO 80303
[7]University of California at Riverside 900 University Ave, Riverside, CA 92521

[8]Paris-Meudon Observatory 61 Avenue de l'Observatoire, 75014 Paris, France

[9]Blue Marble Space Institute of Science 1001 4th Ave Suite 3201, Seattle, WA 98154

[10]University of St. Andrews Department of Earth and Environmental Sciences, Irvine Building, North Street, St Andrews, KY16 9AL, UK

[11]University of Exeter Prince of Wales Road Exeter, Devon UK EX4 4SB

**Abstract**

Recognizing whether a planet can support life is a primary goal of future exoplanet spectral characterization missions, but past research on habitability assessment has largely ignored the vastly different conditions that have existed in our planet's long habitable history. This study presents simulations of a habitable yet dramatically different phase of Earth's history, when the atmosphere contained a Titan-like, organic-rich haze. Prior work has claimed a haze-rich Archean Earth (3.8-2.5 billion years ago) would be frozen due to the haze's cooling effects. However, no previous studies have self-consistently taken into account climate, photochemistry, and fractal hazes. Here, we demonstrate using coupled climate-photochemical-microphysical simulations that hazes can cool the planet's surface by about 20 K, but habitable conditions with liquid surface water could be maintained with a relatively thick haze layer ($\tau \sim 5$ at 200 nm) even with the fainter young sun. We find that optically thicker hazes are self-limiting due to their self-shielding properties, preventing catastrophic cooling of the planet. Hazes may even enhance planetary habitability through UV shielding, reducing surface UV flux by about 97% compared to a haze-free planet, and potentially allowing survival of land-based organisms 2.6-2.7 billion years ago. The broad UV absorption signature produced by this haze may be visible across interstellar distances, allowing characterization of similar hazy exoplanets. The haze in Archean Earth's atmosphere was strongly dependent on biologically-produced methane, and we propose that hydrocarbon haze may be a novel type of spectral biosignature on planets with substantial levels of $CO_2$. Hazy Archean Earth is the most alien world for which we have geochemical constraints on environmental conditions, providing a useful analog for similar habitable, anoxic exoplanets.

**Key Words:** haze, Archean Earth, exoplanets, spectra, biosignatures, planetary habitability

## 1. Introduction

Early in Earth's history, an anoxic atmosphere could have supported the formation of an organic haze (Pavlov et al. 2001a; Trainer et al. 2004; Trainer et al. 2006; DeWitt et al. 2009; Hasenkopf et al. 2010; Zerkle et al. 2012; Kurzweil et al. 2013; Claire et al. 2014; Izon et al. 2015) that strongly interacted with visible and UV radiation, cooling the planet's climate (Pavlov et al. 2001b; Haqq-Misra et al. 2008; Domagal-Goldman et al. 2008; Wolf & Toon 2010; Hasenkopf et al. 2011). This hydrocarbon haze, generated by methane ($CH_4$) photolysis, would have formed when the ratio of $CH_4$ to carbon dioxide ($CO_2$) in the atmosphere exceeded about 0.1 (Trainer et al. 2006).

Unlike the hazes that may exist around exoplanets with thick hydrogen-dominated atmospheres (Sing et al. 2011; Kreidberg et al. 2014; Knutson et al. 2014), the Archean (3.8-2.5 billion years ago) haze was likely biologically-mediated via $CH_4$ produced from methanogenesis, one of the earliest metabolisms (Woese & Fox 1977; Ueno et al. 2006). In addition, several abiotic processes including serpentinization (the hydration of ultramafic rocks, mainly olivine and pyroxenes) can form methane (Kelley et al. 2005; Guzmán-Marmolejo et al. 2013; Etiope & Sherwood Lollar 2013), but the biotic flux of methane to the Archean atmosphere was likely much higher than the abiotic flux (Kharecha et al. 2005), as it is on Earth today. While the climatic effects of this haze have been studied (e.g. Pavlov et al. 2001b), impacts of haze on Archean Earth's habitability have not been previously investigated using tightly coupled climate-photochemical models. This coupling is critical to consider because of potential feedbacks between the impact of temperature on haze formation and the effects of haze on the atmosphere's temperature structure. Additionally, although we anticipate planetary diversity in the exoplanet population, existing spectral studies are largely focused on the observables of modern day Earth (e.g. Sagan et al. 1993; Woolf et al. 2002; Robinson et al. 2011; Robinson et al. 2014a). Those spectral studies that consider Archean Earth and anoxic planets have not examined hazes (Meadows 2006; Kaltenegger et al. 2007; Domagal-Goldman et al. 2011). As we will show, hydrocarbon haze has profound spectral impacts for both reflected light and transit transmission spectra.

**1.1 Evidence for an Archean Haze**

Geochemical data suggest 3-5 distinct intervals of organic haze during the later Archean (Zerkle et al. 2012; Izon et al. 2015), supporting theoretical studies on the causes and consequences of photochemical haze formation in the atmosphere (Pavlov et al. 2001a; Pavlov et al. 2001b; Haqq-Misra et al. 2008; Domagal-Goldman et al. 2008; Kurzweil et al. 2013; Claire et al. 2014) as well as experimental data (Trainer et al. 2004; Trainer et al. 2006; DeWitt et al. 2009; Hasenkopf et al. 2010; Hasenkopf et al. 2011) and theory on their potential radiative effects (Wolf & Toon 2010). The geochemical evidence, described below, implies Neoarchean hazy intervals (Zerkle et al. 2012; Izon et al. 2015) lasting for less than 1 million years. The constraint on the duration of these intervals based on the lower limit of shale sedimentation rates.  In addition, the modeling work of Domagal-Goldman et al. (2008) suggests a longer Meosarchean to Neoarchean hazy period between 3.2 and 2.7 Ga.

Here, we present an overview of the evidence for the Archean haze. The line of evidence most often invoked comes from analyses and modeling of sulfur isotope fractionation data from Earth's rock record. Several studies have proposed links between haze and the mass independent sulfur isotope fractionation signal (S-MIF) (Farquhar et al. 2000) preserved in the geologic record before the Great Oxygenation Event (GOE) at about 2.5 billion years ago (Ga) (Domagal-Goldman et al. 2008; Zerkle et al. 2012; Kurzweil et al. 2013; Claire et al. 2014; Izon et al. 2015). We present a brief review of this evidence here, beginning with an overview of sulfur mass-independent fractionation, on which much of the evidence for an Archean haze is based.

Sulfur has four stable isotopes: $^{32}S$, $^{33}S$, $^{34}S$ and $^{36}S$. Isotope fractionations are reported in part per thousand (‰) using delta notation (δ) such that:

$$\delta\, ^{x}S = \left[\frac{^{x}R_{sample}}{^{x}R_{standard}} - 1\right] \times 10^{3} [‰] \quad (1)$$

Here, $^{x}R_{sample}$ represents isotope ratios of the given minor to major isotope (for sulfur, $^{x}R$ means $^{x}S/^{32}S$ with x = 33, 34, 36) of sampled material. $^{x}R_{standard}$

represents isotope ratios of a standard reference material.

Reactions following classical equilibrium or kinetic behaviors produce isotope fractionation that depend only on the mass differences of the isotopes such that the $\delta^{33}S$ composition of a material is approximately half the $\delta^{34}S$ amount, and the $\delta^{36}S$ composition is roughly twice the $\delta^{34}S$ amount. For elements with more than two stable isotopes, mass dependent fractionation (MDF) quantifies this expected three-isotope relationship, and samples following MDF will have $\delta^{33}S \sim 0.515\times\delta^{34}S$ and $\delta^{36}S \sim 1.89\times\delta^{36}S$.

Mass-independent fractionation (MIF) occurs when samples deviate from this expected three isotope behavior and is quantified with 'capital delta' notation where $\Delta^{33}S = \delta^{33}S - 0.515\times\delta^{34}S$ and $\Delta^{36}S = \delta^{36}S - 1.89\times\delta^{36}S$. MIF in naturally occurring samples is very unusual and is generally diagnostic of quantum chemistry such as can occur in certain atmospheric reactions. While the precise mechanism(s) that produce S-MIF are unknown, photolysis of sulfur gases in an anoxic atmosphere is the only known mechanism that produces large magnitude $\Delta^{33}S$ and $\Delta^{36}S$ seen in the rock record (Farquhar et al. 2001; Farquhar et al. 2007).

The S-MIF signal is variable throughout the Archean and it vanishes completely once $O_2$ builds up to nonnegligible levels in the atmosphere after the great oxygenation event (GOE) at 2.5 Ga. Its recurrence at both ends of the Archean Eon implies that, within 0.8 billion years of Earth's formation, a common mechanism for S-MIF production was already established in the atmosphere (Thomassot et al. 2015). After the GOE, $O_2$ and the ozone ($O_3$) derived from $O_2$ photochemical reactions block the UV photons necessary to photolyze sulfur gases and produce S-MIF. Also, $S_8$ is the most important species to rain out S-MIF from the atmosphere; because a more reducing atmosphere enhances the ability of $S_8$ to polymerize, S-MIF is more easily preserved under reducing conditions (Zahnle et al. 2006). After the GOE, all the sulfur in the atmosphere would have been oxidized into a single exit channel, eliminating any fractionation created in the atmosphere (Pavlov & Kasting 2002). Thus, S-MIF is generally regarded as robust evidence for an anoxic Archean atmosphere.

$\Delta^{33}S$ typically correlates with enrichments in $\delta^{34}S$ and with depletions in $\delta^{36}S$, and variations in magnitude and sign of these isotopic signals in Earth's geologic record hint that strong constraints on Archean atmospheric chemistry will be possible when the precise MIF formation mechanisms are identified (Claire et al. 2014). $\Delta^{36}S/\Delta^{33}S$ in Archean sedimentary rocks is generally around -1, but stratigraphic variations in this slope have been observed in the geological record and interpreted as evidence of changes to the S-MIF production mechanism resulting from changes in atmospheric composition (Zerkle et al. 2012; Kurzweil et al. 2013; Izon et al. 2015), suggesting the influence of haze.

Domagal-Goldman et al. (2008) and Haqq-Misra et al. (2008) studied potential links between S-MIF, hazes, and Archean glaciation. At ~2.9 Ga, there is geological evidence suggesting a glaciation event (Young et al. 1998) may have occurred during the same period when the S-MIF $\Delta^{33}S$ signal dips to lower values. An upper atmosphere haze that decreased tropospheric $SO_2$ photolysis by blocking UV photons and cooled the planet could explain both the glaciation and the decrease in S-MIF. In this conceptual model, the end of the cold period typified by low $\Delta^{33}S$ may be due to a decrease in the atmospheric $CH_4/CO_2$ ratio, which would have cleared any haze present in the atmosphere. If true, this change in atmospheric composition and radiative scattering would have enabled UV photons to penetrate deeper into the atmosphere, interacting with sulfurous gases and affecting their isotopic signatures (Claire et al. 2014). Earth's record of sedimentary sulfates do show a significant change in their minor sulfate isotope behavior between 2.73 and 2.71 Ga (Kurzweil et al. 2013; Izon et al. 2015) that may in fact reflect this change, although predictive models of sulfur isotope fractionation are not yet able to reproduce these trends seen in the rock record (Claire et al. 2014).

Zerkle et al. (2012) discusses the discovery geochemical evidence consistent with the Archean haze hypothesis. The authors analyzed sediments aged 2.5-2.65 Ga collected from the Ghaap Group in South Africa and showed that variations of $\Delta^{36}S/\Delta^{33}S$ associated with changes in atmospheric chemistry were contemporaneous with highly negative excursions of $\delta^{13}C_{org}$ values. Negative values of $\delta^{13}C_{org}$ below -37 ‰ are typically interpreted as evidence for methanogenesis (biological methane production) followed by subsequent incorporation into sediments by methanotrophy (methane consumption) which imparts a strongly negative $\delta^{13}C_{org}$ because organisms preferentially uptake the $^{12}C$ (Urey & Greiff 1935; Schopf 1983; Schidlowski 2001; Eigenbrode & Freeman 2006). The contemporaneous excursions of the sulfur and $C_{org}$ isotopes suggests a close linkage between S-MIF signals and biogenic methane. The links between S-MIF signals and biogenic methane production have been recently expanded over multiple cores and locations, suggesting multiple changes in atmospheric chemistry during the Neoarchean (Izon et al. 2015). Changes observed in the slope of $\Delta^{36}S/\Delta^{33}S$ vary between -1.5 and -0.9 and are interpreted to reflect changes in the S-MIF source reactions driven by varying atmospheric haze thicknesses.

Kurzweil et al. (2013) note that an increase in magnitude of S-MIF signals after 2.73 Ga (Thomazo et al. 2009) occurred during a prolonged negative shift in $\delta^{13}C_{org}$, suggesting enhanced biological methane activity at this time. Similar to Zerkle et al. (2012), they discuss a change in the slope of $\Delta^{36}S/\Delta^{33}S$ from -1.5 to -0.9 at 2.71 Ga and interpret this to be caused by a decrease in the $CH_4/CO_2$ ratio at 2.71 Ga, possibly indicating an organic haze was present for some period of time prior to 2.71 Ga and cleared afterward. In this interpretation, haze-free and reducing atmospheric conditions dominated after 2.71 Ga, with haze reappearing in brief intervals of time as suggested by the Zerkle et al. (2012) and Izon et al. (2015) measurements.

Given the apparent occurrence of haze in the Archean, we investigated the impact of this haze on the climate, spectral appearance and surface UV flux by simulating the hazy Archean environment with boundary conditions consistent with recent geochemical constraints. Unlike previous studies of the Archean climate under a haze, we use realistic fractal (rather than spherical) particles, which have different spectral properties and climatic effects. Our study also represents the first time temperature feedbacks have been investigated in relation to haze production in Archean Earth's atmosphere. Previous studies (Pavlov et al. 2001b; Haqq-Misra et al. 2008; Domagal-Goldman et al. 2008) involving climate modeling have included the haze's impact on temperature, but not corresponding temperature feedbacks on haze formation. Temperature feedbacks have significant impacts on the resultant hazes: as we discuss below, hazes produce stratospheric temperature inversions, and warmer atmospheres produce larger haze particles, so hazes generated by chemistry models without temperature feedbacks may not produce realistic results.

## 2. Models and Methods

To simulate the hazy Archean environment with boundary conditions consistent with recent geochemical constraints, we used a coupled 1D photochemical-climate model we call Atmos and a 1-D radiative transfer model, SMART (Spectral Mapping Atmospheric Radiative Transfer model).

### 2.1 Coupled Photochemical-Climate Model

Our coupled photochemical-climate model, Atmos, is used to simulate Archean Earth's photochemistry and climate. To use Atmos, the photochemical model (which includes particle microphysics) is run first to generate an initial atmospheric state based on user-specified boundary conditions (gas mixing ratios or fluxes, the solar constant at 2.7 Ga (Claire et al. 2012), the stellar spectral type, total atmospheric pressure, the initial temperature-pressure profile). Then, the output files from the photochemical model for altitude, pressure, gas mixing ratios, haze particle sizes and haze number densities are passed into the climate model. The climate model uses the photochemical model's solution as its initial conditions and runs until it reaches a converged state. It then feeds updated temperature and water vapor profiles back into the photochemical model. The models iterate back and forth in this manner until convergence is reached. An example of Atmos finding convergence can be seen in Figure 1.

#### 2.1.1 Photochemical Model

The photochemical portion of the code is based on the 1-D photochemical code developed originally by Kasting et al. (1979), but the version we use here was significantly modernized and updated in Zahnle et al. (2006) and uses the haze formation scheme described in Pavlov et al. (2001b). It was modified by E. Wolf to include fractal hydrocarbon hazes following the methods presented in Wolf & Toon (2010), and was first used to study fractal hazes on Archean Earth in Zerkle et al. (2012). Note that the version of the model used here can simulate atmospheres ranging from extremely anoxic ($pO_2 = 10^{-14}$) to modern-day $O_2$ levels (Zahnle et al. 2006). Subsequent studies using this model or other versions of it to study fractal haze formation include Harman et al. (2013), Kurzweil et al. (2013), and Claire et al. (2014), with the latter two of these studies also derived from the same Zahnle et al. (2006) model branch used here. This model also has a long heritage of being used to study photochemistry in non-hazy atmospheres (e.g. Kasting & Donahue 1980; Pavlov et al. 2002; Ono et al. 2003; Segura et al. 2003; Segura et al. 2005; Zahnle et al. 2006; Grenfell et al. 2007; Segura et al. 2007; Catling et al. 2010; Segura et al. 2010; Domagal-Goldman et al. 2011; Rugheimer et al. 2013; Domagal-Goldman et al. 2014; Rugheimer et al. 2015; Harman et al. 2015; Schwieterman et al. 2016).

The photochemical model parameters are as follows. Our model atmosphere is divided into 200 plane-parallel layers from the surface to 100 km, with a layer spacing of 0.5 km. We show a list of chemical reactions in our Supplementary Table 1. Our Archean scheme includes 76 chemical species, 11 of which are short-lived (Supplementary table 2). Short-lived species are considered in photochemical equilibrium (i.e. their atmospheric transport is neglected) and are not part of the Jacobian solved self-consistently at each timestep. The mixing ratio of each species is found by solving flux and mass continuity equations in each layer simultaneously using a reverse-Euler method, providing exact solutions at steady-state. Vertical transport by molecular and eddy diffusion are included, and boundary conditions which drive the model can be set for each species at the surface and the top of the atmosphere. A δ 2-stream method is used for radiative transfer (Toon et al. 1989). Fixed isoprofiles are assumed for $CO_2$ and $N_2$ in the atmospheres considered here.

Similarly to Zerkle et al. (2012), we set a fixed mixing ratio of $CH_4$ at the surface; the model then calculates the surface flux necessary to maintain this mixing ratio. Since haze formation scales with the $CH_4/CO_2$ ratio, we find this is the most straightforward way to explore haze thicknesses in our atmospheres. Note that when we discuss $CH_4/CO_2$ values in this study, these refer to the ratio at the planetary surface because $CH_4$ does not follow an isoprofile.

Aerosol formation follows the method used in Kasting et al. (1989) and described and updated in Pavlov et al. (2001b). Immediate precursors to haze particles are formed through the reactions $C_2H + C_2H_2 \rightarrow C_4H_2 + H$ and $C_2H + CH_2CCH_2 \rightarrow C_5H_4 + H$. Since the full chemical scheme that leads to aerosol formation is not well understood despite both laboratory and theoretical studies (e.g. Hallquist et al. 2009; Hicks et al. 2015), it is assumed that $C_4H_2$ and $C_5H_4$ condense directly to haze particles (called HCAER and HCAER2 in Supplementary Table 1). In a real atmosphere, the molecules would be larger before aerosols condense, and back-reactions should occur, so this model may overestimate the rate of aerosol formation. In Pavlov et al. (2001b), the authors suggest that if the real aerosol formation rate was slower, the atmosphere would compensate by increasing the $CH_4/CO_2$ ratio, which would increase the polymerization rate. Further discussion of haze formation pathways and caveats of the approach we use here can be found in section 4.4. The model's particles form initially with a radius of 0.001 μm. Each layer of the atmosphere has a monomodal size distribution calculated by comparing the coagulation lifetime to the particle removal lifetime via diffusion into another layer or by sedimentation. The aerosols can grow when the coagulation lifetime is longer than the lifetime for removal in a layer.

The maximum radius of a spherical haze particle (i.e. a haze "monomer") is set to 0.05 μm, the same nominal value used by Wolf and Toon (2010) and similar to the size of the monomers of Titan's fractal haze aggregates (Rannou et al. 1997; Tomasko et al. 2008). Particles larger than this size are treated as fractal agglomerates of $n_{mon}$ spherical monomers of radius $R_{mon}$ that clump into a larger aggregate with an effective geometric radius $R_f$ given by the relation,

$$n_{mon} = \alpha \left(\frac{R_f}{R_{mon}}\right)^{D_f} \quad (2)$$

Here, $\alpha$ represents a dimensionless constant of order unity, and $D_f$ is the "fractal dimension," which can take on values between 1 and 3. $D_f = 3$ represents a spherical (non-fractal or classical Mie) particle, while $D_f = 1$ represents a string of linearly chained monomers. Titan's fractal aggregates are thought to have a fractal dimension of about 2 on average for the aerosol population (Rannou et al. 1997; Larson et al. 2015). Note that the "effective geometric radius" we refer to above is used only to conceptualize the size of a fractal particle and does not indicate that we use Mie scattering for our fractal particles; with the exception of sub-monomer sized particles ($R \leq 0.05$ μm) which remain spherical and thus Mie, we use the mean field approximation for fractal scattering physics for all particles (Botet et al. 1997). The model's fractal production methods are discussed in Zerkle et al. (2012) (including its supplementary online information) where they were first implemented. Additional information about fractal particles and their geometry can be found in, e.g., Köylü et al. (1995) and Brasil et al. (1999). The mean field approximation we use for fractal scattering has been validated against scattering by silica fractal aggregates (Botet et al. 1997) and Titan's hazes (Rannou et al. 1997; Larson et al. 2015).

As in Wolf and Toon (2010), the fractal dimension of our particles varies from 1.5 to 2.4 for aggregate particles, and larger aggregates have a larger fractal dimension to account for folding as the particles coagulate. In general, compared to spherical particles, fractal particles produce *more* extinction in the ultraviolet (UV) but *less* in the visible and near infrared (NIR). In addition, fractals tend to be more forward scattering in the visible and NIR and more isotropically scattering in the UV compared to equal mass spherical particles. Their weakened visible extinction and enhanced forward scattering compared to spherical particles means they produce less cooling since they scatter less incident sunlight back to space (see Figure 3 in Wolf and Toon 2010). Figure 2 shows the extinction efficiency ($Q_{ext}$) and single-scattering albedo of different fractal particle sizes together with the haze optical constants we adopt in this study (Khare et al. 1984a). A discussion of our choice of optical constants and comparison to others in the literature can be found in section 4.5.

In the version of the photochemical model used here, we corrected an error relating to the calculation of the number of $C_5H_4$ molecules composing HCAER2 haze particles. Previously, the model calculated the number of molecules per HCAER2 particle inappropriately using the mass of $C_4H_2$ instead of $C_5H_4$. In addition, we added more particle sizes to the model's scattering grid, increasing the number from 34 particle sizes to 51, and we added options to use different monomer sizes and optical constants than the ones used here for our nominal haze study; how variation of these parameters impact haze formation is a subject of future work. Gas mixing ratios at the surface can be more finely tuned than in previous versions of the model from the addition of a significant figure to the species boundary conditions input file.

The photochemical model is considered converged when redox is conserved and a re-run of the model using last run's output as initial conditions occurs quickly (i.e. < 50 timesteps).

### 2.1.2 Climate Model

Our climate model was originally developed by Kasting and Ackerman (1986). The model we use here has evolved considerably since its first incarnation and versions of it have been applied in subsequent studies on varied topics such as the habitable zones for several stellar spectral types (Kopparapu et al. 2013), the climate of early Mars (Ramirez et al. 2013), the atmospheres of Earth-like planets around various stellar types (Segura et al. 2003; Segura et al. 2010; Segura et al. 2005; Rugheimer et al. 2013), clouds in exoplanet atmospheres (Kitzmann et al. 2010; Kitzmann et al. 2011a), and the climate of early Earth (Haqq-Misra et al. 2008). The version we use here is based directly on that used by Kopparapu et al. (2013). It uses a correlated-k method to compute absorption by spectrally active gases ($O_3$, $CO_2$, $H_2O$, $O_2$, $CH_4$, and $C_2H_6$). This model has $CO_2$ and $H_2O$ correlated k coefficients updated as described in Kopparapu et al. (2013). Our older $CH_4$ coefficients may overestimate the surface temperature by $\lesssim 5$ K at the $CH_4$ mixing ratios used here (Byrne & Goldblatt 2015). However, as we discuss in section 4.2, our model under-predicts the Archean temperature by about 2-5 K compared to 3D climate models with more complete physics describing the planetary system, so these two effects may cancel each other out. The aforementioned gas profiles are passed to the climate model from the photochemical model when running in coupled mode. The net absorbed solar radiation in each layer of the atmosphere is computed using a δ2-stream multiple scattering algorithm (Toon et al. 1989) spanning from $\lambda = 0.2$ to 4.5 μm in 38 spectral intervals. For net outgoing IR radiation, we use a separate set of correlated-k coefficients for each gas in 55 spectral intervals spanning wavenumbers of 0 – 15,000 $cm^{-1}$.

We have made several modifications to the climate model used here. The model previously incorporated the spectral effects of spherical hydrocarbon particles, and it has been updated in our study to include fractal hydrocarbon scattering efficiencies using the mean field approximation of Botet et al. (1997) discussed previously. We have also updated the model so that haze profiles can be passed to it from an input file or by the photochemical code; in previous versions of the climate model, haze distributions were hardcoded and had to be edited manually. We corrected a discrepancy in the spacing between atmospheric layers in the routine that outputs coupling files for the photochemical model: our photochemical model layer spacing is 0.5 km, but a layer spacing of 1 km had been hardcoded. Coupling subroutines have been improved to be able to accept information about atmospheric pressure, stellar parameters, and haze parameters as input from the photochemical model. We also added options to turn ethane opacity and 1D ice-albedo feedbacks (described in section 4.1.1) on or off.

We have been unable to run the climate model to convergence using the same top-of-atmosphere pressure used for the photochemical model: the photochemical model extends to 100 km, but we have only been able to successfully run the climate model up to about 80 km for our 1 bar atmospheres. Thus, when temperature and water profiles are passed from the climate model to the photochemical model, they become isoprofiles above the top of the climate grid based on the highest altitude temperature from the climate grid calculations. At these altitudes the atmosphere is thin, and the particles are very small; both of these effects lead to this portion of the atmosphere having little impact on radiative transfer and climate. We performed a sensitivity test of how the temperature at these altitudes affects the resultant haze distribution in the photochemical model, and the sizes of the largest haze particles produced by an atmosphere that becomes an 80 K isotherm above 80 km versus a 150 K isotherm differ by less than 5%. In the climate model, shifting the particles in figure 1 above 80 km down to lower altitudes alters the surface temperature by < 0.5 K.

The climate model is considered converged when the change in temperature between timesteps and change in flux out the top of the atmosphere are sufficiently small (typically on the order of $1\times10^{-5}$).

**2.2 The SMART Model**

To generate synthetic spectra for the atmospheres we produce with Atmos, we feed outputs from the Atmos model (the temperature-pressure profile, gas mixing ratio profiles, and the haze particle profile), into the SMART code, a 1-D line-by-line fully multiple scattering radiative transfer model (Meadows & Crisp 1996; Crisp 1997). SMART has been validated against observations of multiple solar system planets (Robinson et al. 2011; Arney et al. 2014). The Line-by-Line Absorption Coefficients (LBLABC) code, a companion to SMART, creates line-by-line absorption files for input gas mixing ratios and temperature-pressure profiles using HITRAN 2012 line lists (Rothman et al. 2013). SMART can also incorporate aerosols: as input, it requires "cloud files" with altitude-dependent opacities as well as the particle asymmetry parameter and the extinction, scattering, and absorption efficiencies ($Q_{ext}$, $Q_{scat}$, and $Q_{abs}$). For spherical particles (our small monomers), we use the code "Miescat," to calculate these efficiencies using the indices of refraction measured by Khare et al. (1984a). For fractal hydrocarbon particles, we use scattering inputs from the Wolf and Toon (2010) photochemical study generated with the fractal mean field approximation (Botet et al. 1997). Spherical particles use a full Mie phase function, while fractal particles employ a Henyey-Greenstein phase function (Henyey & Greenstein 1941). To generate transit transmission spectra, we use the SMART-T model (Misra et al. 2014a; Misra et al. 2014b). This version of SMART uses the same inputs as the standard code but simulates the longer path lengths and refraction effects associated with transit transmission observations.

To create SMART cloud files from Atmos haze outputs, we have written a script that bins the haze particles generated by the photochemical model into specified radii (also called particle "modes") while preserving the total mass of each atmospheric layer. The particle mode sizes we use span from 0.001 μm - 2 μm; larger particles do not exist in our atmospheres due to rainout. Spherical modes are R= 0.001 μm, 0.005 μm, 0.01 μm, and 0.05 μm. Fractal modes are R = 0.06 – 2 μm with 4 modes between 0.06 and 0.1, 10 equally spaced modes between 0.1 μm and 1 μm, and 2 μm. In total, this represents 19 particle modes.

In each layer of the SMART cloud files, we include a mixture of two particle modes; the mass density contributed by the two modes is selected based on the distance in log space of the Atmos particle radius to each neighboring SMART size bin. For example, if Atmos produces a particle of radius 0.33 μm in a layer, the corresponding layer in SMART will include 0.3 μm and 0.4 μm particles each comprising 50% of the layer's mass. This binning is necessary because the photochemical model generates many dozens of finely differentiated haze particle radii, but SMART model runtime with this many particle sizes is infeasible.

Once we have binned the Atmos particle radii to our SMART size grid, we must compute the total optical depth from each particle mode at a reference wavelength in each atmospheric layer. We arbitrarily select 1 μm as our reference wavelength. Optical depth in a layer, τ, from particles of a given radius, R, depends the number density of particles per particle size, $n(R)$, the thickness of the atmospheric layer, z, and the wavelength-dependent extinction efficiency, $Q_{ext}$:

$$\tau = z \int_{R_{min}}^{R_{max}} \pi r^2 Q_{ext}(\lambda, R) n(R) dR \ (3)$$

For fractal particles (R > 0.05 μm), the cross sectional area and the corresponding extinction efficiencies are computed relative to the radius of an equal mass spherical particle, following the conventions of mean-field approximation (Botet et al. 1997). Spherical particles in SMART are binned according to log-normal size distributions using the radii mentioned previously and a mode standard deviation of 1.5, which is realistic for an aerosol distribution (Tolfo 1977). For fractal particles, we use a monodisperse distribution, the same size distribution used to compute our inputs from the previous Wolf & Toon (2010) fractal haze study and the same distribution used in the Atmos model.

**2.3 Model Inputs**

In the photochemical model, we set a haze monomer density of 0.64 g/cm$^3$, which is consistent with the laboratory results of Trainer et al. (2006) for early Earth. This density is used in the model to calculate the masses of haze particles and is updated from the value of 1 g/cm$^3$ used by previous studies employing our photochemical model. Hörst & Tolbert (2013) measured a similar effective particle density, 0.65 g/cm$^3$, for a 0.1% $CH_4$ haze experiment using a UV lamp. 0.1% $CH_4$ is consistent with the atmospheres we simulate, although the Hörst and Tolbert hazes were Titan-analog simulants lacking the $CO_2$ present in the Trainer et al. experiments. We apply a Manabe/Wetherald relative humidity model for the troposphere (Manabe & Wetherald 1967) with a surface relative humidity of 0.8 in both the climate and photochemistry models. This humidity parameterization is further described in Pavlov et al. (2000). Our Archean simulations use the solar constant at 2.7 Ga (0.81 = $S/S_0$, where $S_0$ is the modern solar constant and S is the solar constant at 2.7 Ga) modified by a wavelength-dependent solar evolution correction (Claire et al. 2012). We chose this time because it corresponds to the age of the constraints on $CO_2$ used by our study (Driese et al. 2011). We set the mixing ratio of $O_2$ at the surface to $1.0x10^{-8}$, consistent with the Zerkle et al. (2012) study. These conditions reflect the time period after the evolution of oxygenic photosynthesis but prior to Earth's GOE in which substantial biogenic fluxes of both oxygen and methane would have vented into a predominantly reducing atmosphere (Claire et al. 2014). Unless otherwise specified, the surface albedo used by the climate model is 0.32. This includes the effect of clouds, which is standard in this 1D treatment (Kopparapu et al. 2013) and is the albedo that reproduces the average temperature of present day Earth (288 K) with modern atmospheric conditions. Of course, the true cloud distribution on Archean Earth is unknown, and clouds may have had important climatic effects on our early planet (Goldblatt & Zahnle 2011). The solar zenith angles (SZA) used in the climate and photochemical models were chosen to best represent globally averaged behavior of the physics in each specific model, which Segura et al. (2003) finds as SZA = 45° in the photochemical model and SZA = 60° in the climate model. These zenith angles are both tuned to reproduce modern day Earth's average chemical profiles and climate, respectively.

For our SMART spectral simulations, our nominal spectra assume an ocean surface albedo (McLinden et al. 1997). In cases where an icy surface is used, we use an albedo from the USGS Digital Spectral Library (Clark et al. 2007). Our solar spectrum was modeled by Chance & Kurucz (2010), and was scaled by the solar evolution model (Claire et al. 2012) mentioned previously. The solar zenith angle is set at 60° for the reflection spectra, which approximates a planetary disk average near quadrature (planet half illuminated to the observer).

**3. Results**

In this section, we first describe the climate results from Atmos. Following this, we quantify the strength of a hazy UV shield for surface organisms, and we show and describe the spectral consequences of this haze in reflected light and transit transmission spectroscopy.

Recent paleosol measurements have constrained the $CO_2$ partial pressure ($pCO_2$) in the Archean at 2.7 Ga to be between 0.0036-0.018 bars (10-50× the present atmospheric level (PAL)) (Driese et al. 2011), while recent estimates of Archean surface pressure ($P_{surf}$) are consistent with values as low as 0.5 bars (Som et al. 2012; Marty et al. 2013). We simulated four types of atmospheres that span these constraints to examine a range of conditions: $pCO_2$ = 0.01 and $P_{surf}$ = 1 bar of total pressure (Case A), $pCO_2$ = 0.018 and $P_{surf}$ = 1 bar (Case B), $pCO_2$ = 0.01 and $P_{surf}$ = 0.5 bars (Case C), and lastly, $pCO_2$ = 0.0036 and $P_{surf}$ = 0.5 bars (Case D). These are summarized in Table 1. The haze thickness scales with the $CH_4$ abundance relative to $CO_2$, so we investigated a range of $CH_4$ levels for each of these atmospheres. In the sections below, we refer to these Case A-D planets. Figure 3 shows an example of the atmospheric profiles for several gases in atmospheres with two different $CH_4/CO_2$ ratios (0.1 and 0.2), plus the haze number density profiles scaled to fit on the same x-axis. The insignificant haze present in the $CH_4/CO_2$ = 0.1 atmosphere is spectrally indistinguishable from an atmosphere with no haze. The larger amounts of $CH_4$, $C_2H_6$, and $H_2O$ at higher altitudes in the $CH_4/CO_2$ = 0.2 atmosphere illustrates how the haze can shield these gases from photolysis.

Our results presented here required about 60 Atmos model runs. In total, we ran about twice this number for model debugging and testing. Each coupled Atmos run can take between 3-15 hours depending on how many coupling iterations are required. Note that the runtime for the climate model scales nonlinearly with the number of radiatively active gases: a model run that takes less than 20 minutes without $CH_4$ or $C_2H_6$ will require well over an hour with both of these gases turned on. All of the results presented here, except as noted in section 4, were generated with both $CH_4$ and $C_2H_6$.

Note that in the context of the results presented here, a “thick” haze refers to the haze at a $CH_4/CO_2$ ratio ~ 0.2.

### 3.1 Hazy Climates

We find that hazy Archean climates were cold but most likely habitable (Figure 4). Previous 1D climate modeling efforts assumed that planets with globally averaged surface temperatures ($T_{GAT}$, which is equivalent to our 1D surface temperature, $T_{surf}$) below 273 K will experience runaway glaciation (e.g. Haqq-Misra et al. 2008; Domagal-Goldman et al. 2008). However, more recent 3D studies have shown that Archean Earth can maintain an open ocean fraction of > 50% for $T_{GAT}$ >= 260 K and an equatorial open ocean belt for $T_{GAT}$ >= 248 K (Wolf & Toon 2013; Charnay et al. 2013). Furthermore, Abbot et al. (2011) argue that ocean open belts can remain climatologically stable, even if the ice latitude is reduced to 5-15°. Since a planet with any non-zero fraction of open ocean is habitable, we regard these updated globally-averaged temperatures - all of which are significantly below freezing - to be more realistic habitability thresholds than 273 K. We adopt $T_{GAT}$ >= 248 K as our habitability threshold here.

Figure 4 shows that when haze reaches a threshold thickness, further increases in $CH_4$ result in rapid increases in haze thickness, and a corresponding steep falloff in surface temperature. However, at higher $CH_4/CO_2$ ratios, the haze thickness (and the surface temperature) stabilizes because UV self-shielding inhibits methane photolysis, shutting down haze formation. Thus, we find there is a maximum haze optical thickness -- and a minimum temperature from haze-induced cooling -- for each atmosphere. Interestingly, this negative feedback haze self-shielding appears to prevent catastrophic cooling. Note that even using the conventional habitability threshold of 273 K, Cases A-C have a hazy solution space where $T_{surf} > 273$ K, and Case B stabilizes at $T_{surf} = 274$ K with its thickest haze. Using the updated habitability threshold of $T_{surf} > 248$ K, all of our cases even with thick hazes are habitable. Table 2 summarizes these results and includes a sensitivity test of the ice-albedo effect, described below.

Although the cold climates we have simulated are "habitable" in the sense that they have open ocean, a cold climate with extended ice caps ($T_{surf} < 273$ K) from a thick haze may be consistent with a reported glaciation event at 2.9 Ga (Young et al. 1998) as a previous study has suggested (Domagal-Goldman et al. 2008). Later purported hazy periods around 2.7 Ga (Kurzweil et al. 2013; Izon et al. 2015) and between 2.65-2.5 Ga (Zerkle et al. 2012) are not associated with glaciations and may be consistent with the thinner-haze solution space of Cases A and C or even the thickest haze solution space of the warmer Case B.

### 3.1.1 Ice-Albedo Feedback

To test how ice-albedo feedbacks can affect our retrieved temperatures, we tested the influence of these feedbacks on the minimum temperatures reached by our four Cases by parameterizing our model's 1D surface albedo (A) by the relation (based on the results of Charnay et al. (2013)) to include the albedo effect of clouds and ice as a function of the globally averaged temperature:

$$A(T_{GAT}) = 0.65 + (0.3 - 0.65)\times\left(\frac{T_{surf}-240}{290-240}\right)^{0.37} \quad (4)$$

As stated above, the surface albedo used by our nominal model is 0.32. The surface albedos for the Case A, B, C, and D minimum temperatures with this ice-albedo parameterization are 0.39, 0.35, 0.39, and 0.45. The climate model was run to convergence starting with the solution for the minimum stabilized temperature for each Case (i.e. when the haze becomes self-shielding and reaches maximal thickness) as a test of the sensitivity of our minimum temperatures to ice-albedo feedbacks. The temperatures of planets A, B, C, and D with ice albedo feedbacks are 257 K, 271 K, 257 K, and 241 K, a decrease of 3 to 10 K compared with simulations with the nominal albedo. The Bond albedos produced in these cases including haze are 0.26, 0.24, 0.26, and 0.29.

These ice-albedo temperatures may be under-estimates because once haze forms, deposition of dark hydrocarbons onto ice-covered areas will lower the albedo of the ice. This decreased ice albedo may then melt the ice, reverting parts of the surface back to ocean water. Because haze absorbs strongly at blue wavelengths, the radiation that reaches the surface under a haze would have a higher proportion of longer, redder wavelengths compared to shorter, bluer wavelengths. While ice is very reflective at visible wavelengths, it becomes more absorbing at wavelengths > 0.7 μm, changing the true ice-albedo parameterization. Because of this, planets orbiting stars emitting a high proportion of radiation at near-infrared wavelengths are harder to freeze (Shields et al. 2013). Additionally, stratospheric and mesospheric circulation patterns on Earth presently impact high-altitude aerosol distributions by transporting particles preferentially to the poles (Bardeen et al. 2008). In this case, the climatic impact of haze could be reduced with warmer surface temperatures at the equator. On the other hand, hazes can also act as cloud condensation nuclei, enhancing cloud formation (Hasenkopf et al. 2011). This might lead to cooling of the planet or even warming depending on cloud particle size and the altitude – and therefore temperature – of the cloud layer (Goldblatt & Zahnle 2011). A complete treatment of the impact of ice-albedo feedback, haze deposition, haze circulation, and cloud feedbacks is left to future GCM studies better equipped to deal with these inherently 3D issues.

### 3.1.2 Temperature Feedbacks on Haze Production

As the haze gets optically thicker, absorption of UV photons produces an atmospheric temperature inversion (figure 5) similar to that produced by ozone in the modern atmosphere. We find there is a relationship between the size of the haze particles generated and the temperature of the atmosphere. To isolate the effect, we tested haze production by the photochemical model using two completely isothermal temperature profiles of 200 K and 250 K with all other parameters held constant (Figure 6). The largest particles produced by the 250 K atmosphere have a geometric radius of 0.8 μm compared to 0.65 μm radius particles for the 200 K atmosphere. In the photochemical model, when the coagulation timescale ($\tau_{coag}$) is shorter than the timescale for removal in an atmospheric layer, the particles can grow. As temperature increases, $\tau_{coag}$ decreases since particles moving faster collide more frequently (Tolfo 1977). In the hotter atmosphere, $\tau_{coag}$ is smaller than $\tau_{sed}$ through most of the atmospheric column.

**3.2 UV Shielding**

The impact of these hazes on the biosphere goes beyond temperature reduction: their fractal nature makes them strong absorbers at short wavelengths and therefore a potential shield against damaging ultraviolet (UV) radiation for the anoxic Archean (Wolf & Toon 2010) which would have received significantly more UV at the surface than the planet today (Rugheimer et al. 2015). DNA damage is most acute in the UVC ($\lambda < 0.28$ μm) wavelength range (Pierson et al. 1992; Dillon & Castenholz 1999), but in the modern atmosphere, UVC is fully blocked by $O_2$ and ozone. For the haze-free Case B atmosphere ($CH_4/CO_2 = 0.1$), our models calculate the flux of UVC at the surface as about 0.93 W/m$^2$ for a solar zenith angle of 60° and 2.62 W/m$^2$ for SZA = 0°. Both of these values are sufficient for sterilization (Pierson et al. 1992). In contrast, the surface UVC flux under a haze for Case B ($CH_4/CO_2 = 0.21$) would have been about 0.03 W/m$^2$ for SZA = 60° and 0.22 W/m$^2$ for SZA = 0°. We compare these values to the tolerances of *Chloroflexus aurantiacus* (Pierson et al. 1992), a deep-branching, mat-forming anoxygenic phototroph with UV resistance that has been studied as an analog for Archean phototrophs. Our SZA = 60° flux, 0.03 $W/m^2$, is low enough to allow growth of *Chloroflexus aurantiacus* over the length of a day in the late Archean (about 18-19 hours for a day-night cycle; Denis et al. (2002)). Our SZA = 0° flux, 0.22 $W/m^2$, is naturally worse but does not cause immediate sterilization of *Chloroflexus aurantiacus*, allowing modest growth for roughly 10 hours. In a real atmosphere, the UV flux will change with solar zenith angle, but it will not exceed the SZA = 0° flux. At latitudes where the SZA is never 0°, UV survival prospects are better, although these higher latitudes may be icy for our cold planets. Under an Archean haze, it is possible that organisms similar to *Chloroflexus aurantiacus* with robust UV protection mechanisms could have lived at or near the planet's surface. We summarize the UV protection of several types of atmospheres, including ones with water clouds that can confer additional UV protection, in table 3.This table only includes Case B, but the other cases produce similar results for UV shielding because they have similar optical thicknesses.

Possibly, an Archean haze aided the survival of life at or near the surface of our early planet. There is evidence that Archean stromatolitic communities lived in inter- and supratidal zones (Allwood et al. 2006; Noffke & Awramik 2013) experiencing frequent, sometimes extended, exposure to the surface environment, and it has been suggested that microbial mats existed on land as early as 2.6-2.7 Ga (Watanabe et al. 2000). Interestingly, this interval overlaps with periods when haze has been proposed for the Archean atmosphere (Kurzweil et al. 2013; Zerkle et al. 2012; Izon et al. 2015).

It has widely been assumed that Proterozoic Earth's surface received less UV than the Archean due to the rise of oxygen ($O_2$) and ozone (e.g. Rugheimer et al. 2015), but a recent study of chromium isotopes suggests that the Proterozoic $O_2$ mixing ratio was, at most, 0.1% PAL (Planavsky et al. 2014). We tested the strength of an ozone UV shield generated by our photochemical model under these low oxygen conditions against the strength of our hazy UV shield. For the Proterozoic atmospheres, we tested ozone generation at 0.1% and 1% PAL $O_2$ levels (figure 7) with $pCO_2$ fixed at 0.01 bars and $pCH_4$ at 0.0003 bars. Total pressure is set to 1 bar at the surface. According to these assumptions, Proterozoic Earth with 0.1% PAL $O_2$ would have received 0.57 W/m$^2$ of UVC at the surface, so in this case, the Archean hazy UV shield was stronger. Note also that haze is a better shield against UVA ($\lambda$ = 0.315 – 0.400 μm) and UVB ($\lambda$ = 0.280-0.315 μm) than ozone or $O_2$.

### 3.3 Spectra

The strong interaction of haze with radiation means hazes can impact the exoplanet spectra that future space based telescopes will attempt to detect. In figure 8, we show reflectance, thermal emission, and transit transmission spectra for our nominal Case B with an ocean surface; the other Cases produce similar spectra as discussed below. Our predicted spectra of hazy Archean Earth show diagnostic absorption features from $H_2O$, $CO_2$, $CH_4$, $C_2H_6$, CO, and from the haze itself. These features are labeled in figure 8, and another way to show where these gases and the haze absorb is presented in Figures 9 and 10 for reflectance and transit transmission spectra, respectively. Figures 9 and 10 were produced by systematically removing each gas or the haze; in places where a given species absorbs, the original spectrum differs from those with the absorbers removed. To consider the spectral effect of haze without contamination from other atmospheric aerosols, the spectra in this section do not include water clouds, even though cloud albedo is implicit in the parameterization of the Atmos model's surface albedo. This makes the albedos of the planets whose spectra are shown in this section darker than those in the Atmos parameterization. However, since clouds have a major impact on the planet's spectral appearance and albedo (e.g. Kitzmann et al. 2011b), we show spectra with water clouds included in section 3.3.1. The best way to treat the climatic and spectral impact of both clouds and haze would be in a fully-coupled 3D climate-photochemical model that fully considers radiative and photochemical effects of cloud and haze particles, but this is outside the scope of this work. To our knowledge, such a 3D model does not yet exist, but its development would be useful for the comprehensive treatment of this problem.

In reflected light (Figure 8, panel a), the broad UV absorption feature reddens the color of the planet by masking the short-wavelength reflectivity due to Rayleigh scattering. See the bottom section of Figure 8 for the estimated color of the planet to the eye. The planet colors were calculated using the "Spectral Color Spreadsheet" from brucelindbloom.com with the same method used in Charnay et al. (2015) for GJ 1214b. A spectrum can be input to the calculator, which then outputs RGB values. While these colors should be understood as approximations, we tested the colors produced for the modern Earth sky and Titan as a check, and the results appeared reasonable. Colors and photometric bands have been considered as indicators of Earth-like worlds (Traub 2003; Crow et al. 2011; Krissansen-Totton et al. 2016), but hazy Archean Earth suggests that not all Earth-like planets will be pale blue dots. Because methane-producing metabolisms evolved early and Earth's atmosphere was anoxic for about a billion years after the origin of life, pale orange dots may proliferate in the galaxy if other habitable worlds evolve on similar paths to Earth.

Several spectral features are apparent in Figure 8. The haze-mediated stratospheric thermal inversion is clearly seen in thermal emission near 8 μm and 16 μm (Figure 8, panel b). Similar to the Titan transmission spectrum derived from Cassini solar occultation measurements (Robinson et al. 2014b), our simulated hazy transit transmission spectra (Figure 8, panel c) are sloped in the visible and NIR. Gas absorption features in the visible and NIR are muted by the presence of a haze in transit transmission, but mid-IR absorption features are less affected because the haze is relatively transparent at longer wavelengths. In Earthlike clear-sky atmospheres, the minimum atmospheric altitude transit observations are able to probe will typically be limited by refraction (García Muñoz et al. 2012; Bétrémieux & Kaltenegger 2014; Misra et al. 2014a), but in hazy atmospheres, haze controls the minimum effective tangent height, especially at shorter wavelengths where it controls the transit transmission spectral slope. Absorption from the haze itself can be seen as the "bump" in the "thick" haze ($CH_4/CO_2 = 0.21$) transit transmission spectrum at 6 μm, a wavelength region accessible with the James Webb Space Telescope (Wright et al. 2004). There is also a very weak haze feature near 3 μm in transit transmission that can be most easily seen in Figure 10. These features can also be seen in the peaks of the haze imaginary refractive index (Figure 2).

Note the presence of a $C_2H_6$ absorption feature near 12 μm. This $C_2H_6$ forms from photochemistry involving $CH_4$, and its buildup in our spectra is not inconsistent with the results of Domagal-Goldman et al. (2011), which showed much greater $C_2H_6$ accumulation on planets orbiting low-mass stars compared to worlds orbiting the sun. However, the $CH_4$ levels in the Domagal-Goldman et al. solar simulations were an order of magnitude lower than the ones shown here. $C_2H_6$ is a greenhouse gas, and its ability to warm in a $CH_4$- and haze-rich atmosphere has been discussed previously (Haqq-Misra et al. 2008).

Figures 8-10 showed spectra for our Case B planet, but Figures 11 and 12 show representative reflected light and transit transmission spectra for all of our Cases A-D in the visible and near-infrared. The reflectance spectra in Figure 8 and 9 assumed a pure ocean surface albedo to isolate the spectral consequence of atmospheric haze from other spectral changes, but the spectra shown in Figure 11 are constructed from a weighted average of ocean and ice surfaces according to the ice line latitudes reported in Wolf & Toon (2013) for Archean atmospheres with $CO_2$ and $CH_4$. The hazy planets in Figure 11 are more reflective than the spectra shown in Figure 8 due to this ice coverage. Figure 12 shows how thick hazes strongly mute the strength of gaseous absorption features in transit transmission at shorter wavelengths where these hazes are more optically thick.

**3.3.1 Water Clouds**

The goal of the nominal haze spectra we have presented is to show the spectral impact of organic haze independent of any other atmospheric aerosols. However, it is interesting and important to also consider how water clouds affect our hazy spectra. To test the impact of clouds in addition to haze on the spectra of Earth-like planets, we added water clouds to the Case B atmospheres shown in Figure 8. Because these are 1D spectra, we incorporate clouds with a weighted average of cloudy and pure haze spectra where we assume 50% of the planet is covered by haze only, 25% is covered by cirrus clouds (at 10 km altitude) and haze, and 25% by strato cumulus clouds (at 1 km altitude) and haze (Robinson et al. 2011). The resulting spectra are presented in Figure 13. In contrast to hydrocarbon haze particles, which are more transparent in the near infrared compared to shorter wavelengths, water vapor clouds have an approximately gray opacity from the visible into the near infrared. Thus, at longer wavelengths, cloudy worlds are brighter than their haze-only counterparts. Table 4 shows the total integrated brightness of the reflectance spectra for the worlds with clouds divided by their cloud-free counterparts between 0.4-1 μm and between 1-2 μm to quantify the spectral impact of clouds.

The disproportionate increase in brightness from clouds at longer wavelength compared to shorter wavelengths means that the peak of the reflectance spectrum also shifts towards longer wavelengths for the worlds with clouds: for $CH_4/CO_2$ = 0.17, the reflectance spectrum peak shifts from 0.31 μm to 0.38 μm, and for $CH_4/CO_2$ = 0.21, it shifts from 0.56 μm to 0.68 μm. Adding clouds also raises the spectral continuum level, making absorption features appear deeper. This enhanced reflectivity also potentially increases the detectability of water vapor in reflected light spectra, as more reflected flux from the planet reduces noise on the continuum, enhancing the detectability of absorption features that deviate from that continuum. A detailed discussion of the impact of water clouds on the spectra of Earthlike planets for different cloud altitudes and fractional cloud coverages can be found in Kitzmann et al. (2011b).

In transit transmission, water clouds have no spectral impact because they form in the atmosphere at a level below the maximum tangent height set by refraction. The tropopause on Earth is at roughly 10 km, and refraction prohibits transmission of path lengths below about 20 km even for our clear sky worlds. As water vapor is at very low abundance in the Earth's stratosphere, it would be difficult, in general, to see it in transmission observations that can only probe down to stratospheric altitudes. Abundant stratospheric water vapor would imply that the planet is in the midst of a moist or runaway greenhouse state, and thus is not conventionally habitable.

**4. Discussion**

The hazes investigated here have a major spectral impact at short wavelengths due to their strong blue and UV absorption. It has been suggested that the Rayleigh scattering slope could be used to constrain atmospheric pressure on exoplanets (Benneke & Seager 2012), but this would not be possible on planets with hydrocarbon hazes due to these strong short wavelength absorption effects. In reflected light, the haze's broadband UV absorption feature, observed together with methane bands, would strongly imply the existence of hydrocarbon haze in an atmosphere. In the infrared, the diagnostic haze absorption feature at 6 μm (and the weaker one at 3 μm) in transit transmission would allow chemical identification of hydrocarbon haze. Even absent the detection of these specific features in transit transmission, the presence of $CH_4$ bands together with the haze UV-visible-NIR spectral slope would strongly imply the presence of this haze.

**4.1 Haze and biology**

Our study shows how an Archean haze would have profoundly impacted our planet's environment, habitability, and spectrum. It is important to note that geochemical evidence suggests hazy conditions were not present throughout the entire Archean, and its periodic collapse may have put stress on the biosphere if organisms migrated to the surface or near-surface and adapted to lower UV levels created by the haze. On the other hand, if organisms remained protected by some other UV shield such as minerals, layers of overlying microbial mat, or water (Cockell 1998), changes in UV radiation levels should not affect them as strongly, so the colder conditions created by the haze might have been the larger source of stress on organisms. These stressors might have driven evolutionary adaptations as life responded to its changing environment. Note that photosynthetic organisms would not likely have been photon limited by the lower light levels under the haze: the lower light limit for red algae is $6x10^{15}$ photons/$m^2$/s (Littler et al. 1986). Under our Case B $CH_4/CO_2 = 0.21$ haze, total PAR at the surface is $7.1x10^{20}$ photons/$m^2$/s, orders of magnitude above this extreme.

Laboratory experiments on organic haze formation have shown that haze-formation chemistry can involve the formation of important prebiotic molecules such as amino acids and nucleotide bases (Khare et al. 1986; McDonald et al.

1994; DeWitt et al. 2009; Hörst et al. 2012; Trainer 2013) – see also our discussion of haze formation pathways in section 4.4. Although the hazy periods we invoke here occurred hundreds of millions of years after the origin of life on Earth, there may be earlier hazy epochs not yet discovered in the geological record (see Kasting (2005) for a discussion of earlier atmospheric methane), and hazy Titan has been regarded as a type of prebiotic chemical laboratory (Khare et al. 1984b; Clarke & Ferris 1997).

While we know that abiotic hydrocarbon hazes are possible (e.g. on extremely cold worlds like Titan with reducing atmospheres), on a planet like Archean Earth, the presence of hydrocarbon haze may require a higher level of methane production than is possible from abiotic sources alone. The maximum abiotic methane production rate from serpentinization, its primary nonbiological source, has been estimated as $6.8 \times 10^8$ and $1.3 \times 10^9$ molecules/$cm^2$/s for rocky planets of 1 and 5 Earth masses, respectively (Guzmán-Marmolejo et al. 2013), although there has been earlier speculation of higher abiotic production rates (Kasting 2005; Shaw 2008), especially if ancient seafloor spreading rates were faster or the amount of iron-rich ancient seafloor rock was greater. Based on their calculations, Guzmán-Marmolejo et al. (2013) suggest that an atmospheric $CH_4$ concentration greater than 10 ppmv is suggestive of life. At the range of $pCO_2$ allowed by Driese et al. (2011), we find that the $CH_4$ flux needed to initiate haze formation ranges between about $1\text{-}3 \times 10^{11}$ molecules/$cm^2$/s, broadly consistent with estimates for the biological Archean methane flux after the origin of oxygenic photosynthesis (Kharecha et al. 2005; Claire et al. 2014). The higher of the plausible rocky planet abiotic $CH_4$ fluxes from Guzmán-Marmolejo et al., $1.3x10^9$ molecules/cm$^2$/s, will not form a haze in our model even at a $pCO_2$ level four orders of magnitude smaller than the lower limit allowed by Driese et al. (2011), and such a world would be completely frozen given the Archean solar constant. Remote identification of a hydrocarbon haze with a concurrent measurement of $CO_2$ around a planet that absorbs an Earth-like amount of radiation could therefore imply a surface methane flux consistent with biological production. The strength and width of the hydrocarbon haze absorption feature below about 0.5 $\mu$m implies it would be easier to detect than methane itself given sufficient instrumental sensitivity to this range, so the occurrence of haze in the habitable zone may be a way to flag interesting planets for careful follow-up study that would search for other indicators of life and quantify the concentration of $CH_4$ and other gases.

**4.2 Comparison with other climate studies**

To test the robustness of the mean surface temperatures calculated by our computationally efficient 1D climate model, we compared our temperature result for a haze-free Case A atmosphere with $pCO_2 = 0.01$ and $pCH_4 = 0.002$ (but no ethane) and a solar constant for 2.5 billion years ago to the Laboratoire de Météorologie Dynamique (LMD) General Circulation Model (GCM) run with the same inputs. We adopt the same average albedo produced by the LMD model in this simulation, setting $A_{surf}$=0.33 for our planet (as before, this albedo includes the effect of clouds). For an ocean-covered planet with no haze, the LMD model produces a mean surface temperature of 287 K (Charnay et al. 2013). This is comparable to, but 5 K warmer, than our global average 1D result of 282 K. The Charnay et al. results for the same atmospheric properties but with an equatorial supercontinent result in the same overall planetary albedo (0.33) but a lower mean temperature of 285 K, which is closer to our result. We achieve the closest match to the Charnay et al. results for a modern continental land mass arrangement: in the GCM, this yields an average albedo of 0.34 and a temperature of 283.7 K, close to our result of 281.1 K for this configuration.

We also tested our model results against the Community Atmosphere Model (CAM) GCM nominal Archean atmosphere reported in Wolf & Toon (2013). For this planet, the solar constant is 80% modern, $pCO_2$ = 0.06 bars, there is no $CH_4$, no haze, and the planet has an average albedo of 0.317. For this world, the CAM model produces a global average surface temperature of 287.9 K. Our model produces 285.3 K for this configuration, a difference of 2.6 K.

The GCMs we compare to can include a variety of effects our 1D model cannot, including atmospheric circulation, precipitation, cloud formation, and cloud scattering and absorption. Our comparison with these 3D models suggests the temperatures we present in this work are reasonable but may be under-estimates by about 3-5 K. One reason that our 1D results may be colder than the GCM results is that while we have incorporated identical planetary albedos (with clouds), we are still missing the longwave radiative forcing from clouds, which would have a warming effect.

Haqq-Misra et al. (2008) similarly studied the climate of Archean Earth with hydrocarbon hazes and high amounts of $CO_2$, $CH_4$, and $C_2H_6$ with an earlier incarnation of the 1D models we use here. The haze-free surface temperatures we generate are broadly consistent with the Haqq-Misra et al. non-hazy results with $C_2H_6$. Haqq-Misra et al. show that a planet with $pCO_2$ = 0.01 and $CH_4/CO_2$ =0.1 has a surface temperature of about 282 K, which is close to our 283.4 K for a comparable atmosphere. Similar to our study, the Haqq-Misra et al. study found it was difficult to maintain surface temperatures above the freezing point of water with spherical haze particles. However, as we have argued, a mean surface temperature of the freezing point of water is not a useful threshold for global habitability (Shields et al. 2013; Charnay et al. 2013; Wolf & Toon 2013; Kunze et al. 2014), so some of the Haqq-Misra et al. spherical haze results may actually be "habitable". In general, we are able to achieve warmer hazy solutions in our study because, as previously discussed, fractal hydrocarbon hazes produce less extinction of visible wavelengths compared to equal mass spherical haze particles. For example, for a planet with 1 bar of pressure and 2% $CO_2$, the Haqq-Misra et al. spherical haze drops the planet's temperature to below 260 K. The same planet with a fractal haze in our study remains above 273 K (without considering ice-albedo effects not examined in the Haqq-Misra et al. study) after haze self-shielding levels off the temperature. Our results suggest that fractal hazes do indeed produce less antigreenhouse cooling than spherical particles. However, since our non-hazy comparison atmosphere was about 1.4 K warmer than the comparable Haqq-Misra atmosphere, a small component of the warmer temperatures we see here may also result in part from updates to our climate model made by the Kopparapu et al. (2013) study.

**4.3 Potential for $NH_3$ greenhouse gas shielding**

The optical thickness of the haze impacts its ability to shield molecules from photodissociation. Once the UV opacity of the haze exceeds approximately unity, the surface flux of $CH_4$ necessary to maintain a given atmospheric methane mixing ratio drops due to haze-induced $CH_4$ shielding. At higher haze thicknesses, the opacity of the haze levels off because this self-shielding inhibits the methane photolysis needed to initiate haze formation. Wolf & Toon (2010) commented on the possibility of a fractal hydrocarbon haze shielding ammonia ($NH_3$) from photolysis, allowing this greenhouse gas to build up in the Archean atmosphere. Following Sagan & Chyba (1997), Wolf and Toon calculated a $NH_3$ atmospheric lifetime of $7x10^7$ years for a solar incident flux at a 45 degree angle assuming $\tau$~11 at 200 nm. Following Wolf and Toon, we find our maximum haze thickness levels off at $\tau$~5 at 200 nm, which results in a significantly shorter $NH_3$ lifetime of $1x10^4$ years, although we did not include $NH_3$ in our photochemical scheme. Our future work will include $NH_3$ in the photochemical and climate model to study, in a self-consistent atmosphere, how much of this gas can exist in a hazy atmosphere and what its climatic effect could be.

**4.4 Haze formation pathways**

Following the mechanism proposed for the formation of Titan's hydrocarbon haze (Allen et al. 1980; Yung et al. 1984), every model of hydrocarbon haze formation in early Earth's atmosphere - including ours - has assumed that aerosol formation will occur through the formation of acetylene ($C_2H_2$) and its further polymerization to higher polyacetylene chains (Pavlov et al. 2001a; Pavlov et al. 2001b; Domagal-Goldman et al. 2008; Haqq-Misra et al. 2008; Zerkle et al. 2012; Kurzweil et al. 2013; Claire et al. 2014). The two reaction pathways described in Section 2.1.1 provide an initial picture of the process, but haze formation is likely considerably more complex and is still not well understood. Unlike early Earth, we now have access to direct observations of the chemical processes ongoing in Titan's atmosphere. *In situ* measurements by several instruments onboard Cassini have found direct evidence for long hydrocarbons and nitriles chains, benzene ($C_6H_6$) and toluene ($C_6H_5CH_3$), and indirect evidence for Polycyclic Aromatic Hydrocarbons (PAHs) and nitrogen-containing PAHs (PANHs), indicating that these compounds might play a role in the formation of Titan's hazes (Waite et al. 2007; López-Puertas et al. 2013).

Moreover, early Earth's atmosphere was likely not as reducing as Titan's. The chemical pathways for haze formation, including the $C_2H_2$ polymerization pathways, may therefore be inappropriate. Early Earth's atmosphere would have contained negligible $O_2$ but significant amounts of $CO_2$ (e.g. Kasting 1993; Driese et al. 2011) whereas Titan's atmosphere is extremely reducing (de Kok et al. 2007). Even in Titan's highly reducing atmosphere, it was suggested that CO may contribute to oxygen incorporation in the organic aerosols (Hörst et al. 2012). This oxygen incorporation is expected to be much more important to aerosol chemistry in early Earth's far less reducing atmosphere. Using far ultraviolet (FUV) radiation (115-400 nm), organic aerosol production from a $CH_4/CO_2/N_2$ mixture was shown to exceed that from a pure $CH_4/N_2$ mixture (Trainer et al. 2006) and organic aerosol formation was experimentally observed to occur down to C/O ratios as low as 0.1 (Trainer et al. 2006; DeWitt et al. 2009). From the chemical analysis of primary condensed-phase products of photochemistry, it is clear that the composition of the aerosol analogs formed in early Earth-like atmospheres with $C/O < 1$ differs greatly from the aerosol analogs formed in Titan-like atmospheres where $C/O >> 1$. Instead of limiting the formation of organic molecules as initially predicted, the O-atoms released from $CO_2$ photolysis are incorporated into the molecular structure of the organic aerosols. Mass spectrometry of aerosol analogs formed with $C/O = 0.1$ indicates the formation of carbonyl and carboxyl groups rather than aromatic cycles and long-aliphatic chains, and even suggests the formation of organic acids such as succinic acid ($C_4H_6O_4$) (DeWitt et al. 2009).

Finally, haze formation chemistry gets considerably more complex when one considers the coexistence not only of O-heteroatoms but also of N-heteroatoms in aerosol organics. Nitrogen incorporation was recently observed in the aerosols generated by far-UV photolysis of $CH_4/CO_2/N_2$ gas mixtures (Trainer 2013) and in $CH_4/N_2$ mixtures (Sebree et al. 2015). These results bring to light a significant but still unknown mechanism regarding the activation of nitrogen and its inclusion in oxygenated organics, thus providing a new and quantifiable source for these two elements into the early Earth aerosols. Studies have shown the formation of HCN, $CH_3CN$ and other nitrile gas species are formed using the same type of UV source in a $CH_4/N_2$ gas mixture, thus corroborating the indirect nitrogen photochemistry (Trainer et al. 2012; Yoon et al. 2014). These results suggest that $N_2$ chemical activation could be due to its reaction with the methylidyne (CH) radical formed from $CH_4$ photolysis, to form two radical intermediates, diazomethyl HCNN and its isomer HNCN, which might then react to form HCN and other products.

The formation of aerosols in early Earth's atmosphere is thus tightly intertwined with the formation of organic molecules containing more than a few C/H/N/O atoms. These compositional differences should change the properties of the aerosol material sufficiently to be able to distinguish a hazy early Earth from a modern-day Titan (Hasenkopf et al. 2010). For instance, organic molecules with oxygen-containing functional groups (alcohols, carbonyls) tend to have stronger absorbances at longer UV wavelengths as compared to similar hydrocarbon molecules (Workman 2000). The NIR absorption bands of the Archean aerosol analogs would also shift in response to the inclusion of the types of oxygen and nitrogen heteroatom functionalities that have been indicated in the compositional studies.

**4.5 Optical Constants**

The implications of compositional differences of Archean hazes versus Titan hazes for the topics presented in this study and for prebiotic chemistry underscores the need for measurements of Archean Earth analog optical constants as well as a better understanding of the haze formation chemical pathways. Unfortunately, only one study, Hasenkopf et al. (2010), has measured an Archean Earth haze refractive index (as opposed to a Titan haze), and this was only done at a single wavelength (532 nm).

In our study, we have used the hydrocarbon refractive indices from Khare et al. (1984a) to allow us to draw comparisons with previous works involving our suite of models (Domagal-Goldman et al. 2008; Pavlov et al. 2001a; Pavlov et al. 2001b; Zerkle et al. 2012; Haqq-Misra et al. 2008; Kurzweil et al. 2013; Claire et al. 2014), as well as the Wolf and Toon (2010) study which all used the Khare optical constants. An additional advantage of the Khare refractive indices is that they span an extremely wide wavelength range, ranging from 0.02 μm to 920 μm, so only one set of optical constants is needed to cover all the wavelengths relevant to photochemistry, climate, and spectra.

However, more recent measurements of hydrocarbon refractive indices over more restricted wavelength ranges indicate disagreement with the Khare measurements (Imanaka et al. 2012; Mahjoub et al. 2012; Sciamma-O'Brien et al. 2012; Hasenkopf et al. 2010; Ramirez et al. 2002; Tran et al. 2003; Vuitton et al. 2009), although these measurements themselves show considerable variation amongst each other (Figure 14). Differences in the composition of Archean hazes compared to Titan's (and thus differences in their optical constants) are expected as discussed in section 4.4. Again, note the single measurement by Hasenkopf et al. (2010) for an Archean-analog haze; of all of the optical constants plotted in Figure 14, the Khare indices actually produce the closest (although still too low) match to the Hasenkopf Archean real refractive index (n) near 532 nm and produce a reasonable match to the Hasenkopf Archean imaginary refractive index (k), agreeing to within approximately 40% near 532 nm.

As an example and test of the impact different refractive indices have on our spectra, we examined the sensitivity of our nominal spectra to varied refractive indices measured by Mahjoub et al. (2012). The Mahjoub study tested the impact of methane concentration in the gas phase on the resultant hydrocarbon optical properties with gas mixtures containing 1%, 2%, 5% and 10% $CH_4$ in $CH_4$-$N_2$ mixtures. Note the 1% $CH_4$ Mahjoub imaginary refractive index agrees to within 5% of the Hasenkopf Archean measurement near 532 nm. Mahjoub et al. found that refractive indices have a strong dependency on the $CH_4$ concentration over 0.37-1 μm: results indicate that the imaginary index of refraction (k) decreases with increasing $CH_4$ concentration, and the real index of refraction (n) increases with $CH_4$ for the compositions tested. We generated the spectra shown in figure 15 by producing new fractal input files using the Mahjoub optical constants. These files were then used to replace the Khare files in our SMART inputs for the nominal $CH_4/CO_2$ = 0.21 case B spectrum. In addition, we generated a spectrum to test the Hasenkopf Archean haze measurement by applying a scaling factor to the Khare optical constants to match the Hasenkopf n and k values at 532 nm. This spectrum is called “Khare-Hasenkopf” in the figure 15 caption. Of course this does not account for differences expected in the spectral shape of Archean haze analogs across the UV-Visible-IR relative to the Titan haze analogs.

Besides affecting the top-of-atmosphere spectrum, these different optical constants alter how much radiation can reach the surface under a haze. We find that, for the Mahjoub 1%, 2%, 5%, and 10% $CH_4$ optical constants, 0.92, 1.11, 1.13, and 1.16 times the nominal (Khare) total integrated 0.37-1 μm flux reaches the surface of the planet. For the Khare-Hasenkopf spectrum, which shifts both the real and imaginary refractive indices to larger values, this drops to 0.89 times the nominal flux. The Mahjoub constants do not extend shortward of 0.37 μm, but we should anticipate variation at these shorter wavelengths as well.

The variation in surface-incident flux shows us that we should expect differences in the hazy Archean climates we calculate depending on the optical constants used. We tested how the Hasenkopf Archean measurement might impact the climate for a $pCO_2$ = 0.01 and $CH_4/CO_2$ = 0.2 atmosphere. The nominal Khare constants produce a surface temperature of 272 K for this atmosphere. The "Khare-Hasenkopf" optical constants yield a cooler temperature of 267 K, which is expected because these optical constants produce a haze with more efficient scattering and absorption. This difference in temperature is smaller than the difference of using spherical versus fractal particles: our comparison to the Haqq-Misra et al. (2008) study in section 4.2 shows that particle shape can result in temperature differences > 10 K. A full treatment of the impact of varied optical constants using the coupled photochemical-climate model to generate new self-consistent atmospheres and climates is outside the scope of our present study.

Updated haze optical constants generated under Archean Earth-like laboratory conditions (rather than Titan-like conditions) to produce plausible Archean-analog haze compositions would be of immense value to future studies of organic hazes in Earthlike atmospheres, including exoplanets, and would allow updates of the results presented in this study. Due to the properties of fractal hazes, these particles are relatively transparent at wavelengths longer than approximately the visible range, so measurements of refractive indices at visible wavelengths in particular would allow us to improve our estimates of the climatic impacts of this haze. In addition, better constraints on Archean UV refractive indices would allow us to better quantify how good a UV shield these hazes actually are.

**5. Conclusions**

We have shown that a hazy Archean Earth consistent with geochemical constraints on $CO_2$ concentration and geological constraints on surface pressure could have had habitable surface temperatures. Although the fractal hazes simulated here cool the planet by up to ~20 K, these fractal particles produce significantly less cooling than a haze of equivalent mass spherical particles. The climatic effects of this haze could have been part of feedbacks between biological $CH_4$ production, atmospheric chemistry, and surface UV radiation. Haze can cut down the surface-incident UVC radiation on Archean Earth from ~0.9 W/m$^2$ to ~0.03 W/m$^2$ for a solar zenith angle of 60°, and may have allowed survival of otherwise unshielded life at the surface of our Archean planet. The presence of similar hydrocarbon haze on an exoplanet could be observed, as demonstrated by strong features present in synthetic spectra of these worlds. For habitable exoplanets similar to Archean Earth, hydrocarbon haze may be strongly biologically mediated, and serve as a novel non-gaseous biosignature with a strong spectral signature. Discovering habitable exoplanets dissimilar to modern Earth will increase the diversity of known habitable environments. Leveraging our understanding of Earth's history provides us with a variety of analogs with which we can expand our expectations for the "Earth-like" planets beyond our Solar System; future observations of such worlds can provide us with a window into the evolution of terrestrial worlds like our home.

## Tables

Table 1 Atmosphere parameters for Cases A-D

| | **Case A** | **Case B** | **Case C** | **Case D** |
|---|---|---|---|---|
| **$pCO_2$** (bar) | 0.01 | 0.018 | 0.01 | 0.0036 |
| **$P_{surf}$** (bar) | 1 | 1 | 0.5 | 0.5 |

Table 2 Temperature results for Cases A-D

| | $CH_4/CO_2$ to initiate haze formation | Maximum $T_{surf}$ without haze (K) | Stabilized $T_{surf}$ with haze (K) | $T_{surf}$ with ice-albedo feedback (K) |
|---|---|---|---|---|
| Case A | 0.18 | 284 | 263 | 257 |
| Case B | 0.15 | 299 | 274 | 271 |
| Case C | 0.19 | 282 | 262 | 257 |
| Case D | 0.28 | 273 | 251 | 241 |

Table 3. The UV fluxes at the planetary surface for several overlying atmospheres. All values quoted have units of $W/m^2$. The solar constant for geological times has been scaled according to Claire et al. (2012) at 2.5 Ga for the Proterozoic and 2.7 Ga for the Archean. All calculations have been performed assuming that the Sun is either directly overhead (Solar Zenith Angle = 0°) or at a Solar Zenith Angle of 60°. There are three Archean UV fluxes per UV Band and $CH_4/CO_2$ ratio: they refer to haze only (labeled "H"), haze plus cirrus cloud (labeled "H+C"), and haze plus stratocumulus cloud (labeled "H+S"). The Modern Earth and Proterozoic atmospheres are cloud- and haze-free. UVA spans $\lambda = 0.315 – 0.400$ µm. UVB spans $\lambda = 0.280-0.315$ µm. UVC is $\lambda < 0.280$ µm.

| | Modern Earth | Proterozoic 1% PAL $O_2$ | Proterozoic 0.1% PAL $O_2$ | Case B $CH_4/CO_2 = 0.1$ | | | Case B $CH_4/CO_2 = 0.17$ | | | Case B $CH_4/CO_2 = 0.21$ | | |
|---|---|---|---|---|---|---|---|---|---|---|---|---|
| SZA = 0° | | | | | | | | | | | | |
| | | | | H | H+C | H+S | H | H+C | H+S | H | H+C | H+S |
| UVA | 70.5 | 59.1 | 59.3 | 55.5 | 50.84 | 38.1 | 48.8 | 44.3 | 33.2 | 22.8 | 20.2 | 15.0 |
| UVB | 2.49 | 6.18 | 10.6 | 10.2 | 9.32 | 7.26 | 8.11 | 7.38 | 5.76 | 2.19 | 1.96 | 1.52 |
| UVC | ~0 | 0.00764 | 2.03 | 2.62 | 2.41 | 1.95 | 1.87 | 1.71 | 1.38 | 0.216 | 0.196 | 0.158 |
| SZA = 60° | | | | | | | | | | | | |
| UVA | 28.9 | 24.4 | 24.5 | 23.0 | 18.42 | 13.2 | 17.7 | 14.4 | 10.4 | 4.93 | 4.14 | 3.00 |
| UVB | 0.446 | 1.77 | 3.90 | 3.82 | 3.29 | 2.51 | 2.51 | 2.18 | 1.67 | 0.337 | 0.29 | 0.22 |
| UVC | ~0 | $7.29\times10^{-4}$ | 0.565 | 0.932 | 0.841 | 0.673 | 0.512 | 0.471 | 0.376 | 0.0318 | 0.0290 | 0.0252 |

Table 4. The relative brightness of spectra with and without water clouds.

| $CH_4/CO_2$ = | **0.4-1 μm**<br>**With clouds/No clouds** | **1-2 μm**<br>**With clouds/No clouds** |
|---|---|---|
| 0.10 | 2.34 | 4.80 |
| 0.17 | 2.12 | 4.24 |
| 0.21 | 1.56 | 2.24 |

## Figures

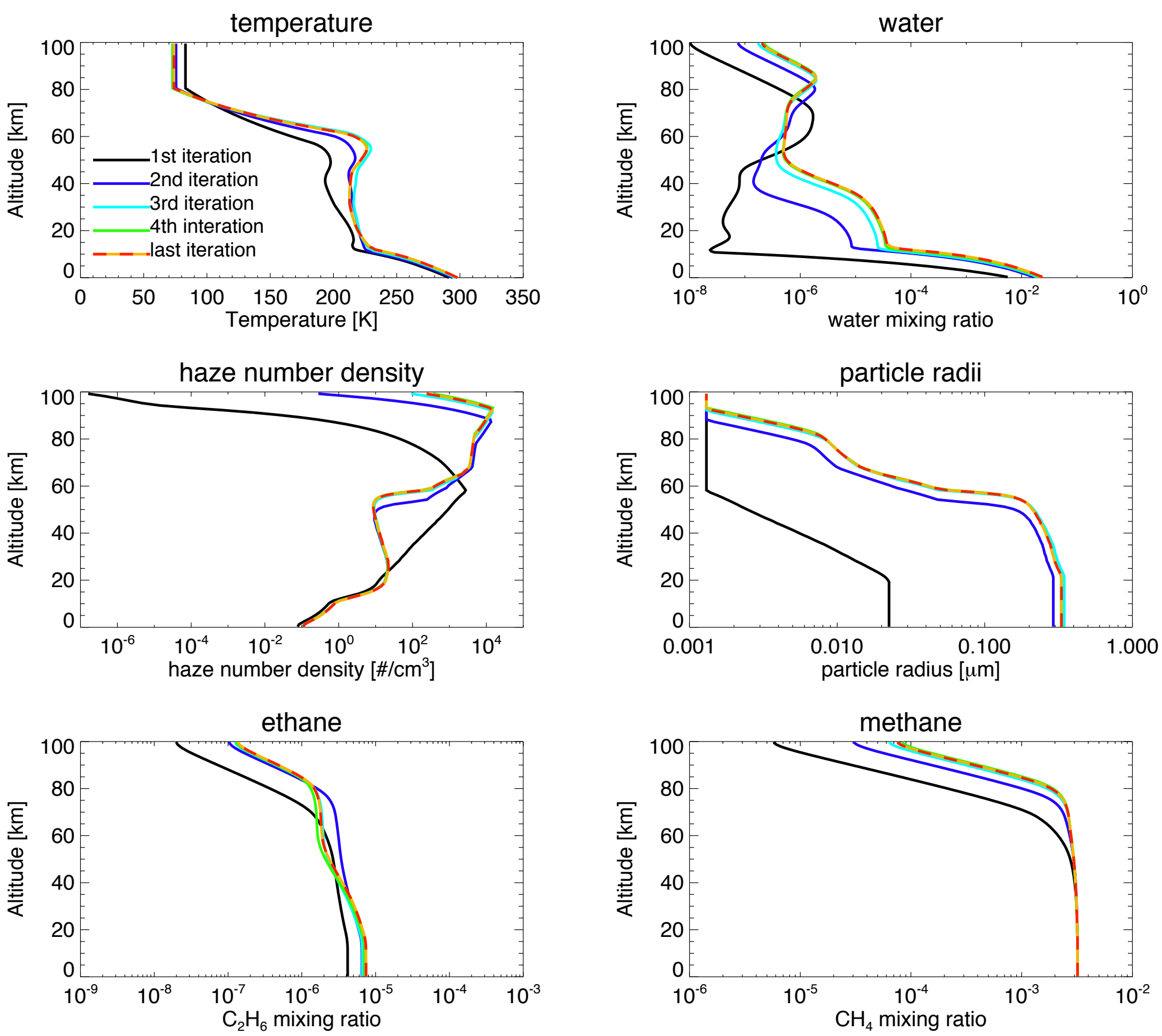

**Figure 1** Shown is an example of the Atmos model convergence process. This atmosphere, which has $CH_4/CO_2 = 0.17$ and $pCO_2 = 0.02$ (total pressure 1 bar) goes through five coupling iterations. The initial temperature profile it uses was stored from a previous similar atmosphere. Here we show the temperature, water, haze number density, haze particle radii, $C_2H_6$ profile, and $CH_4$ profile for each iteration of the coupled model.

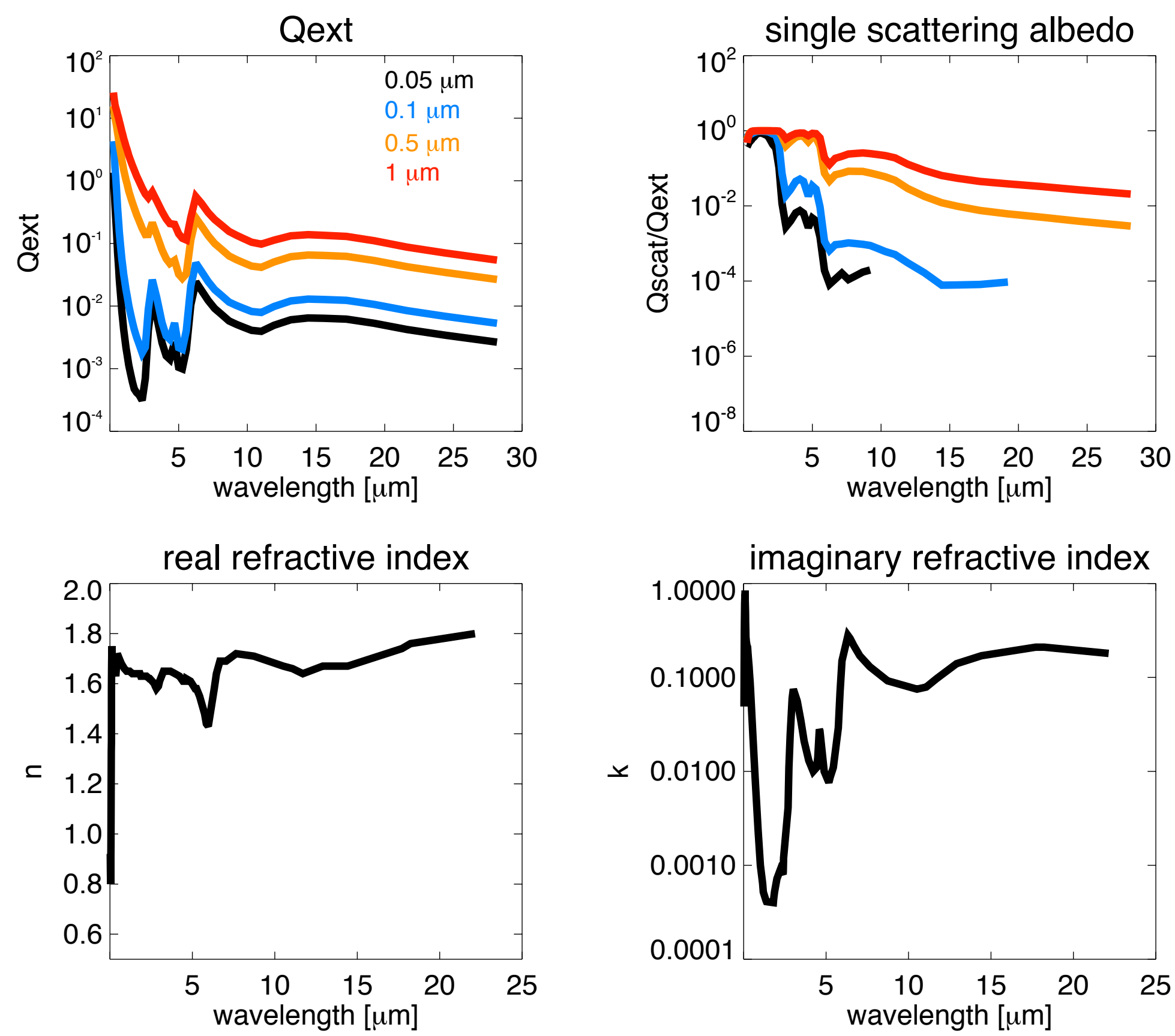

**Figure 2** The top panels present the extinction efficiency (Qext) and single-scattering albedo (= Qscat/Qext) of four sizes of fractal hydrocarbon particles used in this study and in Wolf & Toon (2010). The spherical monomers comprising these particles are 0.05 μm in radius. The radii on the plot correspond to the radii of equivalent mass spherical particles, and the fractal dimensions of these particles, from smallest to largest, are 3 (spherical), 1.51, 2.28, and 2.40. The number of monomers in these particles are one, eight, 1000, and 8000. These particles tend to scatter and absorb light more efficiently at shorter wavelengths,

and larger particles have flatter wavelength dependence for the scattering efficiency.  Refractive indices, shown in the bottom panels, are presented from information in Khare et al. (1984a).

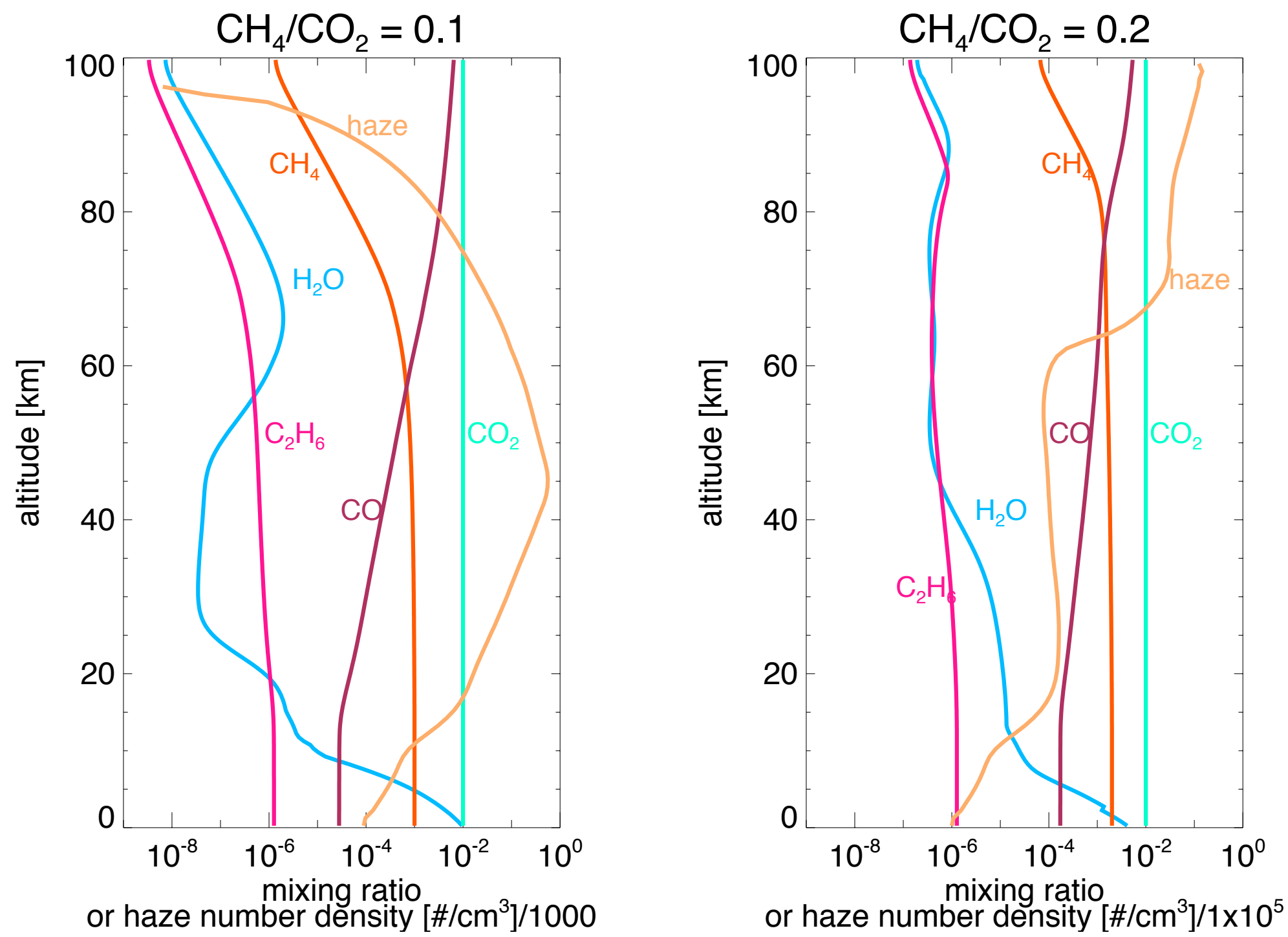

**Figure 3** The gas profiles for $H_2O$, $CH_4$, $CO$, $CO_2$, and $C_2H_6$ for planets with $pCO_2 = 0.01$ bar for $CH_4/CO_2 = 0.1$ (on the left) and $CH_4/CO_2 = 0.2$ (on the right). Also shown are the profiles for the haze particle number density (in pale orange). The $CH_4/CO_2 = 0.1$ haze profile is divided by 1000 and the $CH_4/CO_2 = 0.2$ haze profile is divided by $1\times10^5$ in order to plot it on the same axis as the gases. The profiles in the right panel show larger amounts of $CH_4$, $H_2O$ and $C_2H_6$ above 60 km in altitude, and illustrate how haze-induced shielding can prevent photolysis of these gases. The sharp decrease in haze particle number density between 60 and 70 km in the right panel shows where fractal coagulation occurs.  The atmosphere above the fractal coagulation region is populated by spherical sub-monomer particles.

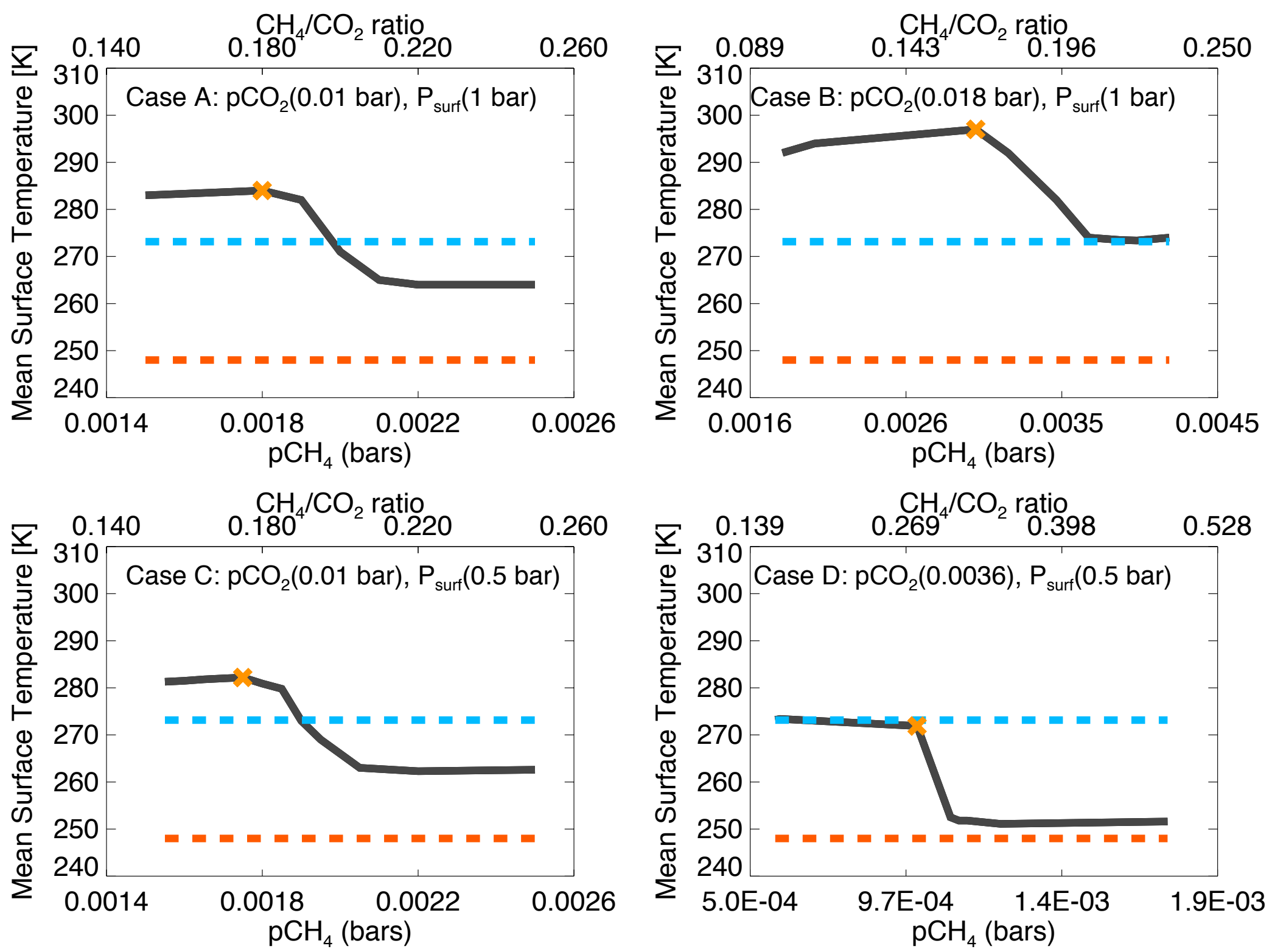

**Figure 4** Mean surface temperatures as a function of $CH_4$ for Archean cases A-D. The dashed blue line shows the freezing point of water (273 K) and the dashed orange line marks our lower threshold of habitability (248 K) for an equatorial ocean belt (Charnay et al. 2013). The "X" in each panel indicates the initiation of haze-induced cooling.

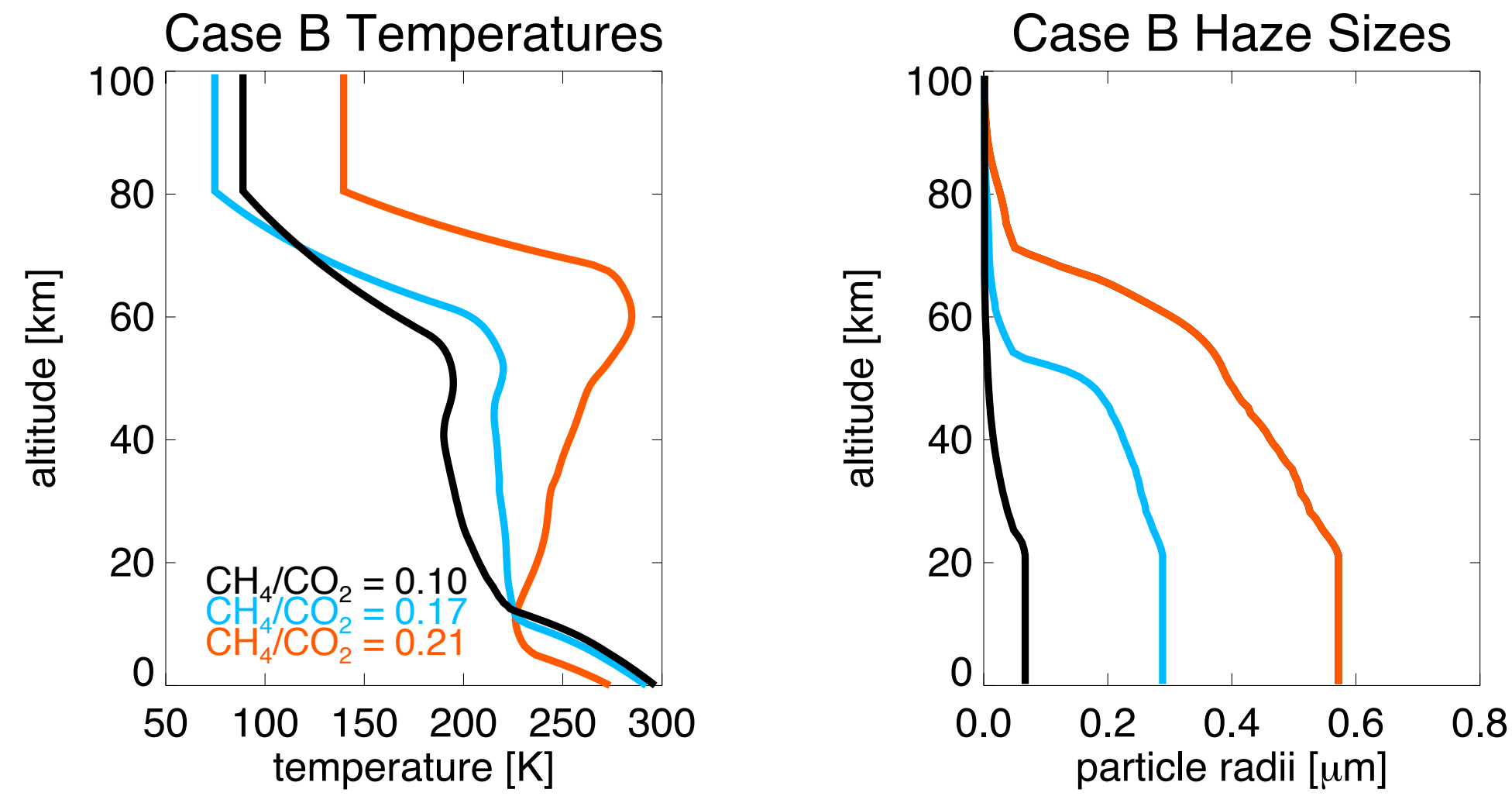

**Figure 5** The left panel presents the temperature profiles of three $CH_4/CO_2$ ratios for the Case B planet. Note the strengthening temperature inversion as the $CH_4$ content of the atmosphere increases. The right panel shows the size of haze particles produced in these three atmospheres, showing the dependence of haze particle size on temperature. From least to most $CH_4$ (and thinnest to thickest haze), the particles reach a maximum radius of 0.067 μm, 0.28 μm, and 0.57 μm. Note that the temperature profiles become isothermal at the top of the climate model grid when transferred to the larger photochemical model grid.

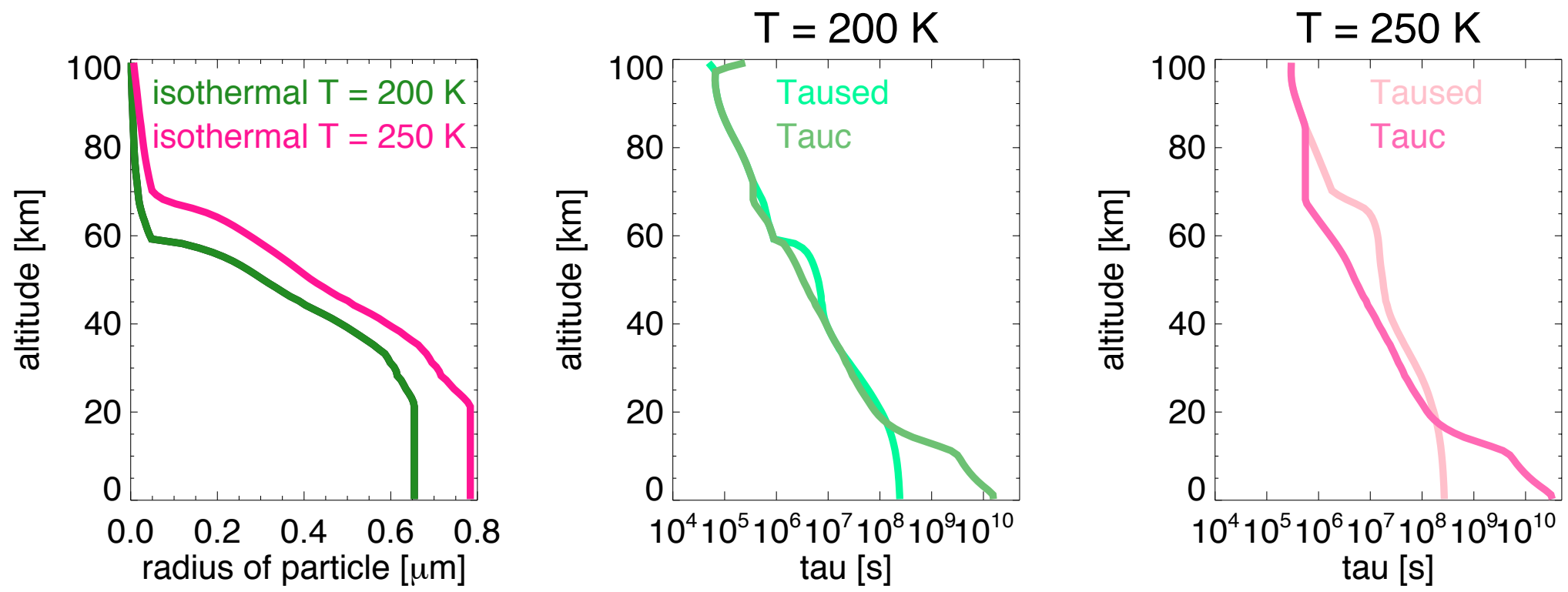

**Figure 6** The haze particle sizes for two completely isothermal atmospheres together with the coagulation and sedimentation timescales for these atmospheres.

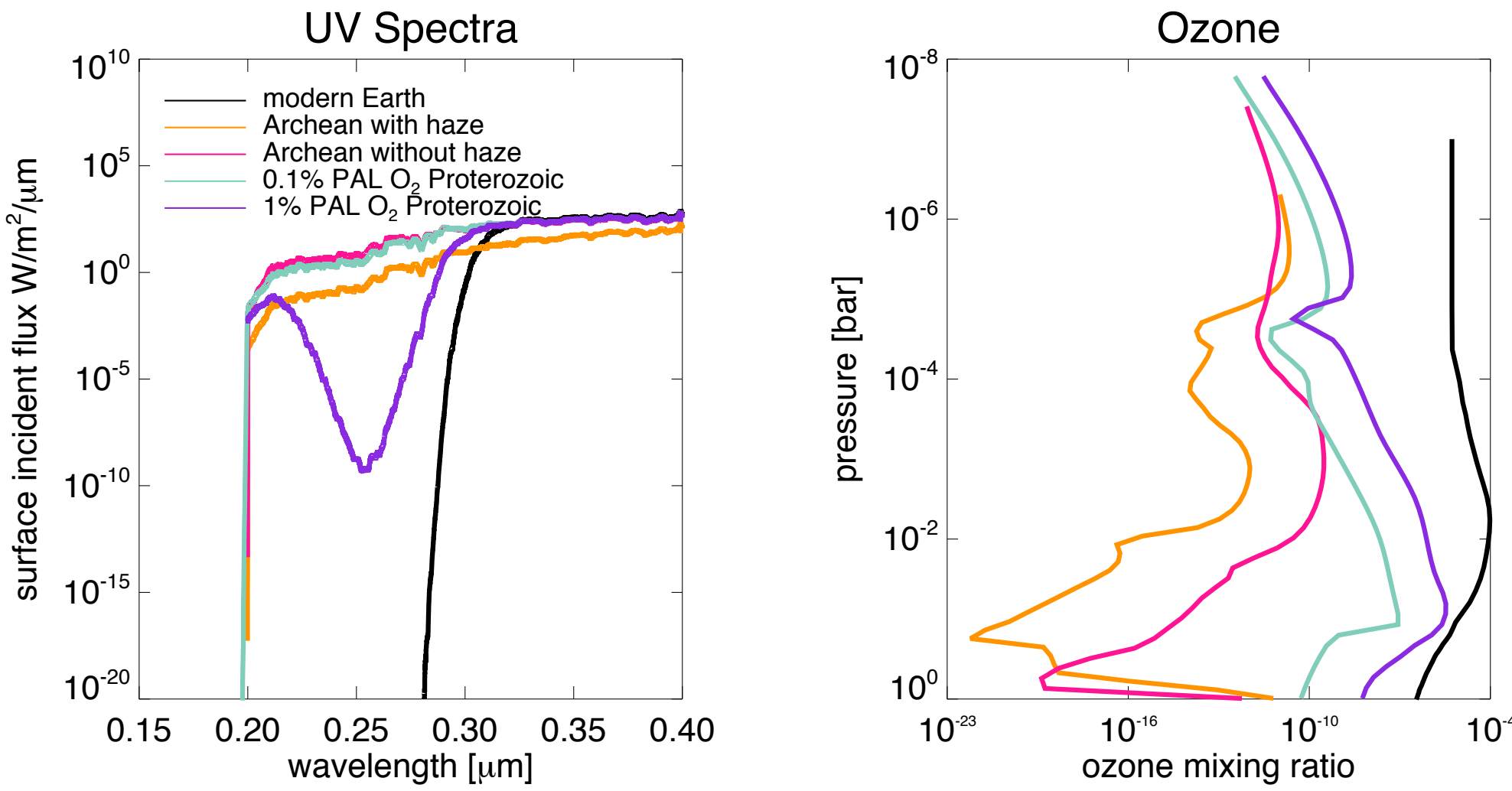

**Figure 7** Shown are surface UV spectra (left) and ozone column abundances (right) for Archean, Proterozoic, and modern Earth atmospheres. A modest amount of $O_2$ in the Proterozoic (1% PAL) produces a stronger UV shield than the Archean haze, but the haze shown here cuts out more UVA (320-400 nm) and UVB (280-320 nm) radiation than ozone in all situations. The haze can produce a stronger UV shield compared to the low $O_2$ atmosphere (0.1% PAL) proposed recently by Planavsky et al. (2014) for our atmospheric assumptions.

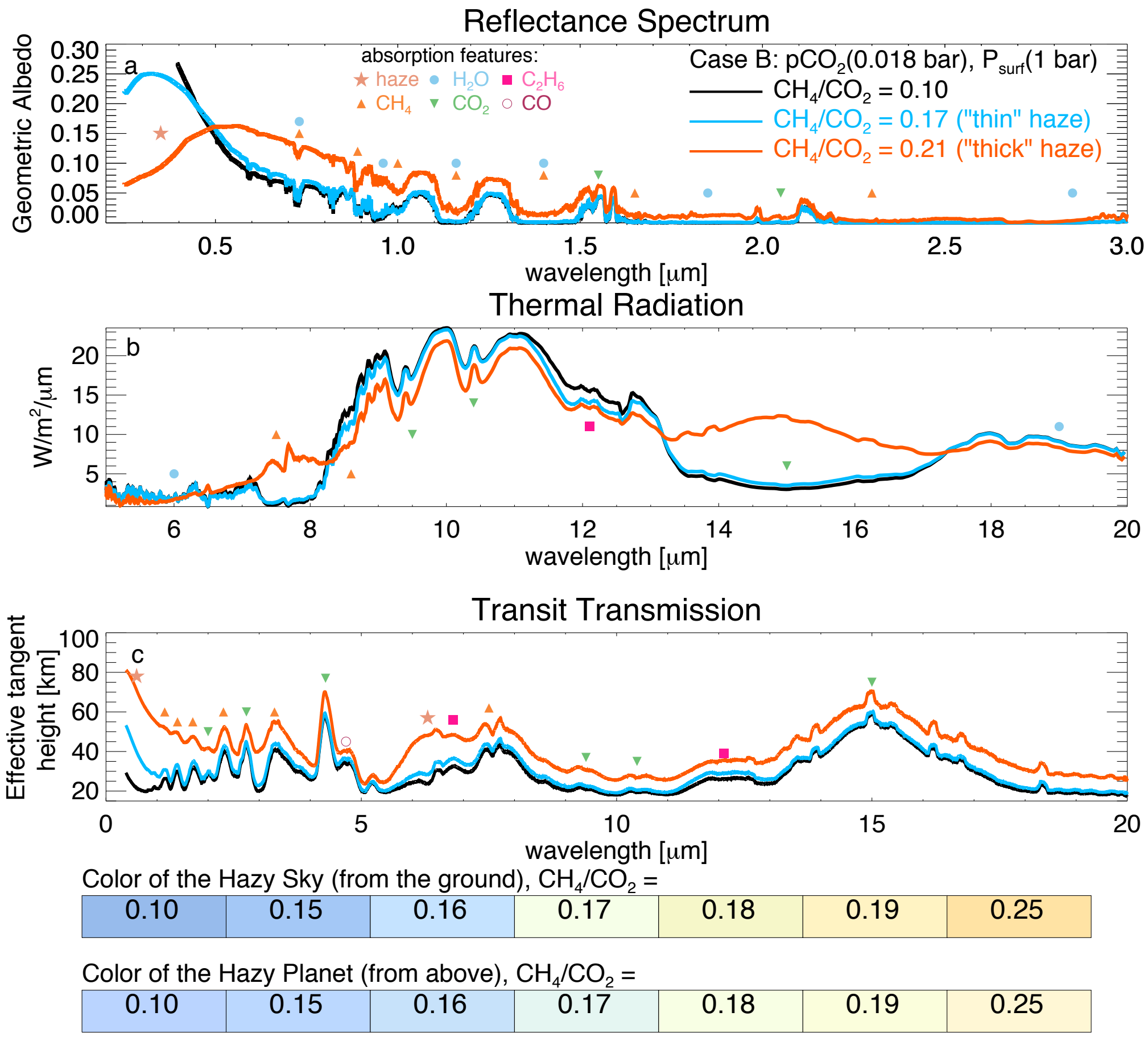

**Figure 8** Shown here are spectra for Case B. Haze and gas absorption features are labeled with the symbols indicated. Panel a: At short wavelengths in direct imaging, haze absorption decreases the planet's brightness; scattering brightens the planet at longer wavelengths. Panel b: Thermal emission from the hot stratosphere of the thickest haze planet ($CH_4/CO_2 = 0.21$) fills in absorption bands near 8 μm and 16 μm. Panel c: The y-axis shows the effective transit height above the planet's surface that light is able to penetrate, and absorption features are inverted compared to panels a) and b) due to an increase in the effective planet

radius during transit resulting from an increase in absorption at these wavelengths. The bottom section shows the approximate color of the hazy sky and planet. Sky colors are computed using the diffuse radiation spectrum at the ground. “Effective tangent height” refers to the minimum altitude above the planet’s surface that light is able to penetrate on transit transmission paths.

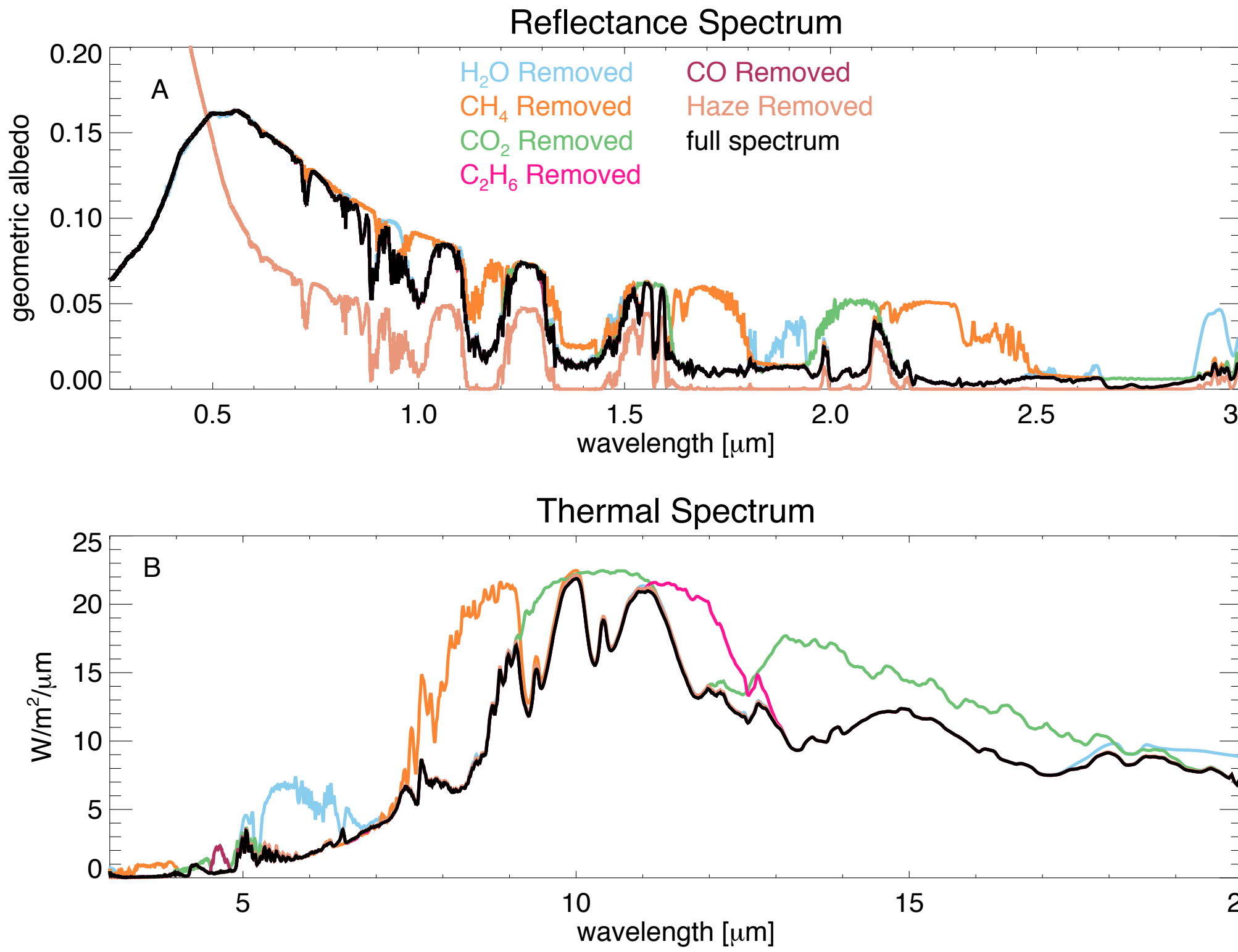

**Figure 9** A reflectance spectrum for a hazy Case B planet in the visible and near-infrared (panel A), and mid-infrared (panel B) is presented with gases and the hydrocarbon haze removed to show where each spectral component interacts with radiation. The full spectrum is shown in black. Places where the black spectrum deviates from the colored spectra indicate where each gas or haze absorbs. For example, the green line shows a spectrum where $CO_2$ is omitted, and a strong $CO_2$ feature is present near 15 μm in panel B as shown by the deviation of the green spectrum from the black spectrum. At some wavelengths, gas and haze absorptions are complex to detangle because multiple species are absorbing: in these cases, the key on figure 8 will indicate which gases are the dominant absorbers in a region.

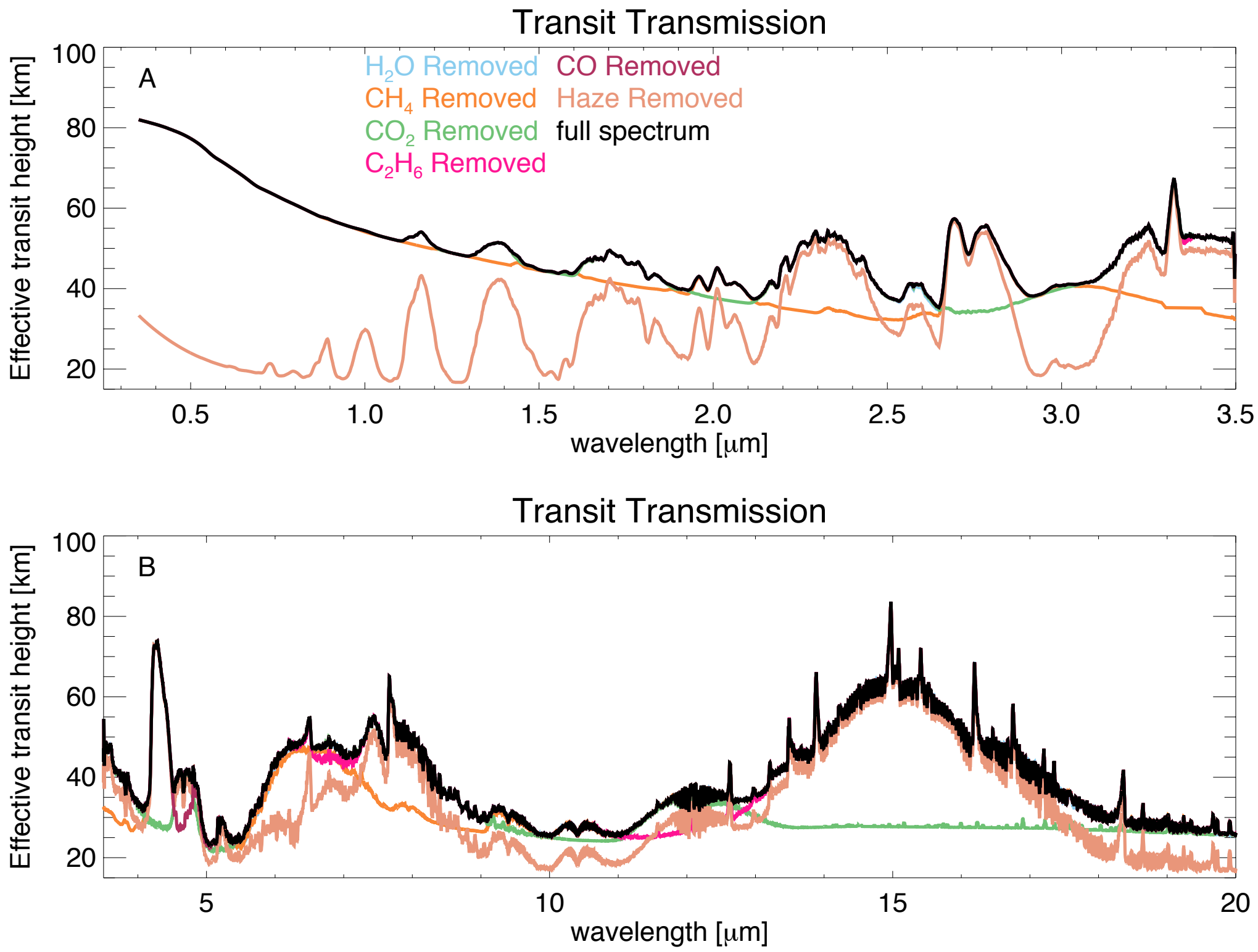

**Figure 10** A transit transmission spectrum for a hazy Case B planet in the visible and near-infrared (panel A), and mid-infrared (panel B) is presented with gases and the hydrocarbon haze removed to show where each spectral component interacts with radiation. The full spectrum is shown in black. Places where the black spectrum deviates from the colored spectra indicate where each gas or haze absorbs. For example, the orange line in panel A indicates $CH_4$ absorption features near 1.15 μm, 1.4 μm, 1.7 μm, 2.3 μm, and 3.3 μm.

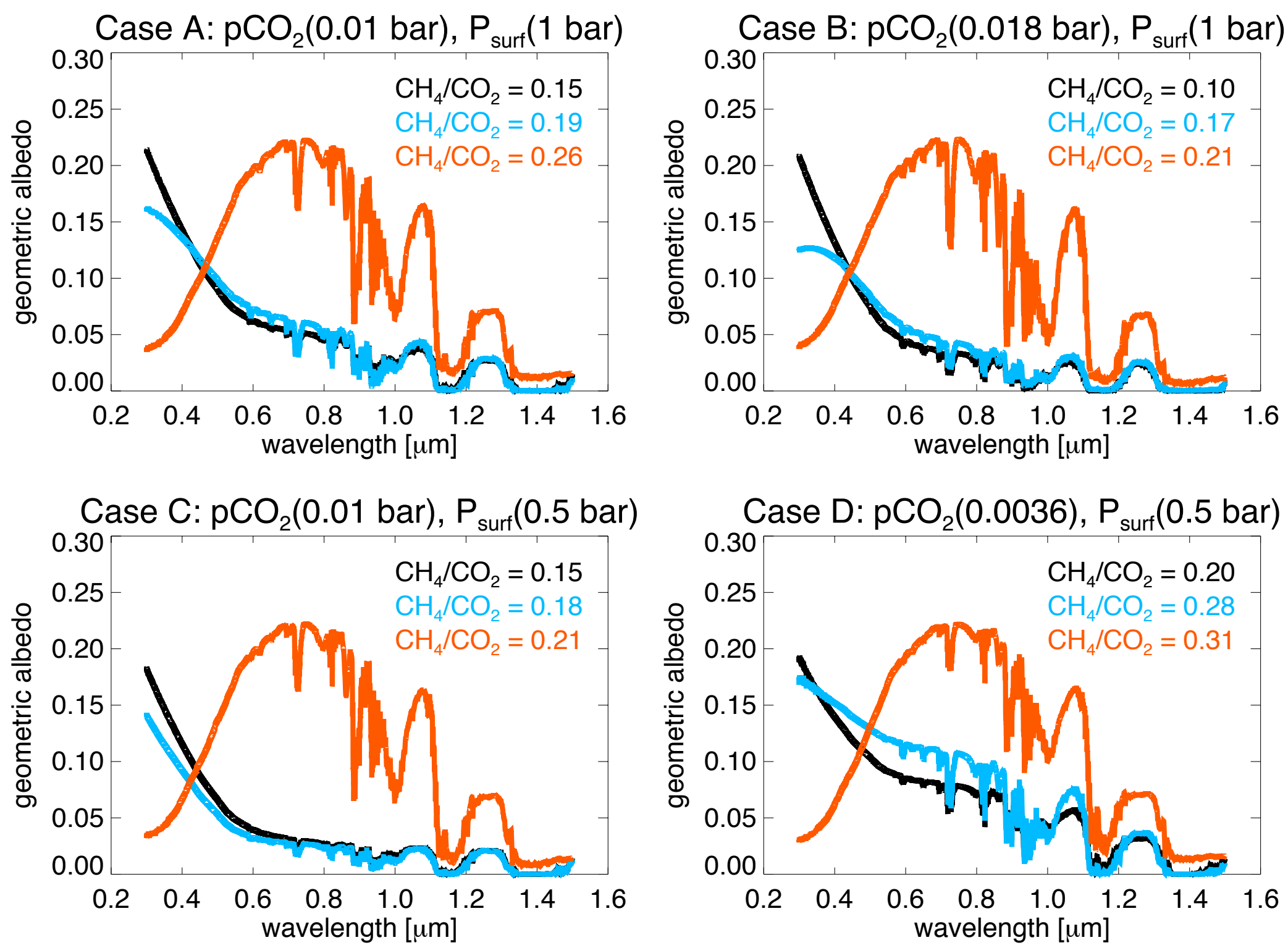

**Figure 11** Example reflectance spectra, intended as analogs for exoplanets like Archean Earth, for all of the types of planets investigated in this study are presented here. Fractional ice coverage is included in these spectra using a weighted average of icy and liquid water surfaces as described in the text.

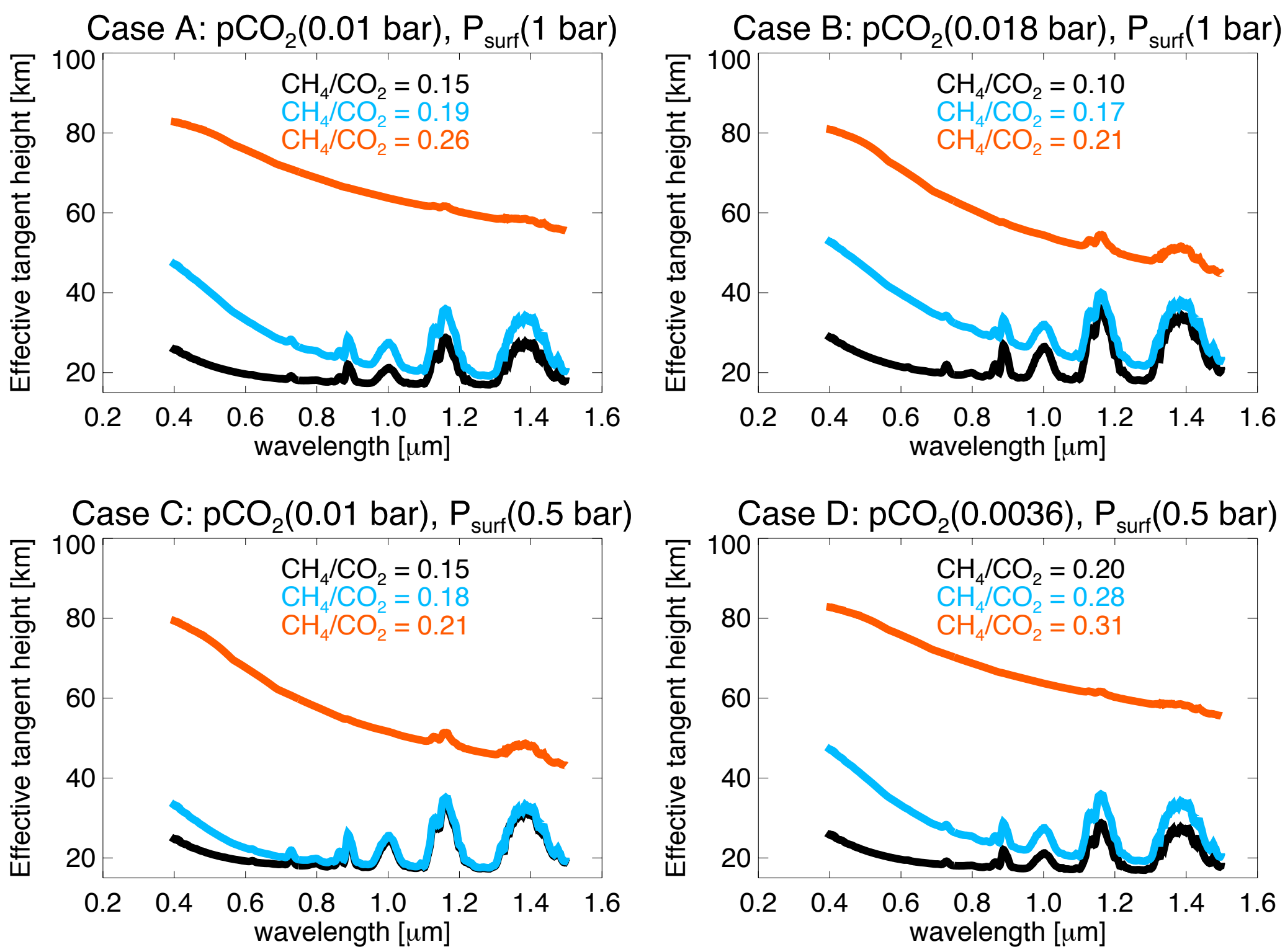

**Figure 12** Transit transmission spectra in the visible and NIR for cases A-D are presented here. For thicker hazes, absorption features shortward of approximately 1 μm vanish. These relatively featureless spectra result because high altitude hazes are effective at obscuring the lower atmosphere with the long path lengths taken by light in transit spectroscopy measurements.

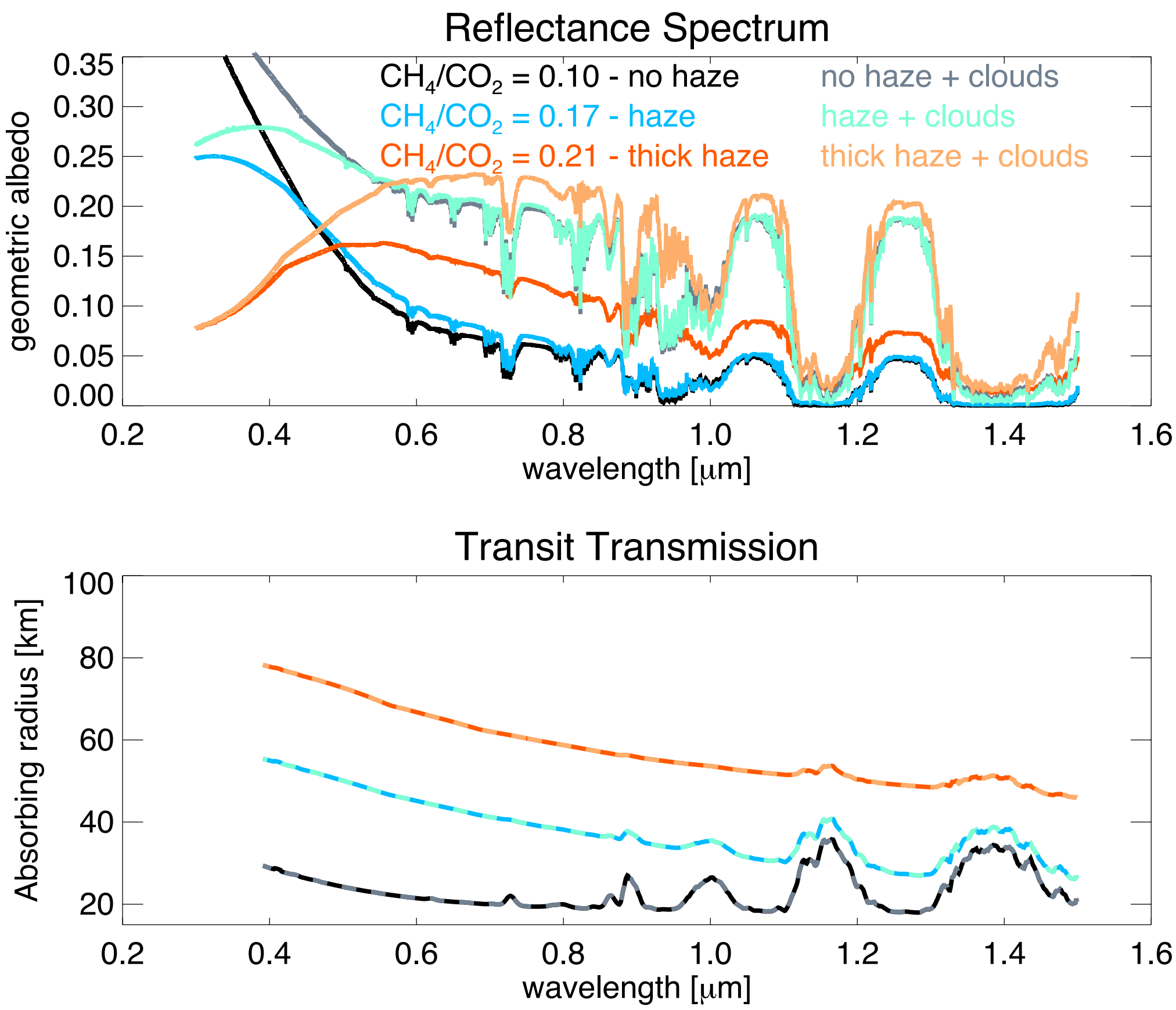

**Figure 13** Here we show the impact of water clouds on our Case B spectra with no haze, a thin haze, and a thick haze. The spectra with cloud and haze are shown in the pale colored lines. The dashed lines over our transit transmission spectra indicate that the spectra with and without water clouds are the same.

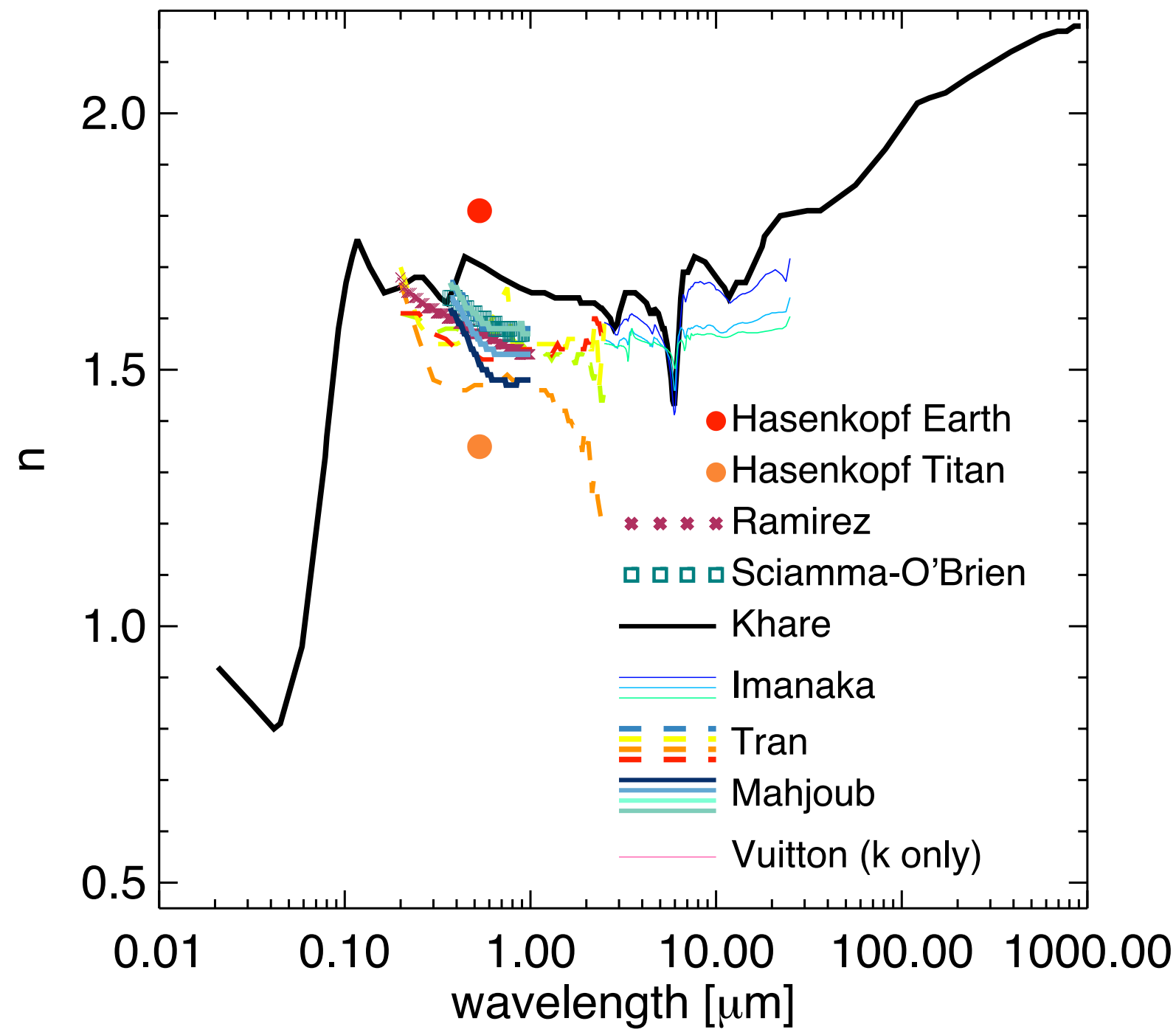
2.0
1.5
1.0
0.5
n
0.01
0.10
1.00
10.00
100.00
1000.00
wavelength [μm]
Hasenkopf Earth
Hasenkopf Titan
Ramirez
Sciamma-O'Brien
Khare
Imanaka
Tran
Mahjoub
Vuitton (k only)

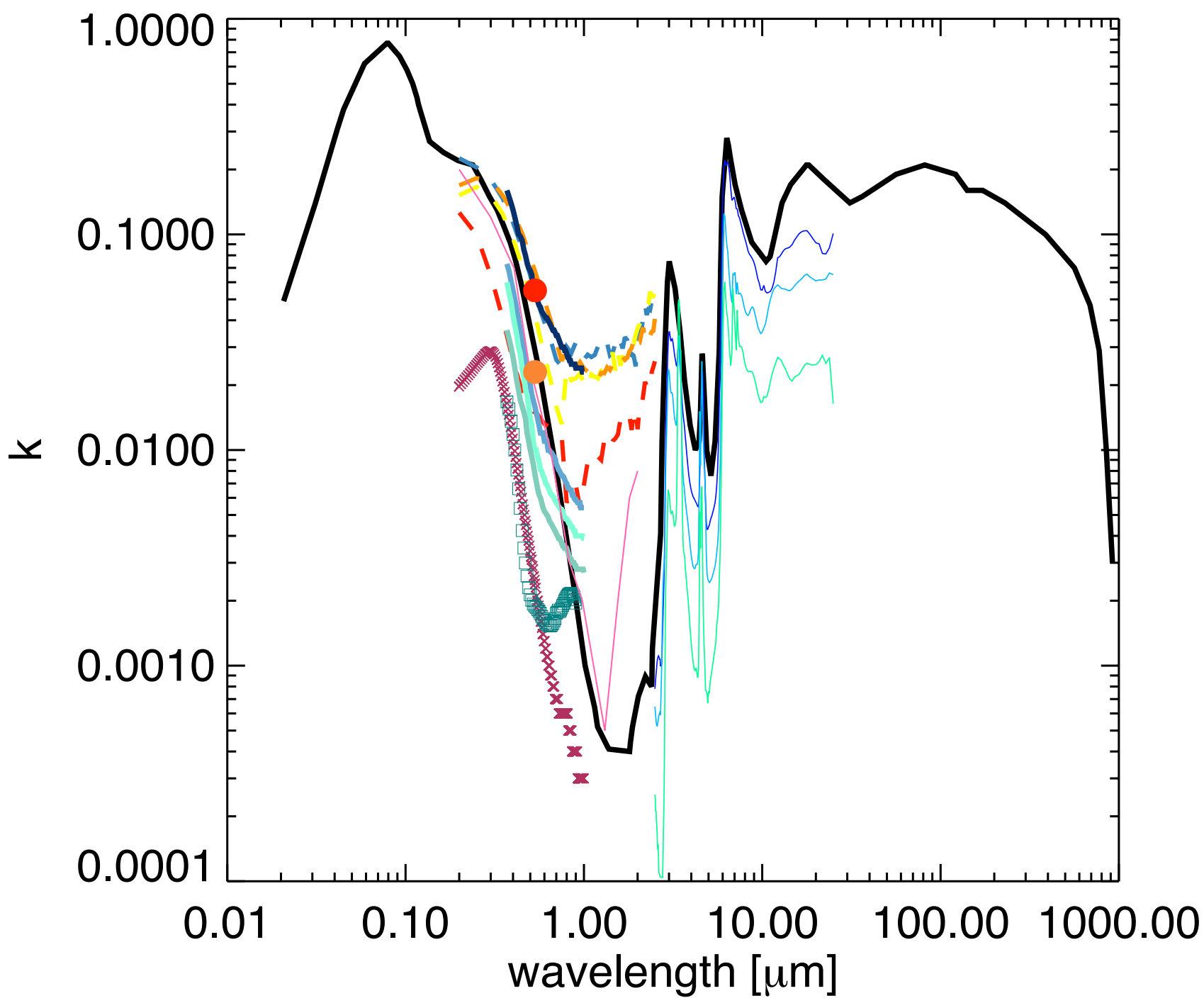
1.0000
0.1000
0.0100
0.0010
0.0001
k
0.01
0.10
1.00
10.00
100.00
1000.00
wavelength [μm]

**Figure 14** This shows the diversity of optical constants measured by several studies. The studies the figure key refers to are: Hasenkopf et al. 2010; Ramirez et al. 2002; Sciamma-O'Brien et al. 2012; Khare et al. 1984a; Imanaka et al. 2012; Tran et al. 2003; Mahjoub et al. 2012; Vuitton et al. 2009. Note in particular the single point measured under Archean Earth-like laboratory conditions by Hasenkopf et al. (2010).

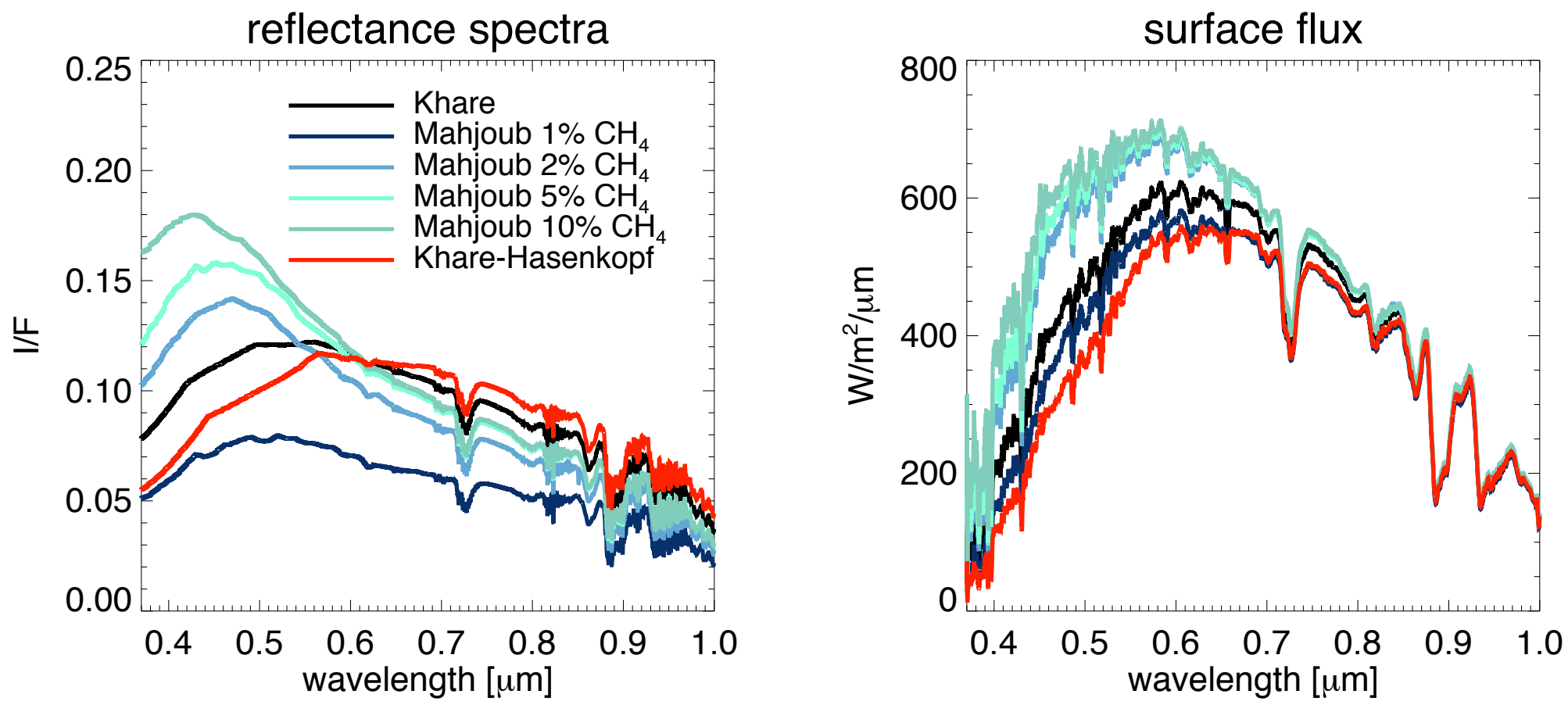

**Figure 15** A comparison of reflectance spectra and surface flux spectra using Khare et al. (1984a) and Mahjoub et al. (2012) optical constants, plus a spectrum generated by shifting the Khare constants to match the Archean haze refractive indices measured by Hasenkopf et al. (2010) at 532 nm (called "Khare Hasenkopf").

**Acknowledgments**

This work was performed as part of the NASA Astrobiology Institute's Virtual Planetary Laboratory, supported by the National Aeronautics and Space Administration through the NASA Astrobiology Institute under solicitation NNH12ZDA002C and Cooperative Agreement Number NNA13AA93A. G. Arney was supported in part by the NASA Astrobiology Institute Early Career Collaboration Award. E.T. Wolf acknowledges NASA Planetary Atmospheres Program award NNH13ZDA001N-PATM and NASA Exobiology Program award NNX10AR17G for financial support. B. Charnay acknowledges support from an appointment to the NASA Postdoctoral Program, administered by Universities Space Research Association. E. Hébrard was supported by an appointment to the NASA Postdoctoral Program at NASA Goddard Space Flight Center, administered by Universities Space Research Association through a contract with NASA. Simulations were facilitated through the use of the Hyak supercomputer system at the University of Washington eScience Institute. We are grateful to C. McKay and three other anonymous reviewers whose comments substantially improved the quality of our manuscript. We thank R. Buick, D. Crisp, N. Kiang, and M. Parenteau for conversations and advice. Spectra shown in this work will be archived at the Virtual Planetary Laboratory online spectral database.

**Author Disclosure Statement**

No competing financial interests exist.

Supplemental Table 1: List of reactions with rate constants and sources for the Archean photochemical code. For photolysis reactions at the bottom of the table, the "Reaction Rate Constant" refers to the reaction rate at the top of the atmosphere during a "standard" simulation for $pCO_2$ = 0.02, $pCH_4$ = 0.0035, 1 bar total pressure (a moderately hazy Case B atmosphere). Refer to Sander et al. (2006) for more information about reaction rate calculations.

| Rxn. # | Reaction | Reaction Rate Constant | Reference |
|---|---|---|---|
| 1. | $OCS + CH \rightarrow CO + HCS$ | $1.99\cdot10^{-10} \times e^{-190/T}$ | (Zabarnick *et al.*, 1989) |
| 2. | $OCS + H \rightarrow CO + HS$ | $9.07\cdot10^{-12} \times e^{-1940/T}$ | (Lee *et al.*, 1977) |
| 3. | $OCS + O \rightarrow S + CO_2$ | $8.3\cdot10^{-11} \times e^{-5530/T}$ | (Singleton and Cvetanovic 1988) |
| 4. | $OCS + O \rightarrow SO + CO$ | $2.1\cdot10^{-11} \times e^{-2200/T}$ | (Toon *et al.*, 1987) |
| 5. | $OCS + OH \rightarrow CO_2 + HS$ | $1.1\cdot10^{-13} \times e^{-1200/T}$ | (Atkinson *et al.*, 2004) |
| 6. | $OCS + S \rightarrow CO + S_2$ | $1.5\cdot10^{-10} \times e^{-1830/T}$ | (Schofield 1973) |
| 7. | $OCS + S + M \rightarrow OCS_2 + M$ | $8.3\cdot10^{-33} \times$ den | (Basco and Pearson 1967) |
| 8. | $OCS_2 + CO \rightarrow OCS + OCS$ | $3.0\cdot10^{-12}$ | (Zahnle *et al.*, 2006) |
| 9. | $OCS_2 + S \rightarrow OCS + S_2$ | $2.0\cdot10^{-11}$ | (Zahnle *et al.*, 2006) |
| 10 | $CH + CS_2 \rightarrow HCS + CS$ | $3.49\cdot10^{-10} \times e^{-40/T}$ | (Zabarnick *et al.*, 1989) |
| 11 | $CS + HS \rightarrow CS_2 + H$ | $1.5\cdot10^{-13} \times (1 + 0.6 \times$ den) | Assumed same as k(CO + OH) |
| 12 | $CS + O \rightarrow CO + S$ | $2.7\cdot10^{-10} \times e^{-760/T}$ | (Atkinson *et al.*, 2004) |
| 13 | $CS + O_2 \rightarrow CO + SO$ | $5\cdot10^{-20}$ | (Wine *et al.*, 1981) |
| 14 | $CS + O_2 \rightarrow OCS + O$ | $4\cdot10^{-19}$ | (Wine *et al.*, 1981) |
| 15 | $CS + O_3 \rightarrow CO + SO_2$ | $3\cdot10^{-12}$ | (Wine *et al.*, 1981) |
| 16 | $CS + O_3 \rightarrow OCS + O_2$ | $3\cdot10^{-12}$ | (Wine *et al.*, 1981) |
| 17 | $CS + O_3 \rightarrow SO + CO_2$ | $3\cdot10^{-12}$ | (Wine *et al.*, 1981) |
| 18 | $CS_2 + O \rightarrow CO + S_2$ | $5.81\cdot10^{-14}$ | (Singleton and Cvetanovic 1988) |
| 19 | $CS_2 + O \rightarrow OCS + S$ | $3\cdot10^{-12} \times e^{-650/T}$ | (Toon *et al.*, 1987) |
| 20 | $CS_2 + O \rightarrow SO + CS$ | $3.2\cdot10^{-11} \times e^{-650/T}$ | (Toon *et al.*, 1987) |
| 21 | $CS_2 + OH \rightarrow OCS + HS$ | $2\cdot10^{-15}$ | (Atkinson *et al.*, 2004) |
| 22 | $CS_2 + S \rightarrow CS + S_2$ | $1.9\cdot10^{-14} \times e^{-580/T} \times (T/300)^{3.97}$ | (Woiki and Roth 1995) |
| 23 | $CS_2 + SO \rightarrow OCS + S_2$ | $2.4\cdot10^{-13} \times e^{-2370/T}$ | Assumed same as k($SO^* + O_2$) |
| 24 | $CS_2^* + CS_2 \rightarrow CS + CS + S_2$ | $1\cdot10^{-12}$ | Assumed same as k($CS_2^* + CS_2$) |
| 25 | $CS_2^* + M \rightarrow CS_2 + M$ | $2.5\cdot10^{-11}$ | (Wine *et al.*, 1981) |
| 26 | $CS_2^* + O_2 \rightarrow CS + SO_2$ | $1\cdot10^{-12}$ | (Wine *et al.*, 1981) |
| 27 | $C + HS \rightarrow CS + H$ | $4\cdot10^{-11}$ | Assumed same as k(C + OH) |
| 28 | $C + S_2 \rightarrow CS + S$ | $3.3 \cdot 10^{-11}$ | Assumed same as k($C + O_2$) |
| 29 | $C_2 + S \rightarrow C + CS$ | $5 \cdot 10^{-11}$ | Assumed same as k($C_2 + O$) |
| 30 | $C_2 + S_2 \rightarrow CS + CS$ | $1.5 \cdot 10^{-11} \times e^{-550/T}$ | Assumed same as k($C_2 + O_2$) |
| 31 | $CH + S \rightarrow CS + H$ | $9.5 \cdot 10^{-11}$ | Assumed same as k($CH + CS_2$) |
| 32 | $CH + S_2 \rightarrow CS + HS$ | $5.9 \cdot 10^{-11}$ | Assumed same as k($CH + O_2$) |
| 33 | $CH_2^1 + S_2 \rightarrow HCS + HS$ | $3 \cdot 10^{-11}$ | Assumed same as k($CH_2^1 + O_2$) |
| 34 | $CH_3 + HCS \rightarrow CH_4 + CS$ | $5.0 \cdot 10^{-11}$ | Assumed same as k($CH_3$ + HCO) |
| 35 | H + CS +M → HCS +M | $2.0 \cdot 10^{-33} \times e^{-850/T} \times$ den | Assumed same as k(H + CO) |
| 36 | $H + HCS \rightarrow H_2 + CS$ | $1.2 \cdot 10^{-10}$ | Assumed same as k(H + HCO) |
| 37 | HS + CO → OCS + H | $4.2 \cdot 10^{-14} \times e^{-7650/T}$ | (Kurbanov and Mamedov 1995) |
| 38 | $HS + HCS \rightarrow H_2S + CS$ | $2.0 \cdot 10^{-11}$ | Assumed same as k(HS + HCO) |
| 39 | OCS + CH → CO + HCS | $1.99 \cdot 10^{-10} \times e^{-190/T}$ | (Zabarnick *et al.,* 1989) |
| 40 | S + CO +M → OCS +M | $6.5 \cdot 10^{-33} \times e^{-2180/T} \times$ den | Assumed same as k(CO + O) |
| 41 | $S + HCS \rightarrow H + CS_2$ | $5.0 \cdot 10^{-11}$ | Assumed same as k(O + HCO → $H + CO_2$) |
| 42 | S + HCS → HS + CS | $5.0 \cdot 10^{-11}$ | Assumed same as k(O + HCO → HS + CO) |
| 43 | $2CH_2^3 \rightarrow C_2H_2 + H_2$ | $5.3 \cdot 10^{-11}$ | (Braun *et al.,* 1970) |
| 44 | $C + H_2 + M \rightarrow CH_2^3 + M$ | $k_0 = 8.75 \cdot 10^{-31} \times e^{524/T}$<br>$k_\infty = 8.3 \cdot 10^{-11}$ | (Zahnle 1986) |
| 45 | $C + O_2 \rightarrow CO + O$ | $3.3 \cdot 10^{-11}$ | (Donovan and Husain 1970) |
| 46 | C + OH → CO + H | $4 \cdot 10^{-11}$ | (Giguere and Huebner 1978) |
| 47 | $C_2 + CH_4 \rightarrow C_2H + CH_3$ | $5.05 \cdot 10^{-11} \times e^{-297/T}$ | (Pitts *et al.,* 1982) |
| 48 | $C_2 + H_2 \rightarrow C_2H + H$ | $1.77 \cdot 10^{-10} \times e^{-1469/T}$ | (Pitts *et al.,* 1982) |
| 49 | $C_2 + O \rightarrow C + CO$ | $5 \cdot 10^{-11}$ | (Prasad and Huntress 1980) |
| 50 | $C_2 + O_2 \rightarrow CO + CO$ | $1.5 \cdot 10^{-11} \times e^{-550/T}$ | (Baughcum and Oldenborg 1984) |
| 51 | $C_2H + C_2H_2 \rightarrow$ HCAER + H | $1.5 \cdot 10^{-10}$ | (Stephens *et al.,* 1987) |
| 52 | $C_2H + C_2H_6 \rightarrow C_2H_2 + C_2H_5$ | $3.6 \cdot 10^{-11}$ | (Lander *et al.,* 1990) |
| 53 | $C_2H + C_3H_8 \rightarrow C_2H_2 + C_3H_7$ | $1.4 \cdot 10^{-11}$ | (Okabe 1983) |
| 54 | $C_2H + CH_2CCH_2 \rightarrow$ HCAER2 + H | $1.5 \cdot 10^{-10}$ | (Pavlov *et al.,* 2001b) |
| 55 | $C_2H + CH_4 \rightarrow C_2H_2 + CH_3$ | $6.94 \cdot 10^{-12} \times e^{-250/T}$ | (Allen *et al.,* 1992; Lander *et al.,* 1990) |
| 56 | $C_2H + H + M \rightarrow C_2H_2 + M$ | $k_0 = 2.64 \cdot 10^{-26} \times e^{-721/T} \times (T/300)^{-3.1}$<br>$k_\infty = 3.0 \cdot 10^{-10}$ | (Tsang and Hampson 1986) |
| 57 | $C_2H + H_2 \rightarrow C_2H_2 + H$ | $5.58 \cdot 10^{-11} \times e^{-1443/T}$ | (Allen *et al.,* 1992; Stephens *et al.,* 1987) |
| 58 | $C_2H + O \rightarrow CO + CH$ | $1\cdot10^{-10} \times e^{-250/T}$ | (Zahnle 1986) |
| 59 | $C_2H + O_2 \rightarrow CO + HCO$ | $2\cdot10^{-11}$ | (Brown and Laufer 1981) |
| 60 | $C_2H_2 + H + M \rightarrow C_2H_3 + M$ | $k_0 = 2.6\cdot10^{-31}$ <br> $k_\infty = 8.3\cdot10^{-11}\times e^{-1374/T}$ | (Romani *et al.,* 1993) |
| 61 | $C_2H_2 + O \rightarrow CH_2^3 + CO$ | $2.9\cdot10^{-11} \times e^{-1600/T}$ | (Zahnle 1986) |
| 62 | $C_2H_2 + OH + M \rightarrow C_2H_2OH + M$ | $k_0 = 5.5\cdot10^{-30}$ + <br> $k_\infty = 8.3\cdot10^{-13} \times (T/300)^{-2}$ | (Sander *et al.,* 2006) |
| 63 | $C_2H_2 + OH + M \rightarrow CH_2CO + H + M$ | $k_0 = 5.8\cdot10^{-31} \times e^{1258/T}$ <br> $k_\infty = 1.4\cdot10^{-12}\times e^{388/T}$ | (Perry and Williamson 1982) |
| 64 | $C_2H_2 + OH \rightarrow CO + CH_3$ | $2.\cdot10^{-12} \times e^{-250/T}$ | (Hampson and Garvin 1977) |
| 65 | $C_2H_2OH + H \rightarrow H_2 + CH_2CO$ | $3.3\cdot10^{-11} \times e^{-2000/T}$ | (Miller *et al.,* 1982) |
| 66 | $C_2H_2OH + H \rightarrow H_2O + C_2H_2$ | $5\cdot10^{-11}$ | (Miller *et al.,* 1982) |
| 67 | $C_2H_2OH + O \rightarrow OH + CH_2CO$ | $3.3\cdot10^{-11} \times e^{-2000/T}$ | (Miller *et al.,* 1982) |
| 68 | $C_2H_2OH + OH \rightarrow H_2O + CH_2CO$ | $1.7\cdot10^{-11} \times e^{-1000/T}$ | (Miller *et al.,* 1982) |
| 69 | $C_2H_3 + C_2H_3 \rightarrow C_2H_4 + C_2H_2$ | $2.4\cdot10^{-11}$ | (Fahr *et al.,* 1991) |
| 70 | $C_2H_3 + C_2H_5 \rightarrow C_2H_4 + C_2H_4$ | $3\cdot10^{-12}$ | (Laufer *et al.,* 1983) |
| 71 | $C_2H_3 + C_2H_5 + M \rightarrow CH_3 + C_3H_5 + M$ | $k_0 = 1.9\cdot10^{-27}$ <br> $k_\infty = 2.5\cdot10^{-11}$ | (Romani *et al.,* 1993) |
| 72 | $C_2H_3 + C_2H_6 \rightarrow C_2H_4 + C_2H_5$ | $3.\cdot10^{-13} \times e^{-5170/T}$ | (Kasting *et al.,* 1983) |
| 73 | $C_2H_3 + CH_3 \rightarrow C_2H_2 + CH_4$ | $3.4\cdot10^{-11}$ | (Fahr *et al.,* 1991) |
| 74 | $C_2H_3 + CH_3 + M \rightarrow C_3H_6 + M$ | $k_0 = 1.3\cdot10^{-22}$ <br> $k_\infty = 1.2\cdot10^{-10}$ | (Raymond *et al.,* 2006) |
| 75 | $C_2H_3 + CH_4 \rightarrow C_2H_4 + CH_3$ | $2.4\cdot10^{-24} \times e^{-2754/T} \times T^{4.02}$ | (Tsang and Hampson 1986) |
| 76 | $C_2H_3 + H \rightarrow C_2H_2 + H_2$ | $3.3\cdot10^{-11}$ | (Warnatz 1984) |
| 77 | $C_2H_3 + H_2 \rightarrow C_2H_4 + H$ | $2.6\cdot10^{-13} \times e^{-2646/T}$ | (Allen *et al.,* 1992) |
| 78 | $C_2H_3 + O \rightarrow CH_2CO + H$ | $5.5\cdot10^{-11}$ | (Hoyermann *et al.,* 1981) |
| 79 | $C_2H_3 + OH \rightarrow C_2H_2 + H_2O$ | $8.3\cdot10^{-12}$ | (Benson and Haugen 1967) |
| 80 | $C_2H_4 + H + M \rightarrow C_2H_5 + M$ | $k_0 = 2.15\cdot10^{-29} \times e^{-349/T}$ <br> $k_\infty = 4.95\cdot10^{-11}\times e^{-1051/T}$ | (Lightfoot and Pilling 1987) |
| 81 | $C_2H_4 + O \rightarrow HCO + CH_3$ | $5.5\cdot10^{-12} \times e^{-565/T}$ | (Hampson and Garvin 1977) |
| 82 | $C_2H_4 + OH + M \rightarrow C_2H_4OH + M$ | $k_0 = 1.0\cdot10^{-28} \times (T/300)^{4.5}$ <br> $k_\infty = 8.8\cdot10^{-12} \times (T/300)^{0.85}$ | (Sander *et al.,* 2006) |
| 83 | $C_2H_4 + OH \rightarrow H_2CO + CH_3$ | $2.2\cdot10^{-12} \times e^{385/T}$ | (Hampson and Garvin 1977) |
| 84 | $C_2H_4OH + H \rightarrow H_2 + CH_3CHO$ | $3.3\cdot10^{-11} \times e^{-2000/T}$ | (Zahnle and Kasting 1986) |
| 85 | $C_2H_4OH + H \rightarrow H_2O + C_2H_4$ | $5\cdot10^{-11}$ | (Miller *et al.,* 1982) |
| 86 | $C_2H_4OH + O \rightarrow OH + CH_3CHO$ | $3.3\cdot10^{-11} \times e^{-2000/T}$ | (Zahnle and Kasting 1986) |
| 87 | $C_2H_4OH + OH \rightarrow H_2O + CH_3CHO$ | $1.7\cdot10^{-11} \times e^{-1000/T}$ | (Zahnle and Kasting 1986) |
| 88 | $C_2H_5 + C_2H_3 \rightarrow C_2H_6 + C_2H_2$ | $6\cdot10^{-12}$ | (Laufer *et al.,* 1983) |
| 89 | $C_2H_5 + C_2H_5 \rightarrow C_2H_6 + C_2H_4$ | $2.3\cdot10^{-12}$ | (Tsang and Hampson 1986) |
| 90 | $C_2H_5 + CH_3 \rightarrow C_2H_4 + CH_4$ | $1.88\cdot10^{-12} \times (T/300)^{-0.5}$ | (Romani *et al.,* 1993) |
| 91 | $C_2H_5 + CH_3 + M \rightarrow C_3H_8 + M$ | $k_0 = 3.9\cdot10^{-10} \times (T/300)^{2.5}$ $k_\infty = 1.4\cdot10^{-8}\times (T/300)^{0.5}$ | (Romani *et al.,* 1993) |
| 92 | $C_2H_5 + H \rightarrow C_2H_4 + H_2$ | $3\cdot10^{-12}$ | (Tsang and Hampson 1986) |
| 93 | $C_2H_5 + H + M \rightarrow C_2H_6 + M$ | $k_0 = 5.5\cdot10^{-23} \times e^{-1040/T}$ $k_\infty = 1.5\cdot10^{-10}$ | (Gladstone *et al.,* 1996) |
| 94 | $C_2H_5 + H \rightarrow CH_3 + CH_3$ | $6.00\cdot10^{-11}$ | (Baluch, 1992) |
| 95 | $C_2H_5 + HCO \rightarrow C_2H_6 + CO$ | $1\cdot10^{-10}$ | (Pavlov *et al.,* 2001b) |
| 96 | $C_2H_5 + HNO \rightarrow C_2H_6 + NO$ | $3\cdot10^{-14}$ | (Pavlov *et al.,* 2001b) |
| 97 | $C_2H_5 + O \rightarrow CH_3 + HCO + H$ | $3.0\cdot10^{-11}$ | (Tsang and Hampson 1986) |
| 98 | $C_2H_5 + O \rightarrow CH_3CHO + H$ | $1.33\cdot10^{-10}$ | (Tsang and Hampson 1986) |
| 99 | $C_2H_5 + O \rightarrow H_2CO + CH_3$ | $2.67\cdot10^{-11}$ | (Tsang and Hampson, 1986) |
| 10 | $C_2H_5 + O_2 + M \rightarrow CH_3 + HCO + OH + M$ | $k_0 = 1.5\cdot10^{-28} \times (T/300)^{3.0}$ $k_\infty = 1.9\cdot10^{-11}$ | (Sander *et al.,* 2006) |
| 10 | $C_2H_5 + OH \rightarrow CH_3CHO + H_2$ | $1\cdot10^{-10}$ | (Pavlov *et al.,* 2001b) |
| 10 | $C_2H_5 + OH \rightarrow C_2H_4 + H_2O$ | $4.0\cdot10^{-11}$ | (Pavlov *et al.,* 2001b) |
| 10 | $C_2H_6 + O \rightarrow C_2H_5 + OH$ | $8.62\cdot10^{-12} \times e^{-2920/T} \times (T/300)^{1.5}$ | (Baulch *et al.,* 1994) |
| 10 | $C_2H_6 + O^1D \rightarrow C_2H_5 + OH$ | $6.29\cdot10^{-10}$ | (Matsumi *et al.,* 1993) |
| 10 | $C_2H_6 + OH \rightarrow C_2H_5 + H_2O$ | $8.7\cdot10^{-12} \times e^{-1070/T}$ | (Sander *et al.,* 2006) |
| 10 | $C_3H_2 + H + M \rightarrow C_3H_3 + M$ | $k_0 = 1.7\cdot10^{-26}$ $k_\infty = 1.5\cdot10^{-10}$ | (Yung *et al.,* 1984) |
| 10 | $C_3H_3 + H + M \rightarrow CH_2CCH_2 + M$ | $k_0 = 1.7\cdot10^{-26}$ $k_\infty = 1.5\cdot10^{-10}$ | (Yung *et al.,* 1984) |
| 10 | $C_3H_3 + H + M \rightarrow CH_3C_2H + M$ | $k_0 = 1.7 \cdot 10^{-26}$ $k_\infty = 1.5 \cdot 10^{-10}$ | (Yung *et al.*, 1984) |
| 10 | $C_3H_5 + CH_3 \rightarrow CH_2CCH_2 + CH_4$ | $4.5 \cdot 10^{-12}$ | (Yung *et al.*, 1984) |
| 11 | $C_3H_5 + CH_3 \rightarrow CH_3C_2H + CH_4$ | $4.5 \cdot 10^{-12}$ | (Yung *et al.*, 1984) |
| 11 | $C_3H_5 + H + M \rightarrow C_3H_6 + M$ | $k_0 = 1.0 \cdot 10^{-28}$ $k_\infty = 1.0 \cdot 10^{-11}$ | (Yung *et al.*, 1984) |
| 11 | $C_3H_5 + H \rightarrow CH_2CCH_2 + H_2$ | $1.5 \cdot 10^{-11}$ | (Yung *et al.*, 1984) |
| 11 | $C_3H_5 + H \rightarrow CH_3C_2H + H_2$ | $1.5 \cdot 10^{-11}$ | (Yung *et al.*, 1984) |
| 11 | $C_3H_5 + H \rightarrow CH_4 + C_2H_2$ | $1.5 \cdot 10^{-11}$ | (Yung *et al.*, 1984) |
| 11 | $C_3H_6 + H + M \rightarrow C_3H_7 + M$ | $k_0 = 2.15 \cdot 10^{-29} \times e^{-349/T}$ $k_\infty = 4.95 \cdot 10^{-11} \times e^{-1051/T}$ | (Pavlov *et al.*, 2001b) assumed same as $k(C_2H_4 + H)$ |
| 11 | $C_3H_6 + O \rightarrow CH_3 + CH_3CO$ | $4.1 \cdot 10^{-12} \times e^{-38/T}$ | (Hampson and Garvin 1977) |
| 11 | $C_3H_6 + O \rightarrow CH_3 + CH_3 + CO$ | $4.1 \cdot 10^{-12}\ e^{-38/T}$ | Hampson and Garvin (1977) |
| 11 | $C_3H_6 + OH \rightarrow CH_3CHO + CH_3$ | $4.1 \cdot 10^{-12} \times e^{540/T}$ | (Hampson and Garvin 1977) |
| 11 | $C_3H_7 + CH_3 \rightarrow C_3H_6 + CH_4$ | $2.5 \cdot 10^{-12} \times e^{-200/T}$ | (Yung *et al.*, 1984) |
| 12 | $C_3H_7 + H \rightarrow CH_3 + C_2H_5$ | $7.95 \cdot 10^{-11} \times e^{-127/T}$ | (Pavlov *et al.*, 2001b) |
| 12 | $C_3H_7 + O \rightarrow C_2H_5CHO + H$ | $1.1 \cdot 10^{-10}$ | (Pavlov *et al.*, 2001b) |
| 12 | $C_3H_7 + OH \rightarrow C_2H_5CHO + H_2$ | $1.1 \cdot 10^{-10}$ | (Pavlov *et al.*, 2001b) |
| 12 | $C_3H_8 + O + M \rightarrow C_3H_7 + OH + M$ | $k_0 = 1.6 \cdot 10^{-11} \times e^{-2900/T}$ $k_\infty = 2.2 \cdot 10^{-11} \times e^{-2200/T}$ | (Hampson and Garvin 1977) |
| 12 | $C_3H_8 + O^1D \rightarrow C_3H_7 + OH$ | $1.4 \cdot 10^{-10}$ | (Pavlov *et al.*, 2001b) |
| 12 | $C_3H_8 + OH \rightarrow C_3H_7 + H_2O$ | $1.1 \cdot 10^{-11} \times e^{-700/T}$ | (DeMore *et al.*, 1992) |
| 12 | $CH + C_2H_2 + M \rightarrow C_3H_2 + H + M$ | $k_0 = 2.15 \cdot 10^{-29} \times e^{-349/T}$ $k_\infty = 4.95 \cdot 10^{-11} \times e^{-1051/T}$ | (Romani *et al.*, 1993) |
| 12 | $CH + C_2H_4 + M \rightarrow CH_2CCH_2 + H + M$ | $k_0 = 1.75 \cdot 10^{-10} \times e^{61/T}$ $k_\infty = 5.3 \cdot 10^{-10}$ | (Romani *et al.*, 1993) |
| 12 | $CH + C_2H_4 + M \rightarrow CH_3C_2H + H + M$ | $k_0 = 1.75 \cdot 10^{-10} \times e^{61/T}$ $k_\infty = 5.3 \cdot 10^{-10}$ | (Romani *et al.*, 1993) |
| 12 | $CH + CH_4 + M \rightarrow C_2H_4 + H + M$ | $k_0 = 2.5 \cdot 10^{-11} \times e^{200/T}$ $k_\infty = 1.7 \cdot 10^{-10}$ | (Romani *et al.*, 1993) |
| 13 | $CH + CO_2 \rightarrow HCO + CO$ | $5.9 \cdot 10^{-12} \times e^{-350/T}$ | (Berman *et al.*, 1982) |
| 13 | $CH + H \rightarrow C + H_2$ | $1.4 \cdot 10^{-11}$ | (Becker *et al.*, 1989) |
| 13 | $CH + H_2 \rightarrow CH_2^3 + H$ | $2.38 \cdot 10^{-10} \times e^{-1760/T}$ | (Zabarnick *et al.*, 1986) |
| 13 | $CH + H_2 + M \rightarrow CH_3 + M$ | $k_0 = 8.75 \cdot 10^{-31} \times e^{524/T}$ $k_\infty = 8.3 \cdot 10^{-11}$ | (Romani *et al.*, 1993) |
| 13 | $CH + O \rightarrow CO + H$ | $9.5 \cdot 10^{-11}$ | (Messing *et al.*, 1981) |
| 13 | $CH + O_2 \rightarrow CO + OH$ | $5.9\cdot10^{-11}$ | (Butler *et al.,* 1981) |
| 13 | $CH_2^1 + CH_4 \rightarrow CH_3 + CH_3$ | $7.14\cdot10^{-12} \times e^{-5050/T}$ | (Böhland *et al.,* 1985) |
| 13 | $CH_2^1 + CO_2 \rightarrow H_2CO + CO$ | $1\cdot10^{-12}$ | (Zahnle 1986) |
| 13 | $CH_2^1 + H_2 \rightarrow CH_2^3 + H_2$ | $1.26\cdot10^{-11}$ | (Romani *et al.,* 1993) |
| 13 | $CH_2^1 + H_2 \rightarrow CH_3 + H$ | $5\cdot10^{-15}$ | (Tsang and Hampson 1986) |
| 14 | $CH_2^1 + M \rightarrow CH_2^3 + M$ | $8.8\cdot10^{-12}$ | (Ashfold *et al.,* 1981) |
| 14 | $CH_2^1 + O_2 \rightarrow HCO + OH$ | $3\cdot10^{-11}$ | (Ashfold *et al.,* 1981) |
| 14 | $CH_2^3 + C_2H_2 + M \rightarrow CH_2CCH_2 + M$ | $k_0 = 3.8\cdot10^{-25}$<br>$k_\infty = 3.7\cdot10^{-12}$ | (Laufer 1981; Laufer *et al.,* 1983) |
| 14 | $CH_2^3 + C_2H_2 + M \rightarrow CH_3C_2H + M$ | $k_0 = 3.8\cdot10^{-25}$<br>$k_\infty = 2.2\cdot10^{-12}$ | (Laufer 1981; Laufer *et al.,* 1983) |
| 14 | $CH_2^3 + C_2H_3 \rightarrow CH_3 + C_2H_2$ | $3\cdot10^{-11}$ | (Tsang and Hampson 1986) |
| 14 | $CH_2^3 + C_2H_5 \rightarrow CH_3 + C_2H_4$ | $3\cdot10^{-11}$ | (Tsang and Hampson 1986) |
| 14 | $CH_2^3 + CH_3 \rightarrow C_2H_4 + H$ | $7\cdot10^{-11}$ | (Tsang and Hampson 1986) |
| 14 | $CH_2^3 + CO + M \rightarrow CH_2CO + M$ | $k_0 = 1.0\cdot10^{-28}$<br>$k_\infty = 1.0\cdot10^{-15}$ | (Yung *et al.,* 1984) |
| 14 | $CH_2^3 + CO_2 \rightarrow H_2CO + CO$ | $1.0\cdot10^{-14}$ | (Laufer 1981) |
| 14 | $CH_2^3 + H \rightarrow CH + H_2$ | $4.7\cdot10^{-10} \times e^{-370/T}$ | (Zabarnick *et al.,* 1986) |
| 15 | $CH_2^3 + H + M \rightarrow CH_3 + M$ | $k_0 = 3.1\cdot10^{-30} \times e^{457/T}$<br>$k_\infty = 1.5\cdot10^{-10}$ | (Gladstone *et al.,* 1996) |
| 15 | $CH_2^3 + O \rightarrow CH + OH$ | $8\cdot10^{-12}$ | (Huebner and Giguere 1980) |
| 15 | $CH_2^3 + O \rightarrow CO + HH$ | $8.3\cdot10^{-11}$ | (Homann and Wellmann 1983) |
| 15 | $CH_2^3 + O \rightarrow HCO + H$ | $1\cdot10^{-11}$ | (Huebner and Giguere 1980) |
| 15 | $CH_2^3 + O_2 \rightarrow HCO + OH$ | $4.1\cdot10^{-11} \times e^{-750/T}$ | (Baulch *et al.,* 1994) |
| 15 | $CH_2^3 + S_2 \rightarrow HCS + HS$ | $4.1\cdot10^{-11}\ e^{-750/T}$ | Assumed same as $k(CH_2^3 + O_2)$ |
| 15 | $CH_2CCH_2 + H \rightarrow C_3H_5$ | $k_0 = 8.9\cdot10^{-29} \times e^{-1225/T} \times (T/300)^{-2.0}$<br>$k_\infty = 1.4\cdot10^{-11} \times e^{-1000/T}$ | (Yung *et al.,* 1984) |
| 15 | $CH_2CCH_2 + H \rightarrow CH_3 + C_2H_2$ | $k_0 = 8.9\cdot10^{-29} \times e^{-1225/T} \times (T/300)^{-2.0}$<br>$k_\infty = 9.7\cdot10^{-13} \times e^{-1550/T}$ | (Yung *et al.,* 1984) |
| 15 | $CH_2CCH_2 + H \rightarrow CH_3C_2H + H$ | $1\cdot10^{-11} \times e^{-1000/T}$ | (Yung *et al.,* 1984) |
| 15 | $CH_2CO + H \rightarrow CH_3 + CO$ | $1.9\cdot10^{-11} \times e^{-1725/T}$ | (Michael *et al.,* 1979) |
| 16 | $CH_2CO + O \rightarrow H_2CO + CO$ | $3.3\cdot10^{-11}$ | (Lee 1980; Miller *et al.,* 1982) |
| 16 | $CH_3 + C_2H_3 \rightarrow C_3H_5 + H$ | $2.4\cdot10^{-13}$ | (Romani *et al.,* 1993) |
| 16 | $CH_3 + CH_3 + M \rightarrow C_2H_6 + M$ | $k_0 = 4.0\cdot10^{-24} \times e^{-1390/T} \times (T/300)^{-7.0}$<br>$k_\infty = 1.79\cdot10^{-10} \times e^{-329/T}$ | (Wagner and Wardlaw 1988) |
| 16 | $CH_3 + CO + M \rightarrow CH_3CO + M$ | $1.4 \cdot 10^{-32} \times e^{-3000/T} \times$ den | (Watkins and Word 1974) |
| 16 | $CH_3 + H + M \rightarrow CH_4 + M$ | $k_0 = 1.0 \cdot 10^{-28} \times (T/298)^{-1.80}$ $k_\infty = 2.0 \cdot 10^{-10} \times (T/298)^{-0.40}$ | (Baulch *et al.,* 1994; Tsang and Hampson 1986) |
| 16 | $CH_3 + H_2CO \rightarrow CH_4 + HCO$ | $1.60 \cdot 10^{-16} \times e^{899/T} \times (T/298)^{6.10}$ | (Baulch *et al.,* 1994) |
| 16 | $CH_3 + HCO \rightarrow CH_4 + CO$ | $5.0 \cdot 10^{-11}$ | (Tsang and Hampson 1986) |
| 16 | $CH_3 + HNO \rightarrow CH_4 + NO$ | $3.3 \cdot 10^{-12} \times e^{-1000/T}$ | (Choi and Lin 2005) |
| 16 | $CH_3 + O \rightarrow H_2CO + H$ | $1.1 \cdot 10^{-10}$ | (Sander *et al.,* 2006) |
| 16 | $CH_3 + O_2 \rightarrow H_2CO + OH$ | $k_0 = 4.5 \cdot 10^{-31} \times (T/300)^{-3.0}$ $k_\infty = 1.8 \cdot 10^{-12} \times (T/300)^{-1.7}$ | (Sander *et al.,* 2006) |
| 17 | $CH_3 + O_3 \rightarrow H_2CO + HO_2$ | $5.4 \cdot 10^{-12} \times e^{-220/T}$ | (Sander *et al.,* 2006) |
| 17 | $CH_3 + O_3 \rightarrow CH_3O + O_2$ | $5.4 \cdot 10^{-12}\ e^{-220/T}$ | (Sander *et al.,* 2006) |
| 17 | $CH_2^3 + C_2H^3 \rightarrow CH_3 + C_2H_2$ | $3 \cdot 10^{-11}$ | Tsang and Hampson (1986) |
| 17 | $CH_3 + OH \rightarrow CH_3O + H$ | $9.3 \cdot 10^{-11} \times e^{-1606/T} \times (T/298)$ | (Jasper *et al.,* 2007) |
| 17 | $CH_3 + OH \rightarrow CO + H_2 + H_2$ | $6.7 \cdot 10^{-12}$ | |
| 17 | $CH_3C_2H + H + M \rightarrow C_3H_5 + M$ | $k_0 = 8.88 \cdot 10^{-29} \times e^{-1225/T} \times (T/300)^{-2}$ $k_\infty = 9.7 \cdot 10^{-12} \times e^{-1550/T}$ | (Yung *et al.,* 1984) |
| 17 | $CH_3C_2H + H \rightarrow CH_3 + C_2H_2$ | $k_0 = 8.88 \cdot 10^{-29} \times e^{-1225/T} \times (T/300)^{-2}$ $k_\infty = 9.7 \cdot 10^{-12} \times e^{-1550/T}$ | (Whytock *et al.,* 1976) |
| 17 | $CH_3CHO + CH_3 \rightarrow CH_3CO + CH_4$ | $2.8 \cdot 10^{-11} \times e^{-1540/T}$ | (Zahnle 1986) |
| 17 | $CH_3CHO + H \rightarrow CH_3CO + H_2$ | $2.8 \cdot 10^{-11} \times e^{-1540/T}$ | (Zahnle 1986) |
| 17 | $CH_3CHO + O \rightarrow CH_3CO + OH$ | $5.8 \cdot 10^{-13}$ | (Washida 1981) |
| 18 | $CH_3CHO + OH \rightarrow CH_3CO + H_2O$ | $1.6 \cdot 10^{-11}$ | (Niki *et al.,* 1978) |
| 18 | $CH_3CO + CH_3 \rightarrow C_2H_6 + CO$ | $5.4 \cdot 10^{-11}$ | (Adachi *et al.,* 1981) |
| 18 | $CH_3CO + CH_3 \rightarrow CH_4 + CH_2CO$ | $8.6 \cdot 10^{-11}$ | (Adachi *et al.,* 1981) |
| 18 | $CH_3CO + H \rightarrow CH_4 + CO$ | $1 \cdot 10^{-10}$ | (Zahnle 1986) |
| 18 | $CH_3CO + O \rightarrow H_2CO + HCO$ | $5 \cdot 10^{-11}$ | (Zahnle 1986) |
| 18 | $CH_3O + CO \rightarrow CH_3 +$ | $2.6 \cdot 10^{-11} \times e^{-5940/T}$ | (Wen *et al.,* 1989) |
| | $CO_2$ | | |
| 18 | $CH_3O_2 + H \rightarrow CH_4 + O_2$ | $1.4\cdot10^{-11}$ | (Tsang and Hampson 1986) |
| 18 | $CH_3O_2 + H \rightarrow H_2O + H_2CO$ | $1\cdot10^{-11}$ | (Zahnle *et al.,* 2006) |
| 18 | $CH_3O + NO \rightarrow HNO + H_2CO$ | $2.3\cdot10^{-12} \times (300/T)^{0.7}$ | IUPAC datasheet |
| 18 | $NO_2 + CH_3O \rightarrow H_2CO + HNO_2$ | $9.6\cdot10^{-12}\ e^{-1150/T}$ | IUPAC datasheet |
| 19 | $CH_3O_2 + O \rightarrow H_2CO + HO_2$ | $1\cdot10^{-11}$ | (Vaghjiani and Ravishankara 1990) |
| 19 | $CH_3O_2 + NO \rightarrow CH_3O + NO_2$ | $2.8\cdot10^{-12}\ e^{-300/T}$ | (Sander *et al.,* 2006) |
| 19 | $CH_4 + HS \rightarrow CH_3 + H_2S$ | $2.99\cdot10^{-31}$ | (Kerr and Trotman-Dickenson 1957) |
| 19 | $CH_4 + O \rightarrow CH_3 + OH$ | $8.75\cdot10^{-12} \times e^{-4330/T} \times (T/298)^{1.5}$ | (Tsang and Hampson 1986) |
| 19 | $CH_4 + O^1D \rightarrow CH_3 + OH$ | $1.125\cdot10^{-10}$ | (Sander *et al.,* 2006) |
| 19 | $CH_4 + O^1D \rightarrow H_2CO + H_2$ | $7.5\cdot10^{-12}$ | (Sander *et al.,* 2006) |
| 19 | $CH_4 + O^1D \rightarrow CH_3O + H$ | $3.0\cdot10^{-11}$ | (Sander *et al.,* 2006) |
| 19 | $CH_4 + OH \rightarrow CH_3 + H_2O$ | $2.45\cdot10^{-12} \times e^{-1775/T}$ | (Sander *et al.,* 2006) |
| 19 | $CO + O + M \rightarrow CO_2 + M$ | $2.2\cdot10^{-33} \times e^{-1780/T} \times$ den | (Tsang and Hampson 1986) |
| 19 | $CO + OH \rightarrow CO_2 + H$ | $1.5\cdot10^{-13} \times (1 + 0.6 \times$ den) | (Sander *et al.,* 2006) |
| 20 | $CO + O^1D \rightarrow CO + O$ | $7.0\cdot10^{-11}$ | (Sander *et al.,* 2006) |
| 20 | $H + CO + M \rightarrow HCO + M$ | $1.4\cdot10^{-34} \times e^{-100/T} \times$ den | (Baulch *et al.,* 1994) |
| 20 | $H + H + M \rightarrow H_2 + M$ | $8.85\cdot10^{-33} \times (T/298)^{-0.6} \times$ den | (Baulch *et al.,* 1994) |
| 20 | $H + HCO \rightarrow H_2 + CO$ | $1.8\cdot10^{-10}$ | (Baulch *et al.,* 1992) |
| 20 | $H + HNO \rightarrow H_2 + NO$ | $3.01\cdot10^{-11} \times e^{500/T}$ | (Tsang and Herron 1991) |
| 20 | $H + HO_2 \rightarrow H_2 + O_2$ | $7.2\cdot10^{-12}$ | (Sander *et al.,* 2006) |
| 20 | $H + HO_2 \rightarrow H_2O + O$ | $1.60\cdot10^{-12}$ | (Sander *et al.,* 2006) |
| 20 | $H + HO_2 \rightarrow OH + OH$ | $7.12\cdot10^{-11}$ | (Sander *et al.,* 2006) |
| 20 | $H + NO + M \rightarrow HNO + M$ | $2.1\cdot10^{-32} \times (T/298)^{1.00} \times$ den | (Hampson and Garvin 1977) |
| 20 | $H + O_2 + M \rightarrow HO_2 + M$ | $5.7\cdot10^{-32} \times 7.5\cdot10^{-11} \times (T/298)^{1.6}$ | (Sander *et al.,* 2006) |
| 21 | $H + O_3 \rightarrow OH + O_2$ | $1.4\cdot10^{-10} \times e^{-470/T}$ | (Sander *et al.,* 2006) |
| 21 | $H + OH + M \rightarrow H_2O + M$ | $6.8\cdot10^{-31} \times (T/300)^{-2} \times$ den | (McEwan and Phillips 1975) |
| 21 | $H + SO + M \rightarrow HSO + M$ | $k_0 = 5.7\cdot10^{-32} \times (T/298)^{1.6}$ $k_\infty = 7.5\cdot10^{-11}$ | (Kasting 1990) |
| 21 | $H_2 + O \rightarrow OH + H$ | $1.34\cdot10^{-15} \times e^{-1460/T} \times (T/298)^{6.52}$ | (Robie *et al.,* 1990) |
| 21 | $H_2 + O^1D \rightarrow OH + H$ | $1.1\cdot10^{-10}$ | (Sander *et al.,* 2006) |
| 21 | $H_2 + OH \rightarrow H_2O + H$ | $5.5\cdot10^{-12} \times e^{-2000/T}$ | (Sander *et al.,* 2006) |
| 21 | $H_2CO + H \rightarrow H_2 + HCO$ | $2.14^{-12} \times e^{-1090/T} \times (T/298)^{1.62}$ | (Baulch *et al.,* 1994) |
| 21 | $H_2CO + O \rightarrow HCO + OH$ | $3.4\cdot10^{-11} \times e^{-1600/T}$ | (Sander *et al.,* 2006) |
| 21 | $H_2CO + OH \rightarrow H_2O + HCO$ | $5.5\cdot10^{-12} \times e^{125/T}$ | (Sander *et al.,* 2006) |
| 21 | $H_2O + O^1D \rightarrow OH + OH$ | $2.2\cdot10^{-10}$ | (Sander *et al.,* 2006) |
| 22 | $H_2O_2 + O \rightarrow OH + HO_2$ | $1.4\cdot10^{-12} \times e^{-2000/T}$ | (Sander *et al.,* 2006) |
| 22 | $H_2O_2 + OH \rightarrow HO_2 + H_2O$ | $2.9\cdot10^{-12} \times e^{-160/T}$ | (Sander *et al.,* 2006) |
| 22 | $H_2S + H \rightarrow H_2 + HS$ | $3.66\cdot10^{-12} \times e^{-455/T} \times (T/298)^{1.94}$ | (Peng *et al.,* 1999) |
| 22 | $H_2S + O \rightarrow OH + HS$ | $9.2\cdot10^{-12} \times e^{-1800/T}$ | (Sander *et al.,* 2006) |
| 22 | $H_2S + OH \rightarrow H_2O + HS$ | $6.0\cdot10^{-12} \times e^{-70/T}$ | (Sander *et al.,* 2006) |
| 22 | $HCO + H_2CO \rightarrow CH_3O + CO$ | $3.8\cdot10^{-17}$ | (Wen *et al.,* 1989) |
| 22 | $HCO + HCO \rightarrow H_2CO + CO$ | $4.5\cdot10^{-11}$ | (Tsang and Hampson 1986) |
| 22 | $HCO + NO \rightarrow HNO + CO$ | $1.3\cdot10^{-11}$ | (Tsang and Hampson 1986) |
| 22 | $HCO + O_2 \rightarrow HO_2 + CO$ | $5.2\cdot10^{-12}$ | (Sander *et al.,* 2006) |
| 22 | $HNO_2 + OH \rightarrow H_2O + NO_2$ | $1.8\cdot10^{-11} \times e^{-390/T}$ | (Sander *et al.,* 2006) |
| 23 | $HNO_3 + OH \rightarrow H_2O + NO_2 + O$ | $7.2\cdot10^{-15} \times e^{-785/T} + (1.9\cdot10^{-33} \times e^{725/T} \times den)/(1 + 4.6\cdot10^{-16} \times e^{-715/T} \times den)$ | (Sander *et al.,* 2006) |
| 23 | $HO_2 + HO_2 \rightarrow H_2O_2 + O_2$ | $k_0 = 2.3\cdot10^{-13} \times e^{590/T}$<br>$k_\infty = 1.7\cdot10^{-33} \times e^{1000/T}$ | (Sander *et al.,* 2006) |
| 23 | $HO_2 + O \rightarrow OH + O_2$ | $3.0\cdot10^{-11} \times e^{200/T}$ | (Sander *et al.,* 2006) |
| 23 | $HO_2 + O_3 \rightarrow OH + O_2 + O_2$ | $1.0\cdot10^{-14} \times e^{-490/T}$ | (Sander *et al.,* 2006) |
| 23 | $HO_2 + NO_2 \rightarrow HNO_2 + O_2$ | $5.0\cdot10^{-16}$ | (Sander *et al.,* 2006) |
| 23 | $HS + H \rightarrow H_2 + S$ | $2.0\cdot10^{-11}$ | (Schofield 1973) |
| 23 | $HS + HCO \rightarrow H_2S + CO$ | $2.0\cdot10^{-11}$ | (Kasting 1990) |
| 23 | $HS + HO_2 \rightarrow H_2S + O_2$ | $1.0\cdot10^{-11}$ | (Stachnik and Molina 1987) |
| 23 | $HS + HS \rightarrow H_2S + S$ | $2.0^{-11}$ | (Schofield 1973) |
| 23 | $HS + NO_2 \rightarrow HSO + NO$ | $2.9\cdot10^{-11} \times e^{240/T}$ | (Sander *et al.,* 2006) |
| 24 | $HS + O \rightarrow H + SO$ | $7.0\cdot10^{-11}$ | (Sander *et al.,* 2006) |
| 24 | $HS + O_3 \rightarrow HSO + O_2$ | $9.0\cdot10^{-12} \times e^{-280/T}$ | (Sander *et al.,* 2006) |
| 24 | $HS + S \rightarrow H + S_2$ | $1.0\cdot10^{-11}$ | (Kasting 1990) |
| 24 | $HSO + H \rightarrow H_2 + SO$ | $1.0 \cdot 10^{-11}$ | (Kasting 1990) |
| 24 | $HSO + H \rightarrow HS + OH$ | $2.0 \cdot 10^{-11}$ | (Kasting 1990) |
| 24 | $HSO + HS \rightarrow H_2S + SO$ | $3.0 \cdot 10^{-12}$ | (Kasting 1990) |
| 24 | $HSO + O \rightarrow OH + SO$ | $3.0 \cdot 10^{-11}$ | (Kasting 1990) |
| 24 | $HSO + OH \rightarrow H_2O + SO$ | $3.0 \cdot 10^{-11}$ | (Kasting 1990) |
| 24 | $HSO + S \rightarrow HS + SO$ | $1.0 \cdot 10^{-11}$ | (Kasting 1990) |
| 24 | $HSO_3 + O_2 \rightarrow HO_2 + SO_3$ | $1.3 \cdot 10^{-12} \times e^{-330/T}$ | (Sander *et al.,* 2006) |
| 25 | $N + NO \rightarrow N_2 + O$ | $2.1 \cdot 10^{-11} \times e^{-100/T}$ | (Sander *et al.,* 2006) |
| 25 | $N + O_2 \rightarrow NO + O$ | $1.5 \cdot 10^{-12} \times e^{-3600/T}$ | (Sander *et al.,* 2006) |
| 25 | $N + OH \rightarrow NO + H$ | $3.8 \cdot 10^{-11} \times e^{85/T}$ | (Atkinson *et al.,* 1989) |
| 25 | $N + HO_2 \rightarrow NO + OH$ | $2.2 \cdot 10^{-11}$ | (Brune et al. 1983) |
| 25 | $NO + HO_2 \rightarrow NO_2 + OH$ | $3.5 \cdot 10^{-12} \times e^{250/T}$ | (Sander *et al.,* 2006) |
| 25 | $NO + O + M \rightarrow NO_2 + M$ | $9 \cdot 10^{-31} 3. \cdot 10^{-11} \times (T/298)^{1.5}$ | (Sander *et al.,* 2006) |
| 25 | $NO + O_3 \rightarrow NO_2 + O_2$ | $2.0 \cdot 10^{-12} \times e^{-1500/T}$ | (Sander *et al.,* 2006) |
| 25 | $NO + OH + M \rightarrow HNO_2 + M$ | $k_0 = 7 \cdot 10^{-31} \times (T/298)^{2.6}$ $k_\infty = 3.6 \cdot 10^{-11} \times (T/298)^{0.1}$ | (Sander *et al.,* 2006) |
| 25 | $NO_2 + H \rightarrow NO + OH$ | $4. \cdot 10^{-10} \times e^{-340/T}$ | (Sander *et al.,* 2006) |
| 25 | $NO_2 + O \rightarrow NO + O_2$ | $5.6 \cdot 10^{-12} \times e^{180/T}$ | (Sander *et al.,* 2006) |
| 26 | $NO_2 + OH + M \rightarrow HNO_3 + M$ | $k_0 = 2.0 \cdot 10^{-30} \times (T/298)^{3.0}$ $k_\infty = 2.5 \cdot 10^{-11}$ | (Sander *et al.,* 2006) |
| 26 | $O + HCO \rightarrow H + CO_2$ | $5.0 \cdot 10^{-11}$ | (Tsang and Hampson 1986) |
| 26 | $O + HCO \rightarrow OH + CO$ | $1.0 \cdot 10^{-10}$ | (Hampson and Garvin 1977) |
| 26 | $O + HNO \rightarrow OH + NO$ | $3.8 \cdot 10^{-11}$ | (Tsang and Hampson 1986) |
| 26 | $O + O + M \rightarrow O_2 + M$ | $9.46 \cdot 10^{-34} \times e^{480/T} \times$ den | (Campbell and Gray 1973) |
| 26 | $O + O_2 + M \rightarrow O_3 + M$ | $6 \cdot 10^{-34} \times 3 \cdot 10^{-11} \times (T/298)^{2.40}$ | (Sander *et al.,* 2006) |
| 26 | $O + O_3 \rightarrow O_2 + O_2$ | $8.0 \cdot 10^{-12} \times e^{-2060/T}$ | (Sander *et al.,* 2006) |
| 26 | $O^1D + M \rightarrow O + M$ | $1.8 \cdot 10^{-11} \times e^{110/T}$ | (Sander *et al.,* 2006) |
| 26 | $O^1D + O_2 \rightarrow O + O_2$ | $3.2 \cdot 10^{-11} \times e^{70/T}$ | (Sander *et al.,* 2006) |
| 26 | $OH + HCO \rightarrow H_2O + CO$ | $1.0 \cdot 10^{-10}$ | (Baulch *et al.,* 1992) |
| 27 | $OH + HNO \rightarrow H_2O + NO$ | $5 \cdot 10^{-11}$ | (Sun *et al.,* 2001) |
| 27 | $OH + HO_2 \rightarrow H_2O + O_2$ | $4.8 \cdot 10^{-11} \times e^{250/T}$ | (Sander *et al.,* 2006) |
| 27 | $OH + O \rightarrow H + O_2$ | $2.2 \cdot 10^{-11} \times e^{120/T}$ | (Sander *et al.,* 2006) |
| 27 | $OH + O_3 \rightarrow HO_2 + O_2$ | $1.6 \cdot 10^{-12} \times e^{-940/T}$ | (Sander *et al.,* 2006) |
| 27 | $OH + OH \rightarrow H_2O + O$ | $4.2 \cdot 10^{-12} \times e^{-240/T}$ | (Sander *et al.,* 2006) |
| 27 | $OH + OH \rightarrow H_2O_2$ | $6.9\cdot10^{-31} \times 2.6\cdot10^{-11} \times (T/298)^{1.00}$ | (Sander *et al.,* 2006) |
| 27 | $S + HCO \rightarrow HS + CO$ | $1.0\cdot10^{-11}$ | (Kasting 1990) |
| 27 | $S + HO_2 \rightarrow HS + O_2$ | $5.0\cdot10^{-12}$ | (Kasting 1990) |
| 27 | $S + HO_2 \rightarrow SO + OH$ | $5.0\cdot10^{-12}$ | (Kasting 1990) |
| 27 | $S + O_2 \rightarrow SO + O$ | $2.3\cdot10^{-12}$ | (Sander *et al.,* 2006) |
| 28 | $S + O_3 \rightarrow SO + O_2$ | $1.2\cdot10^{-11}$ | (Sander *et al.,* 2006) |
| 28 | $S + OH \rightarrow SO + H$ | $6.6\cdot10^{-11}$ | (Sander *et al.,* 2006) |
| 28 | $S + S + M \rightarrow S_2 + M$ | $1.98\cdot10^{-33} \times e^{-206/T} \times$ den | (Du *et al.,* 2008) |
| 28 | $S + S_2 + M \rightarrow S_3 + M$ | $2.8\cdot10^{-32} \times$ den | (Kasting 1990) |
| 28 | $S + S_3 + M \rightarrow S_4 + M$ | $2.8\cdot10^{-31} \times$ den | (Kasting 1990) |
| 28 | $S_2 + O \rightarrow S + SO$ | $1.1\cdot10^{-11}$ | (Hills *et al.,* 1987) |
| 28 | $S_2 + S_2 + M \rightarrow S_4 + M$ | $2.8\cdot10^{-31} \times$ den | (Baulch *et al.,* 1976) |
| 28 | $S_4 + S_4 + M \rightarrow S_8AER + M$ | $2.8\cdot10^{-31} \times$ den | (Kasting 1990) |
| 28 | $SO + HCO \rightarrow HSO + CO$ | $5.6\cdot10^{-12} \times (T/298)^{-0.4}$ | (Kasting 1990) |
| 28 | $SO + NO_2 \rightarrow SO_2 + NO$ | $1.4\cdot10^{-11}$ | (Sander *et al.,* 2006) |
| 29 | $SO + O + M \rightarrow SO_2 + M$ | $5.1\cdot10^{-31} \times$ den | (Sander *et al.,* 2006) |
| 29 | $SO + O_2 \rightarrow O + SO_2$ | $2.6\cdot10^{-13} \times e^{-2400/T}$ | (Sander *et al.,* 2006) |
| 29 | $SO + O_3 \rightarrow SO_2 + O_2$ | $4.5\cdot10^{-12} \times e^{-1170/T}$ | (Atkinson *et al.,* 2004) |
| 29 | $SO + OH \rightarrow SO_2 + H$ | $8.6\cdot10^{-11}$ | (Sander *et al.,* 2006) |
| 29 | $SO + SO \rightarrow SO_2 + S$ | $3.5\cdot10^{-15}$ | (Martinez and Herron 1983) |
| 29 | $SO_2 + HO_2 \rightarrow SO_3 + OH$ | $8.63\cdot10^{-16}$ | (Lloyd 1974) |
| 29 | $SO_2 + O + M \rightarrow SO_3 + M$ | $k_0 = 1.3\cdot10^{-33} \times (T/298)^{-3.6}$ $k_\infty = 1.5\cdot10^{-11}$ | (Sander *et al.,* 2006) |
| 29 | $SO_2 + OH + M \rightarrow HSO_3 + M$ | $k_0 = 3\cdot10^{-31} \times (T/298)^{3.3}$ $k_\infty = 1.5\cdot10^{-12}$ | (Sander *et al.,* 2006) |
| 29 | $SO_2^1 + O_2 \rightarrow SO_3 + O$ | $1.0\cdot10^{-16}$ | (Turco *et al.,* 1982) |
| 29 | $SO_2^1 + SO_2 \rightarrow SO_3 + SO$ | $4.0\cdot10^{-12}$ | (Turco *et al.,* 1982) |
| 30 | $SO_2^3 + SO_2 \rightarrow SO_3 + SO$ | $7.0\cdot10^{-14}$ | (Turco *et al.,* 1982) |
| 30 | $SO_3 + H_2O \rightarrow H_2SO_4$ | $1.2\cdot10^{-15}$ | (Sander *et al.,* 2006) |
| 30 | $SO_3 + SO \rightarrow SO_2 + SO_2$ | $2.0\cdot10^{-15}$ | (Chung *et al.,* 1975) |
| 30 | $SO_2^1 + h\nu \rightarrow SO_2 + h\nu$ | $0.0\cdot10^{0}$ | (Turco *et al.,* 1982) |
| 30 | $SO_2^1 + h\nu \rightarrow SO_2^3 + h\nu$ | $0.0\cdot10^{0}$ | (Turco *et al.,* 1982) |
| 30 | $SO_2^3 + h\nu \rightarrow SO_2 + h\nu$ | $0.0\cdot10^{0}$ | (Turco *et al.,* 1982) |
| 30 | $O_2 + h\nu \rightarrow O + O^1D$ | $2.38\cdot10^{-06}$ | |
| 30 | $O_2 + h\nu \rightarrow O + O$ | $4.77\cdot10^{-08}$ | |

| | | |
|---|---|---|
| 30 | $H_2O + h\nu \rightarrow H + OH$ | $8.25 \cdot 10^{-06}$ |
| 30 | $O_3 + h\nu \rightarrow O_2 + O^1D$ | $2.47 \cdot 10^{-03}$ |
| 31 | $O_3 + h\nu \rightarrow O_2 + O$ | $7.37 \cdot 10^{-04}$ |
| 31 | $H_2O_2 + h\nu \rightarrow OH + OH$ | $3.65 \cdot 10^{-05}$ |
| 31 | $CO_2 + h\nu \rightarrow CO + O$ | $1.00 \cdot 10^{-09}$ |
| 31 | $H_2CO + h\nu \rightarrow H_2 + CO$ | $2.51 \cdot 10^{-05}$ |
| 31 | $H_2CO + h\nu \rightarrow HCO + H$ | $2.86 \cdot 10^{-05}$ |
| 31 | $CO_2 + h\nu \rightarrow CO + O^1D$ | $2.90 \cdot 10^{-07}$ |
| 31 | $HO_2 + h\nu \rightarrow OH + O$ | $2.17 \cdot 10^{-04}$ |
| 31 | $CH_4 + h\nu \rightarrow CH_2^1 + H_2$ | $2.08 \cdot 10^{-06}$ |
| 31 | $C_2H_6 + h\nu \rightarrow CH_4 + CH_2^1$ | $1.34 \cdot 10^{-06}$ |
| 31 | $HNO_2 + h\nu \rightarrow NO + OH$ | $1.58 \cdot 10^{-09}$ |
| 32 | $HNO_3 + h\nu \rightarrow NO_2 + OH$ | $7.40 \cdot 10^{-05}$ |
| 32 | $HNO + h\nu \rightarrow NO + N$ | $7.0 \cdot 10^{-04}$ |
| 32 | $HCO + h\nu \rightarrow H + CO$ | $1.0 \cdot 10^{-02}$ |
| 32 | $NO + h\nu \rightarrow N + O$ | $1.92 \cdot 10^{-06}$ |
| 32 | $NO_2 + h\nu \rightarrow NO + O$ | $3.23 \cdot 10^{-03}$ |
| 32 | $CH_3 + h\nu \rightarrow CH_2^1 + H$ | $1.64 \cdot 10^{-01}$ |
| 32 | $SO + h\nu \rightarrow S + O$ | $1.65 \cdot 10^{-04}$ |
| 32 | $SO_2 + h\nu \rightarrow SO + O$ | $7.27 \cdot 10^{-05}$ |
| 32 | $H_2S + h\nu \rightarrow HS + H$ | $1.02 \cdot 10^{-04}$ |
| 32 | $SO_2 + h\nu \rightarrow SO_2^1$ | $7.14 \cdot 10^{-04}$ |
| 33 | $SO_2 + h\nu \rightarrow SO_2^3$ | $4.94 \cdot 10^{-07}$ |
| 33 | $S_2 + h\nu \rightarrow S + S$ | $4.56 \cdot 10^{-04}$ |
| 33 | $SO_3 + h\nu \rightarrow SO_2 + O$ | $1.57 \cdot 10^{-05}$ |
| 33 | $SO_2^1 + h\nu \rightarrow SO_2^3 + h\nu$ | $0.00 \cdot 10^{0}$ |
| 33 | $SO_2^1 + h\nu \rightarrow SO_2 + h\nu$ | $0.00 \cdot 10^{0}$ |
| 33 | $SO_2^3 + h\nu \rightarrow SO_2 + h\nu$ | $0.00 \cdot 10^{0}$ |
| 33 | $HSO + h\nu \rightarrow HS + O$ | $2.17 \cdot 10^{-04}$ |
| 33 | $S_4 + h\nu \rightarrow S_2 + S_2$ | $4.56 \cdot 10^{-04}$ |
| 33 | $S_3 + h\nu \rightarrow S_2 + S$ | $4.45 \cdot 10^{-04}$ |
| 33 | $C_2H_2 + h\nu \rightarrow C_2H + H$ | $1.02 \cdot 10^{-06}$ |
| 34 | $C_2H_2 + h\nu \rightarrow C_2 + H_2$ | $4.65 \cdot 10^{-07}$ |
| 34 | $C_2H_4 + h\nu \rightarrow C_2H_2 + H_2$ | $1.60 \cdot 10^{-05}$ |
| 34 | $C_3H_8 + h\nu \rightarrow C_3H_6 + H_2$ | $0.00 \cdot 10^{-00}$ |
| 34 | $C_3H_8 + h\nu \rightarrow C_2H_6 + CH_2^1$ | $1.43 \cdot 10^{-06}$ |
| 34 | $C_3H_8 + h\nu \rightarrow C_2H_4 +$ | $6.98 \cdot 10^{-06}$ |
| | $CH_4$ | |
| 34 | $C_3H_8 + h\nu \rightarrow C_2H_5 + CH_3$ | $3.69 \cdot 10^{-06}$ |
| 34 | $C_2H_6 + h\nu \rightarrow C_2H_2 + H_2 + H_2$ | $1.46 \cdot 10^{-06}$ |
| 34 | $C_2H_6 + h\nu \rightarrow C_2H_4 + H + H$ | $1.67 \cdot 10^{-06}$ |
| 34 | $C_2H_6 + h\nu \rightarrow C_2H_4 + H_2$ | $9.15 \cdot 10^{-07}$ |
| 34 | $C_2H_6 + h\nu \rightarrow CH_3 + CH_3$ | $4.31 \cdot 10^{-07}$ |
| 35 | $C_2H_4 + h\nu \rightarrow C_2H_2 + H + H$ | $1.67 \cdot 10^{-05}$ |
| 35 | $C_3H_6 + h\nu \rightarrow C_2H_2 + CH_3 + H$ | $1.07 \cdot 10^{-05}$ |
| 35 | $CH_4 + h\nu \rightarrow CH_2^3 + H + H$ | $3.94 \cdot 10^{-06}$ |
| 35 | $CH_4 + h\nu \rightarrow CH_3 + H$ | $1.93 \cdot 10^{-06}$ |
| 35 | $CH + h\nu \rightarrow C + H$ | $3.27 \cdot 10^{-05}$ |
| 35 | $CH_2CO + h\nu \rightarrow CH_2^3 + CO$ | $1.53 \cdot 10^{-04}$ |
| 35 | $CH_3CHO + h\nu \rightarrow CH_3 + HCO$ | $3.25 \cdot 10^{-05}$ |
| 35 | $CH_3CHO + h\nu \rightarrow CH_4 + CO$ | $3.25 \cdot 10^{-05}$ |
| 35 | $C_2H_5CHO + h\nu \rightarrow C_2H_5 + HCO$ | $7.77 \cdot 10^{-05}$ |
| 35 | $C_3H_3 + h\nu \rightarrow C_3H_2 + H$ | $7.16 \cdot 10^{-04}$ |
| 36 | $CH_3C_2H + h\nu \rightarrow C_3H_3 + H$ | $1.75 \cdot 10^{-05}$ |
| 36 | $CH_3C_2H + h\nu \rightarrow C_3H_2 + H_2$ | $6.57 \cdot 10^{-06}$ |
| 36 | $CH_3C_2H + h\nu \rightarrow CH_3 + C_2H$ | $8.75 \cdot 10^{-07}$ |
| 36 | $CH_2CCH_2 + h\nu \rightarrow C_3H_3 + H$ | $1.91 \cdot 10^{-11}$ |
| 36 | $CH_2CCH_2 + h\nu \rightarrow C_3H_2 + H_2$ | $7.16 \cdot 10^{-12}$ |
| 36 | $CH_2CCH_2 + h\nu \rightarrow C_2H_2 + CH_2^3$ | $2.87 \cdot 10^{-12}$ |
| 36 | $C_3H_6 + h\nu \rightarrow CH_2CCH_2 + H_2$ | $1.80 \cdot 10^{-05}$ |
| 36 | $C_3H_6 + h\nu \rightarrow C_2H_4 + CH_2^3$ | $6.30 \cdot 10^{-07}$ |
| 36 | $C_3H_6 + h\nu \rightarrow C_2H + CH_4 + H$ | $1.58 \cdot 10^{-06}$ |
| 36 | $OCS + h\nu \rightarrow CO + S$ | $8.71 \cdot 10^{-06}$ |
| 37 | $CS_2 + h\nu \rightarrow CS + S$ | $9.33 \cdot 10^{-04}$ |
| 37 | $CS_2 + h\nu \rightarrow CS_2^*$ | $9.71 \cdot 10^{-05}$ |

Supplemental Table 2. Atmospheric species in the Archean photochemical code with lower boundary condition type and values. Lower boundary conditions are given in cm/s for deposition velocity (Vdep), a dimensionless mixing ratio by volume for fixed concentration ($f_o$), and molecules/cm2/s for flux (flux). Species names ending in "AER" are types of aerosols.

| Species | Lower Boundary Type | Vdep/$f_0$/flux |
|---|---|---|
| | Long-Lived Species | |
| O | constant deposition velocity | 1 |
| $O_2$ | constant mixing ratio | $1{\cdot}10^{-08}$ |
| $H_2O$ | constant deposition velocity | 0 |
| H | constant deposition velocity | 1 |
| OH | constant deposition velocity | 1 |
| $HO_2$ | constant deposition velocity | 1 |
| $H_2O_2$ | constant deposition velocity | $2{\cdot}10^{-01}$ |
| $H_2$ | constant deposition velocity* | $2.4{\cdot}10^{-04}$ |
| CO | constant deposition velocity | $1.2{\cdot}10^{-04}$ |
| HCO | constant deposition velocity | 1 |
| $H_2CO$ | constant deposition velocity | $2{\cdot}10^{-01}$ |
| $CH_4$ | constant mixing ratio | variable† |
| $CH_3$ | constant deposition velocity | 1 |
| $C_2H_6$ | constant deposition velocity | 0 |
| NO | constant deposition velocity | $3{\cdot}10^{-04}$ |
| $NO_2$ | constant deposition velocity | $3{\cdot}10^{-03}$ |
| HNO | constant deposition velocity | 1 |
| $O_3$ | constant deposition velocity | $7{\cdot}10^{-02}$ |
| $HNO_3$ | constant deposition velocity | $2{\cdot}10^{-01}$ |
| N | constant deposition velocity | 0 |
| $H_2S$ | constant deposition velocity* | $2{\cdot}10^{-02}$ |
| HS | constant deposition velocity | 0 |
| S | constant deposition velocity | 0 |
| SO | constant deposition velocity | 0 |
| $SO_2$ | constant deposition velocity* | 1 |
| $SO_3$ | constant deposition velocity | 0 |
| $H_2SO_4$ | constant deposition velocity | 1 |
| HSO | constant deposition velocity | 1 |
| $S_2$ | constant deposition velocity | 0 |
| $C_2$ | constant deposition velocity | 0 |
| CH | constant deposition velocity | 0 |
| $C_2H$ | constant deposition velocity | 0 |
| $CH_2^3$ | constant deposition velocity | 0 |
| $C_2H_5$ | constant deposition velocity | 0 |
| $C_2H_2$ | constant deposition velocity | 0 |
| $C_2H_4$ | constant deposition velocity | 0 |
| $C_3H_8$ | constant deposition velocity | 0 |
| $C_3H_7$ | constant deposition velocity | 0 |
| $C_3H_5$ | constant deposition velocity | 0 |
| $C_2H_3$ | constant deposition velocity | 0 |
| $C_3H_6$ | constant deposition velocity | 0 |
| $C_3H_2$ | constant deposition velocity | 0 |
| $C_3H_3$ | constant deposition velocity | 0 |
| $CH_2CCH_2$ | constant deposition velocity | 0 |
| $CH_2CO$ | constant deposition velocity | 0 |
| $CH_3CO$ | constant deposition velocity | 0 |
| $CH_3CHO$ | constant deposition velocity | 0 |
| $CH_3O$ | constant deposition velocity | 0 |
| $CH_3O_2$ | constant deposition velocity | 0 |
| $C_2H_4OH$ | constant deposition velocity | 0 |
| $C_2H_2OH$ | constant deposition velocity | 0 |
| $C_2H_5CHO$ | constant deposition velocity | 0 |
| $CH_3C_2H$ | constant deposition velocity | 0 |
| $CS_2$ | constant deposition velocity | 0 |
| HCS | constant deposition velocity | 0 |
| OCS | constant deposition velocity | 0 |
| CS | constant deposition velocity | 0 |
| $SO_4$AER | constant deposition velocity | $1 \cdot 10^{-02}$ |
| $S_8$AER | constant deposition velocity | $1 \cdot 10^{-02}$ |
| HCAER | constant deposition velocity | $1 \cdot 10^{-02}$ |
| HCAER2 | constant deposition velocity | $1 \cdot 10^{-02}$ |
| | Short-Lived Species | |
| $HNO_2$ | constant deposition velocity | 0 |
| $O^1D$ | constant deposition velocity | 0 |
| $CH_2^1$ | constant deposition velocity | 0 |
| C | constant deposition velocity | 0 |
| $SO_2^1$ | constant deposition velocity | 0 |
| $SO_2^3$ | constant deposition velocity | 0 |
| $HSO_3$ | constant deposition velocity | 0 |
| $OCS_2$ | constant deposition velocity | 0 |
| $CS_2$* | constant deposition velocity | 0 |
| $S_3$ | constant deposition velocity | 0 |
| $S_4$ | constant deposition velocity | 0 |
| | Inert Species | |
| $CO_2$ | constant mixing ratio | variable† |
| $N_2$ | constant mixing ratio | remainder‡ |

* - In addition to a constant deposition velocity, we also use a volcanic flux for these gases. Specifically, we used volcanic fluxes of $3.5 \cdot 10^9$ molecules/cm$^2$/s of $H_2$, $1 \cdot 10^{10}$ molecules/cm$^2$/s of $SO_2$, and $3.5 \cdot 10^8$ molecules/cm$^2$/s of $H_2S$.
† - See text for information on these mixing ratios.
‡ - $N_2$ fills the remainder of the atmosphere